# Quantum Error Correcting Codes and the Security Proof of the BB84 Protocol

Ramesh Bhandari

Laboratory for Telecommunication Sciences

8080 Greenmead Drive, College Park, Maryland 20740, USA

rbhandari617@gmail.com

(Dated: December 2011)

We describe the popular BB84 protocol and critically examine its security proof as presented by Shor and Preskill. The proof requires the use of quantum error-correcting codes called the Calderbank-Shor-Steanne (CSS) quantum codes. These quantum codes are constructed in the quantum domain from two suitable classical linear codes, one used to correct for bit-flip errors and the other for phase-flip errors. Consequently, as a prelude to the security proof, the report reviews the essential properties of linear codes, especially the concept of cosets, before building the quantum codes that are utilized in the proof. The proof considers a security entanglement-based protocol, which is subsequently reduced to a "Prepare and Measure" protocol similar in structure to the BB84 protocol, thus establishing the security of the BB84 protocol. The proof, however, is not without assumptions, which are also enumerated. The treatment throughout is pedagogical, and this report, therefore, serves as a useful tutorial for researchers, practitioners and students, new to the field of quantum information science, in particular quantum cryptography, as it develops the proof in a systematic manner, starting from the properties of linear codes, and then advancing to the quantum error-correcting codes, which are critical to the understanding of the security proof.

# Contents

# Contents (continued)

# 1  Introduction

## 1.1 Motivation for Quantum Key Distribution (QKD)

Quantum Key Distribution (QKD), an application of quantum cryptography, refers to the generation of a cryptographic key with unconditional security guaranteed by the laws of physics.  The key is generated by two parties, Alice and Bob, using quantum signals, called quantum bits, or simply qubits.  Any attempt to obtain information on the key by an eavesdropper (say, Eve) leads to a disruption of the quantum signal, leading to errors and thus Eve's detection.  Consideration of such consequences of the laws of physics then helps to establish a quantum key in a secure manner. Once the (quantum) key is established in a secure manner, it can then be used in encrypting data to be transmitted.

The encryption scheme employed by Alice in transmitting data is called the one-time pad (also known as Vernam cipher) [3]. In one-time pad encryption, Alice and Bob begin with identical and private $n$-bit secret bit strings (keys), Alice then encodes her $n$-bit message by adding the key and message together, and sends the resulting $n$-bit string to Bob, who then decodes by subtracting the identical $n$-bit secret key he is in possession of, to obtain the message Alice sent. Note that the length of the message that is combined with the key is as long as the length of the key, which is $n$-bits here.  The key used is then discarded (used one-time only), a requirement of the Vernam cipher[1], and a new key generated for new data to be encrypted. Cryptosystems employing the one-time pad encryption scheme are called *private cryptosystems*.

In *private key cyptosystems*, a major requirement is ensuring the key at both parties (Alice and Bob) are not just identical, but completely private.  In the current  non-QKD environment, in order for Bob to have the same key as Alice, one may use trusted couriers to deliver to Bob a copy of a random key generated by Alice or one may have it electronically transmitted to Bob on  private secured communication links, but the possibility that someone has copied the key material en route to Bob or it has been compromised while in safekeeping before its use, or someone has  tapped the private classical communication link to gain information on the key material (if transmitted electronically) without anyone knowing about it, cannot be completely eliminated.  QKD provides the hope that the key generated and shared between Alice and  Bob will not only be random, but completely private, with no one else having any knowledge of the contents (except Alice and Bob), as its generation (which is simultaneous at both ends as opposed to generation at one end and its delivery at the other end by current classical means)  is protected by the laws of physics and any attempt to gain information on it results in a detectable disturbance of the quantum signal, indicating eavesdropper's presence. Protected by the laws of physics, key generation can occur over public channels, and in a continuous manner, allowing Alice for continual encryption of the data as it comes in and its near simultaneous decryption at the other end by Bob (the recipient) who is in possession of the same shared key. This is to be contrasted with the existing situation in private cryptosystems where each time encrypted message has to be sent, one has to ensure in advance the availability of an identical, private (and different random) key at the other end.

---

[1] If the same key were to be used more than once, Eve, the eavesdropper, could record all of the encrypted messages and start to build a picture of the plain texts and thus also of the key.

Another motivation for QKD stems from the discovery that a quantum computer can factor large numbers in polynomial time. The factorization of large numbers and certain other mathematical problems (like the discrete log problem) are considered hard to solve on classical computers, and for this reason are employed in a class of cryptographic schemes [15] (comprising *public key cryptography*) upon which *public key cryptosystems* are based. Such cryptosystems, deployed since the late 1970's, are popularly and widely used currently as an alternative to the private key cryptosystems because they eliminate the inconvenience of delivering a private key physically (or electronically) from one end to the other. Although realizing a practical quantum computer is somewhat faraway, clearly the fact it could be built one day then threatens the security of these public key cryptosystems, which then gives QKD further boost as being the means for secure key generation for *real* secure communication in the future.

The first QKD protocol was presented by Charles H. Bennett of IBM and Gilles Brassard of University of Montreal in 1984 and is popularly called the BB84 protocol. Ever since then, QKD has expanded into an active area of research, both theoretical and practical. Several new protocols along with their security proofs have since being given [3]. But the BB84 protocol remains the most popular protocol used within practical QKD systems.

In this report, we describe the BB84 protocol and focus on its security proof. There are several proofs for the BB84 protocol [3]. The most widely cited security proof, however, is by Shor and Preskill [4]. This proof involves the use of quantum error-correcting codes, called the Calderbank-Shor-Steane (CSS) codes [6,7]. Based on the properties of the linear codes, these quantum code allow multiple, arbitrary quantum errors such as bit-flip, phase-flip, etc. to be corrected. Basically, Shor and Preskill start with an existing secure entanglement-based protocol [5] and perform judicious reduction to arrive at a version of the BB84 protocol, which is very similar to the standard version. The standard BB84 version is then proven secure by virtue of this reduction and its similarity to the Shor-Preskill version.

## 1.2 Organization of the Report

The report is organized as follows: in Chapter 2, we start with a general QKD setting, discuss the features of the QKD protocol such as the BB84 protocol and give a detailed description of the standard BB84 protocol. In Section 3.2 of the next chapter, we describe linear codes, which are the basic building blocks of the CSS quantum codes used in the BB84 proof. Subsequently, in Section 3.3, we describe in detail the quantum error correcting codes, starting from the single error correcting codes, and progressing to the more practically useful quantum error-correcting codes, called the CSS codes, which are capable of correcting more than a single error. Equipped with these tools, we describe in detail in Section 3.4 the reduction of the secure entanglement based protocol of Lo and Chau [5] to the BB84 protocol. Chapter 4 is the summary and conclusion.

# 2  QKD and the BB84 Protocol

## 2.1 The General QKD Setting

As mentioned earlier, the QKD key is not generated at one end (say by Alice) and then transported to the other end, where Bob, the receiver, is located. Rather, the key is created simultaneously at both ends by generation and transmission of appropriate quantum signals called the qubits (by Alice), and their measurements by Bob, followed by analysis of information shared by both parties. The qubits traverse a public channel, which we call the quantum channel (see Figure 1). The qubits may be photons, in which case the quantum channel physically may be a fiber or simply free-space. To generate the key, Alice and Bob communicate over another public channel to share information on qubit preparation, transmission, and measurement; we call this channel a classical channel, which physically could be in the same fiber as the quantum channel or another fiber connecting Alice and Bob, or simply in free-space. This QKD setting is shown in Figure 1.

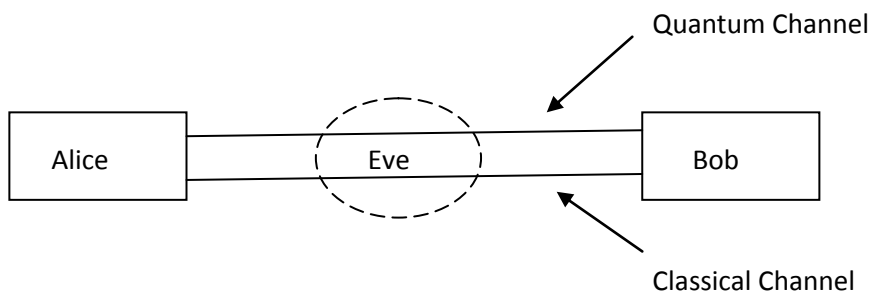

**Figure 1**  Alice sends Bob  qubits over the quantum channel; Alice and Bob communicate with each other over the classical channel regarding Alice's qubit  state preparation and  Bob's measurement results; Eve is an eavesdropper located between Alice and Bob.

In this scenario, Alice prepares qubits (representative of the classical key bits)  in certain quantum states and sends them to Bob over the quantum channel, which Eve (an eavesdropper) can tap, restricted only by the laws of physics.  For example, Eve can perform an <u>Intercept and Resend</u> attack in the hope of gaining information on the key bits; in this attack (described in more detail below), she intercepts qubits sent by Alice, makes a measurement of each intercepted qubit, makes a copy of the measured qubit, which she then resends to Bob, but her interference results in detectable errors, unraveling her presence, which is in sharp contrast to electronic communication, where any tapping can go undetected. It is important to mention here also that the quantum channel is in general noisy (see Chapter 3.3) and is also a source of errors in qubits.

Along with the quantum channel, on which qubits are transmitted, Alice and Bob are also connected by an authenticated classical channel to communicate during the key generation process. On this classical channel, Eve can only listen to classical messages exchanges between Alice and Bob; she cannot alter or send messages pretending to be Alice or Bob.

Below we describe the basics of private random key generation within the framework of QKD.

## 2.2 Basic Concept of Key Generation in QKD

We initially assume Eve is not present and the quantum channel is noiseless, i.e., the qubits traveling on this channel undergo no errors. Thus, the qubits received by Bob are the same qubits that Alice sent.

Suppose now Alice creates a string of random classical bits; the first bit may be a 0 or 1 (with equal probability); the second bit may be a 0 or 1(with equal probability), and so on. Suppose this string of random 0's and 1's is of length $n$. Suppose further Alice encodes these (random) classical bits into quantum bits called qubits. For example, bit 0 will get encoded into qubit $|0>$, which is a quantum state; similarly, the classical bit 1 gets encoded into qubit $|1>$, which is a quantum state orthogonal to $|0>$ (the most general state of a qubit is $a|0> + b|1>$, where the arbitrary coefficients $a$ and $b$ are complex, see Appendix A). Encoding the classical bits 0 and 1 as $|0>$ and $|1>$, respectively, is equivalent to preparing the qubits corresponding to bits 0 and 1in the (0, 1) basis; this basis is also sometimes called the $Z$ basis. Alice transmits to Bob each qubit prepared in the (0, 1) basis in accordance with the random classical bits of the $n$-bit string.  Bob also measures the received qubits in the (0, 1) basis, which means, if he received $|0>$, he will measure $|0>$, which would be decoded to classical bit 0, and similarly, if he received $|1>$, he will measure it as $|1>$, which would be decoded to classical bit 1. Thus, Bob obtains the same classical bit string, which serves as the shared private, random key.

### Presence of Eve

 We now assume that Eve is present and is aware that Alice is sending qubits prepared in the (0, 1) basis. If she intercepts the qubits sent by Alice and measures them in the (0, 1) basis, she will obtain exactly the same information  as Bob, on the classical bits sent by Alice, i.e., $|0>$ sent by Alice will be measured as $|0>$ and decoded to classical bit 0, and similarly, the qubit $|1>$ sent by Alice will be measured as $|1>$ and decoded to bit 1.  After intercepting and making a measurement on each qubit, Eve makes a copy of the measured qubit, and forwards it to Bob (this is the heart of the Intercept and Resend attack). Bob receives the replicated qubits, which are the same qubits that Alice had prepared and transmitted over the quantum channel. After measurement in the (0, 1) basis and decoding, Bob obtains the key. But both Alice and Bob have no way of knowing that Eve had intercepted the qubits and has complete information on the key as well.

### Detecting Eve's Presence

To detect Eve's presence, Alice now encodes bits 0 and 1 not just in the (0, 1) basis but also in the (+, -) basis (also called the $X$  basis), i.e., bits 0 and 1 are also encoded into the states $|+>$ and $|->$, respectively, where $|+> = 1/\sqrt{2}(|0> + |1>)$ and $|-> = 1/\sqrt{2}(|0> - |1>)$; $|+>$ and $|->$ are obtained from $|0>$ and $|1>$, respectively,  via Hadamard transform (see Appendix B). These states also form an orthonormal basis. The $Z$ and the $X$ bases are chosen randomly by Alice.

Suppose Alice sends in bit 0 encoded as |+>. Eve, who intercepts this qubit, does not know whether it was prepared in the (0, 1) basis or the (+, -) basis. So she randomly selects one of the two bases ((0, 1) and (+, -)) to make her measurement in. There is a 50% chance she selects the correct basis, which is the (+, -) basis. Suppose she measures in the correct basis; the result is |+>, which she decodes correctly to bit 0; she resends |+> to Bob as part of her Intercept and Resend attack. Bob also does not know *a priori* which basis Alice prepared the qubit in, so he also randomly selects the (0, 1) basis or the (+,-) basis for the measurement. Suppose Bob correctly chooses (+,-). Then he also obtains the result |+>, which is decoded correctly to the classical bit 0, the bit Alice sent encoded as |+>. Thus, when Eve chooses the right basis (and Bob also chooses the right basis), Bob measures exactly what Alice sent without any hint of Eve's presence, i.e., Eve is still invisible to Alice and Bob (in the Intercept and Resend attack).

Suppose Eve measures |+> (the state prepared by Alice) in the wrong basis, i.e., basis (0, 1). There are two possible (equally probable) outcomes of the measurement: |0> and |1>. Consequently, she will measure |0> or |1>[2]. Regardless of the outcome, she makes a copy (|0> or |1>), which she resends to Bob. We assume, as before, that Bob selects the same basis for his measurement as Alice prepared the qubit in, which is the (+, -) basis. Now $|0> = 1/\sqrt{2}(|+> + |->)$ and $|1> = 1/\sqrt{2}(|+> - |->)$. Thus, there is a 50% chance that Bob measures the qubit sent by Eve as |->, which he decodes to 1 (which is different from the bit 0 Alice encoded and sent). *Thus, Eve's interference causes an error, which is due to the fact that Eve measured in the wrong basis*.

How is the assumption that Bob measures a received qubit in the same basis as the basis Alice prepared the qubit in, implemented within a QKD protocol? After Bob has received all the qubits, *n* in number, he announces this fact to Alice on the classical channel. Alice shares with Bob the information on which bases she prepared the qubits in (not their bit values), and Bob also communicates to Alice which bases he measured the qubits in. They quietly throw away the bits, for which the preparation and measurement bases were different, i.e., these bits are no longer part of key generation process. Thus, for the remaining bits, the bases Bob measured in and Alice prepared in are identical, i.e., in the absence of errors, the keys Alice and Bob generate from the remaining bits are identical.

*Important Remark*: Because Bob is equally likely to select the wrong basis as the right basis, the expectation is that roughly half of the bits in Alice's random *n*-bit string will be thrown away after communication over the classical channel. The bits remaining are said to constitute a **sifted** key. So, to have a sifted key which is at least *n*-bits long, Alice actually selects a random string, which is slightly more than 2*n* bits long. This initial selection of the random string is called the **raw** key.

How many errors does Eve's interference in the Intercept and Resend attack cause? Recalling from above that errors due to Eve's interference are caused only when she measures in the wrong basis, the **probability that Eve causes an error =** probability that Eve measures in the wrong basis (= 1/2) times the probability that Bob's measurement of the qubit sent by Eve yields the incorrect bit value, given Eve measured in the wrong basis (=1/2), which equals 1/2 x 1/2 = **1/4.**   A 25% error rate is high enough to deduce that privacy has been compromised and the key bits cannot be trusted anymore and should be discarded completely.   In practice, to be on the safe side, the error threshold for aborting the key generation protocol is set much lower. If the total measured error rate (which includes errors due to quantum noise also) is below a certain threshold, then only is error-correction performed with the expectation that the final keys Alice and Bob have would be identical (see Section 2.3); all measured errors are ascribed to Eve's presence as there is no way of distinguishing the error caused by the quantum channel and the error caused by Eve's interference.

---

[2] Note that Eve's interference (measurement) here has changed the quantum signal from |+> to |0> or |1>; this disturbance of the quantum signal is in accordance with quantum physics and is the central basis of the QKD protocol.

To ascertain the error-rate, the size of the raw key is further doubled to more than 4$n$, so that the sifted key is of length slightly greater than 2$n$; subsequently half of this sifted key's bits are used as check bits to determine the error-rate (see Section 2.3); if the error-rate is acceptable, the approximately $n$ remaining bits are subjected to **error-correction** (also called **information reconciliation**).

## Eve's information

In the Intercept and Resend attack, Eve's interference causes 25% of errors in Bob's measurement results. How much bit information has she obtained during this attack? The answer is very simple. When she chooses a basis for her measurement after intercepting the qubit, the probability she guesses the right basis is 1/2, in which case she has full information on the qubit. On the other hand, if she chooses the wrong basis, her measurement yields either 0 or 1 with equal probability, in which case she has no information (see also Renyi information in Ref. [15]).   In other words, the probability she learns a bit value in her measurement is 1/2; she either learns about a bit value with certainty or not at all.

Additionally, Eve can also acquire information related to key bits from the public channel, e.g., during error correction performed by Alice and Bob classically using the classical channel. She can acquire information by mounting other attacks as well [3, 9].   After estimating Eve's information from all possible sources/attacks, Alice and Bob subject the resulting key bits after error-correction to a process called **privacy amplification** to obtain a random, private key of length $k$ (<$n$).   The process of privacy amplification reduces the information that Eve is estimated to have, the reduction being larger, smaller the value of $k$ (see Section 2.3).

## No Cloning

At this point, one might ask the question: to gain full information on all the key bits, why cannot Eve simply make copies of the incoming qubits, keep the copies with herself, while forwarding the original qubits to Bob, and measure her copies only after she learns from the communication between Alice and Bob which sifted bits they are going to keep? This way she learns the entire key[3] without causing errors. Unfortunately, this type of an attack by Eve does not succeed as it violates the No-Cloning Theorem of physics [10], which essentially states that a quantum state from a given set of states can be cloned only if the states within the set are mutually orthogonal. But in the above scheme, Alice is preparing and sending randomly four states: |0>, |1>, |+>, |->, which are not mutually orthogonal. Thus, a single

---

[3] We are assuming no errors are caused by the quantum channel itself.

mechanism that can replicate any of the above four states is not possible. Eve will have to have a separate cloning mechanism for the |0> and |1> states (which are mutually orthogonal)[4] and a separate cloning mechanism for the other pair of mutually of orthogonal states: |+> and |->. Because Alice prepares the qubit state randomly in the (0, 1) basis and the (+, -) basis, Eve does not know which cloning mechanism to apply when she intercepts a qubit; the situation is akin to the Intercept and Resend attack case where Eve does not know which basis Alice prepared the qubit states in. Thus, this attack fails!

In the next section, we summarize the major generic steps in the quantum-key distribution process.

## 2.3 Major Steps in QKD

Based on our analysis and discussion above, we describe the major steps that comprise a QKD protocol such as the BB84 protocol:

1) **Raw Key Generation**

    Here Alice generates $(4 + \delta)$ $n$ random bits[5], encodes each one of them (0 or 1) as a qubit in the (0,1) basis or the (+, -) basis, selected randomly. She then transmits these $(4 + \delta)$ $n$ qubits to Bob on the quantum channel.

    Bob measures the received qubits in one of the two bases, the (0, 1) basis and the (+, -) basis, also selected randomly by him. He obtains a binary string of length $(4 + \delta)$ $n$.

2) **Sifted Key**

    There is a 50% chance that the basis Bob selects randomly to measure the received qubit is not the same as the one Alice prepared the qubit in; if Bob's basis is not the same as Alice's basis, then there is the potential of error in Bob's decoded bit. Therefore, in order to generate identical keys, Alice and Bob want to keep only those bits in their bit strings for which Alice's preparation basis and Bob's measurement basis are identical. Consequently, they share the information on their bases (not the bit values) on the available classical channel, and discard those bits from their bit strings for which the bases were not identical.

    Since for every qubit received, the probability of selecting the incorrect basis is 1/2, and bits corresponding to incorrect bases are discarded, Alice and Bob are left with a bit string that is roughly half the size of the raw key. This is sometimes called the **sifted** key.

---

[4] If we know that the state is either |0> or |1>, then cloning is accomplished by simply measuring the state in the (0, 1) basis, and replicating the result (|0> or |1>), and similarly, when the state is |+> or |->.
[5] $\delta$ (a small fraction less than 1) is added to ensure that at least $2n$ bits are left after key sifting.

### 3) Error Correction (or Information Reconciliation)

Even though bits corresponding to different bases are discarded to obtain the sifted key, Bob's bit string will have errors due to noise in the quantum channel and Eve's interference. Since there is no way of knowing which errors were caused by the quantum channel noise and which by Eve's interference, Alice and Bob take the conservative approach of ascribing all errors to Eve's interference. They have decided that, if the errors are large, meaning Eve has interfered too much, they would abort the key generation process.

So Alice randomly selects $n$ bits from the slightly greater than $2n$ bits she has left (after sifting) to serve as check bits, i.e., bits to be used to check for errors. Subsequently, she announces to Bob which $n$ bits she selected as check bits and their values. Bob then compares the bit values that he measured for the $n$ check bits selected by Alice and announces the bits in error. If the errors exceed a certain agreed-upon limit, they abort the protocol. Otherwise, they proceed with the error-correction process, discarding first the $n$ check bits and performing error correction on a randomly set of $n$ bits selected from the remaining slightly more than $n$ bits. They use the classical channel for error-correction, using appropriate error-correcting algorithms. After error-correction, Bob is left with an identical string[6].

### 4) Eve's Information

Eve, all this while during the key generation process, has gleaned information on the key bits through her attempts to interfere on the quantum channel and by listening on the classical channel, especially during the error-correction process.  Note also that, in addition to the Intercept and Resend attack, Eve can carry out a number of other types of attacks [3,9] such as 1) side-channel attacks, where information leaks out of devices that Eve then collects 2) probing attacks, where information is collected by attacking Alice's and Bob's equipment directly such as probing the settings of Alice's and Bob's devices by sending some light into them and collecting the reflected signal, and so on.

From all possible Eve's attacks, including her listening on the classical channel, Alice and Bob estimate the information that Eve might have obtained. This information is quantified as *Renyi* information, which is measured in bits [15]. If Eve is said to have $t$ bits of information, it does not necessarily mean that she has information on $t$ specific key bits; rather, she has knowledge, which is the equivalent of $t$ bits of the key.

---

[6] How much identical Bob's string is will depend upon the extent of error-correction; because of the random nature of errors, error-correction, especially when $n$ is large, can never be guaranteed to be 100% complete without spending an enormous time in computing; e.g., a popularly referenced error-correcting algorithm, called the CASCADE algorithm [22], employs an iterative procedure that can be very time-consuming, depending upon the required accuracy of error correction; clearly, slow error correction slows down the time-rate of key generation in practical QKD systems.

**5) Privacy Amplification**

The $n$-bit long strings Alice and Bob have after error-correction are subjected to *privacy amplification* to reduce Eve's information. Privacy amplification is generally accomplished through the use of suitable hashing functions [23, 10]. In the use of such functions, the input is the $n$-bit string and the output is a $k$ ($<n$) bit string. This hashing then reduces the information Eve might have. The greater the amount of information Eve has, smaller the value of $k$ you want.

Specifically, if the string is $n$-bits long after error-correction, and Eve has estimated $t$ bits of information, then in order to achieve security, Alice and Bob will have to discard at least $t$ bits from the $n$ bits. The more bits they can discard in excess of $t$ bits, the more secret their final shared key string would be.

For example, if Alice and Bob can shed a further $s$ bits from their $n$-$t$ bit long string, then according to a theorem by Bennett et al. [21], Eve's information about the $k$-bit string ($k = n$-$t$-$s$) is bounded by $2^{-s}/\ln 2$, i.e., Eve's information shrinks exponentially with increasing $s$; each additional bit Alice and Bob can sacrifice in their final key reduces the upper bound by a factor of 2.

The above 5 steps can be rewritten into an expanded form, leading to the BB84 protocol [1], named after the inventors.

## 2.4 The BB84 Protocol

This protocol can be stated as follows:

1. Alice generates a bit string $d$ of (4+$\delta$) $n$ random bits[7].
2. Alice chooses another random-bit string $b$ of length (4+$\delta$) $n$. For each bit of the string $d$, she creates a qubit in the $Z$ basis or the $X$ basis according to the bit values of the random string, $b$.
3. Alice sends the resulting (4+$\delta$) $n$ qubits to Bob (one at a time).
4. Bob receives the (4+$\delta$) $n$ qubits, publicly announces this fact, and measures each in the $Z$ or $X$ basis at random.
5. Alice announces the string $b$ that determined the basis she used to encode each bit in.
6. Bob discards the measured qubit values obtained if he measured in a basis different from the one that Alice prepared in. He tells Alice which measurements (but not their results) he discarded. Alice then discards the same set of bits.
   - With high probability, there are at least $2n$ bits left; if not, abort the protocol.
7. Alice randomly selects $2n$ bits from the remaining ($\geq 2n$) bits and announces which $2n$ bits she selected (but not their values).

---

[7] The fraction $\delta$ is added to ensure that after discarding bits in Step 6, at least $2n$ bits are left.

8. Alice randomly selects $n$ of the $2n$ bits to use as check bits, and announces this selection of the $n$ bits and their bit values.

9. Bob compares the bit values he measured for the $n$ check bits selected by Alice and announces the bits where they disagree.  If more than an acceptable number of these check bit values disagree, they abort the protocol.

10. Alice now has an $n$ bit string $x$, and Bob has an $n$-bit string $x + e_1$, where $e_1$ is the error caused by Eve's interference and/or channel noise.

11. Alice and Bob perform information reconciliation, i.e., error correction whereby Bob's string is corrected to $x$.

12.  Alice and Bob further perform privacy amplification on their $n$-bit strings to obtain $k$ shared key bits.

In the next Chapter, we provide a proof of the above protocol; as the proof involves the use of quantum error-correcting codes, the focus is first on describing these codes.

# 3  Quantum Error Correcting Codes and the BB84 Security Proof

## 3.1 Introduction

The BB84 protocol was presented in 1984 [1].  A proof of the security of the BB84 protocol was first given by Meyers [2] in 1996.  Additional proofs [3], including the one by Shor and Preskill [4], have since been presented.  The proof by Shor and Preskill is the most popular and widely cited proof because of its relative simplicity. Therefore, in this chapter, we present this proof for the security of the BB84 protocol. This proof is based on the reduction of an entanglement based QKD protocol of Lo and Chau [5] to a "Prepare and Measure" BB84 protocol.  As a result, the security proof of the Lo-Chau protocol translates to the security proof of the derived BB84 protocol. The reduction involves the use of a quantum error-correcting code, called the Calderbank-Shor-Steane (CSS) code [6, 7].  This quantum code is constructed from two classical linear codes applied in a quantum setting. Based on the error-correcting properties of the linear codes, the derived quantum code allows multiple, arbitrary errors such as bit-flip, phase-flip, etc. to be corrected.

Starting from the entanglement-based QKD protocol of Lo and Chau, Shor and Preskill first replace entanglement distillation using quantum computers [5, 8] with entanglement distillation based on the CSS codes [4];  entanglement distillation refers to the extraction of high-fidelity  entangled pairs of qubits ($k$ in number) from the original set of  $n$ pairs of maximally entangled pairs, which have been corrupted due to Eve's interference and/or channel noise;  the key rate obtained is then $k/n$.

Shor and Preskill subsequently reduce the ensuing protocol, called the Modified Lo-Chau protocol, to one devoid of the requirement of entangled pairs of qubits.  They then perform another reduction in which quantum decoding is replaced with classical decoding, employing the classical linear codes that comprised the CSS code. In their final reduced form, which is the "Prepare and Measure" BB84 protocol, one linear code performs error correction, and the second one (in conjunction with the first one) performs privacy amplification to reduce the amount of information Eve might have obtained during the establishment of the private, random key.  Except for minor differences, the Shor-Preskill version obtained from the reduction is very similar structurally to the standard version of the BB84 protocol given in Section 2.3. Thus, the Shor-Preskill security proof of the BB84 protocol applies not only to the derived version but also to the original version. Nevertheless, there are assumptions and shortcomings of the security proof, which we point out in Section 3.4.5.

In Section 3.2, we describe the linear codes (the basic ingredients of the quantum CSS code), followed by the CSS code in Section 3.3. Subsequently, starting with the description of the entanglement based protocol in Section 3.4, we show its reduction to the Shor-Preskill version of the BB84 protocol. We then compare the Shor-Preskill version to the standard version, and also cite the assumptions of the security proof.

## 3.2 Classical Linear Codes

In telecommunication and classical coding theory, error-correcting codes (of which linear code is an example) are codes in which redundant data is added to the intended message by the sender to enable error-correction by the recipient, the errors occurring due to noise in the transmission channel; the carefully designed redundancy allows the receiver to detect and correct a limited number of errors occurring anywhere in the message  without the need to ask the sender for any additional information. The combination of the redundant data and the actual message results in an entity called a codeword.

To understand the basic concept behind coding theory and how redundancy facilitates error detection and correction, consider the following scenario:

Suppose the sender (Alice) and the receiver (Bob) agree on a simple code where the 8 English letters *a*, *b*, *c*, *d*, *e*, *f*, *g*, *h* are represented by the 8 three-bit blocks 000,001,010,011, 100, 101, 110, 111, respectively. Suppose further Alice wants to send Bob the message *bad* ; therefore, in accordance with the above coding scheme she sends out the bit string 001000011, which is the encoding of  *bad*. Suppose Bob receives the encoded message as 101000011, i.e., an error has occurred during transmission with the first bit 0 flipping to bit 1. Chopping the received message into 3 blocks: 101, 001, 011, Bob decodes the received message as *fad*, which is different from the transmitted message *bad*. So this is not a good coding scheme!

Alice and Bob at least want Bob to be able to detect an error

In order for Bob to able to detect an error (bit flip: 0 → 1 or 1→ 0), Alice and Bob decide to use a *3-bit string repetition* scheme, repeating each 3-bit block twice. In other words, letters *a*, *b*, *c*, *d*, *e*, *f*, *g*, and *h* are now to be coded as 000000, 001001, 010010, 011011, 100100, 101101, 110110, 111111, respectively (note that in this scheme the first three bits of each 6-bit block, which constitutes a codeword, may be treated as the message bits with the remaining 3 bits then acting as the redundant bits). Accordingly, Alice now sends *bad* encoded as 001001000000011011. Suppose Bob receives the encoded message as 101001000000011011 (error in the first bit). He chops the received codeword into three blocks: 101001, 00000, and 011011. He correctly interprets the second and third blocks (codewords) as *a* and *d*, respectively, and recognizes (correctly also) that an error has occurred in the first block, since it cannot be identified with any of the codewords of the coding scheme.  Unfortunately, he cannot correct the error, since 101001 can be corrected to 101101 (letter *f*) or 001001 (letter *b*) (assuming a single error has occurred).  In this situation, Bob can ask Alice to resend the message, hoping there would be no channel error this time.

Alice and Bob want Bob to be able to correct any single bit error

 This error-correction can be achieved if Alice and Bob agree on a coding scheme in which each 3-bit block is repeated thrice. The message (*bad*) is now encoded as 001001001000000000011011011 by Alice and sent to Bob. Suppose now Bob receives this codeword as 101001001000000000011011011 (again error in the first bit).  To decode, he breaks this string into blocks of nine, namely, 101001001, 0000000, 011011011, and interprets the second and third blocks correctly as *a* and *d*. But he cannot identify the first block 101001001 with any codeword, so knows that an error has occurred. He then breaks this block into three strings: 101, 001, 001, and using the "best two out of three" approach, declares that the error must have been in the first string 101, since it does not agree with the remaining two identical strings: 001, 001; furthermore, he concludes that the string 101 must have been 001 originally, so he corrects it to 001, and the received word[8] to 001001001, which is the letter ***b*** (the transmitted letter encoded as 001001001). Thus, Bob is able to correct single bit errors, when 9-bit codewords are used.

The above repetition schemes, using 6-bit codewords for single bit error detection and 9-bit codewords for single error-correction, are very inefficient; in the former case, the number of redundant bits is 3, while in the second case, the number of redundant bits is 6. Hamming, an American mathematician working at AT&T Bell Laboratories, invented in 1950 a ground –breaking code, now famously called the *[7, 4]* Hamming code. This code, belonging to the new class of error-correcting codes called the linear codes, could correct single bit errors in messages of length 4 bits with only 3 (=7-4) redundant bits, in sharp contrast to the above repetition code of length 9-bits, which has 6 redundant bits. Hamming essentially pioneered the field of efficient error-correcting codes [11]. Recent breakthroughs such as the introduction of the turbo codes [12] and low density parity check codes [13, 14] have further increased the efficiency of error-correction algorithms.

In this section, we focus on classical linear codes, which are used in the construction of the quantum codes, called the CSS codes [6, 7]; as mentioned earlier, these codes play a critical role in proving the security of the BB84 protocol. In what follows, we systematically describe the properties of the linear codes and their construction, and introduce the concepts of dual code and syndrome measurement that are essential to understanding the classical code error correction process, which then serves as a precursor to the quantum error-correction process described in Section 3.2. The number of errors that a CSS code can correct is dependent upon the choice of linear codes and their error-correcting properties, hence their understanding is vital. We also define and discuss *cosets*, upon whose properties the construction of the CSS codes is based.

### 3.2.1 Definition of a Linear Code

In general, a linear code can be defined over an arbitrary finite field $F_q$ of *q* elements [15]. Since we are dealing with data communication, in the definitions and discussions below, the field $F_q$ will always be the binary field, $F_2$ with elements {0, 1}. Some definitions and properties are:

a) An *[n, k]* linear code *C* is a subspace *C* of dimensionality *k* of the binary vector space $F_2^n$, i.e., $C \subseteq F_2^n$. $F_2^n$ is a field of the two binary elements (0, 1); $F_2^n$ contains $2^n$ *n*-bit elements (or words), where *n* denotes the dimensionality of the vector space $F_2^n$; the codespace *C* contains $2^k$ *n*-bit elements (or codewords); *n* is also called the length of the codeword (or code *C)*; an *n*-bit codeword may also be denoted by an *n*-bit vector, $\vec{c} \in C$.

---

[8] The term *word* is used to indicate a codeword, possibly, with errors.

b) $C$ being a subspace of $F_2^n$ implies the rules of the vector space apply to the subspace $C$ as well [16], i.e.,

    i)      The set $C$ is nonempty
    ii)     For every $\vec{x}$ , $\vec{y}$ ∈ $C$, $\vec{x} + \vec{y}$ ∈ $C$ *(sum of two codewords is a codeword itself)*
    iii)    For scalar $\alpha$ ∈ *{0,1}*, and every $\vec{x}$ ∈ $C$, $\alpha \vec{x}$ ∈ $C$

The code $C$ is closed under addition and scalar multiplication; all arithmetic operations are performed modulo 2. Property iii) implies that the *n*-bit vector (with all bits zero) $\vec{0}$ ∈ $C$.

c) $k$, the dimensionality of code $C$, is the number of *information* (or *message*) bits encoded into a codeword of $n$ bits; there are $2^k$ such possible $k$-bit message strings, and hence $2^k$ codewords in the set $C$. The remaining $n-k$ bits of a codeword are the redundant bits, called *check* bits.

The efficiency of the code, or *code rate, R = k/n*.

d) Hamming weight of a codeword is the number of 1's in the codeword (in what follows, we will use the term *weight* for *Hamming weight*); the minimum weight of code $C$, denoted $d$, is the weight of the codeword with the minimum weight (excluding $\vec{0}$).

e) An *[n, k]* linear code $C$ <u>detects</u> $u = d - 1$ errors and <u>corrects</u> $t = \lfloor (d\text{-}1)/2 \rfloor$ errors [15], where $d$ is the minimum weight of the code, and $\lfloor a \rfloor$ is the floor of $a$, i.e., the largest integer less than or equal to $a$; *<u>by a code C detecting u errors</u>, we mean that whenever u or fewer errors occur during transmission, no matter what codeword Alice sent or where the errors occurred within the codeword, Bob (the receiver) is guaranteed to tell accurately whether or not an error has occurred; similarly, <u>by a code C correcting t errors</u>, we mean that whenever t or fewer errors occur during transmission, no matter what codeword Alice sent or where the errors occurred within the codeword, Bob is guaranteed to accurately correct the errors*.

<u>*Example 1:*</u> Binary Parity Check Code of length $n$ = 4 and $k$ =3

The binary parity check code of length $n$ = 4 and $k$ = 3 is a *[4,3]* linear code, containing $2^3$ (= 8) four-bit elements (or codewords) that comprise the code $C \subseteq F_2^4$:

$$C = \{0000, 0011, 0101, 0110, 1001, 1010, 1100, 1111\} \qquad (3.2.1)$$

$F_2^4$ has a total of $2^4$ (=16) elements. All the 8 elements of $C$ have even umber of 1's. The other 8 elements of $F_2^4$ that are not in $C$ are 0001, 0010, 0100, 1000, 1101, 1011, 1110, and 0111, each with odd # of 1's.

Note that $\vec{0}$ ≡ 0000 is an element of $C$ in Eq. (3.2.1), and the sum of any two elements of $C$ is an element of $C$, e.g., 0101 + 0110 = 0011 (modulo 2).

<u>*k*= 3</u> implies a set of 8 three-bit message vectors: {000, 001, 010, 011, 100, 101, 110, 111}, which are then encoded into the 8 four-bit codewords comprising $C$ above, e.g., 001 message is encoded into the codeword 0011, 011 message is encoded into the codeword 0110, and so on. The fourth bit here is the check bit added to make the weight (or parity) of the generated codeword even.

The weight of the codeword 0011 in Eq. (3.2.1) is equal to 2, since there are two 1's in 0011. In fact, for our example, the weight of all codewords in $C$ is equal to 2. Consequently, the minimum weight $d$ of the code is equal to 2. From property e) above, this implies that the number of errors $u$ the code can detect = is equal to 2 -1 = 1 and the number of errors $t$ it can correct is equal to 0.

If, during transmission, a codeword in $C$ in Eq. (3.2.1) undergoes a single error, i.e., a single bit flip (0 → 1 or 1 → 0) anywhere within the codeword, then the parity of the codeword changes from *even* to *odd*. Odd parity codewords do not belong to $C$, so a single error is detectable. For example, 0110 becomes 0111 upon the 4th bit flipping. Thus, a single bit-flip is detectable at Bob's (receiver's) end because 0111 does not belong to the codespace $C$ (see Eq. (3.2.1)). Two bit flips anywhere in a codeword keep the parity of the codeword *even*, so the received word belongs to $C$; thus two errors are not detectable. For example, if the first and the fourth bits flip in 0110, the received word is $1111 \in C$, so Bob, the receiver, does not detect an error, thinking that the original transmitted codeword is indeed 1111, when actually it was 0110. Note that even though three-bit flips, i.e., three errors (which lead to odd weight codewords), are similarly detectable, we say that the above code detects a single error because it fails for two errors.

The code rate $R = k/n = $ ¾ = 0.75, which is to be contrasted with the code rate of 3/6 = 0.5 for the repetition code used to detect a single bit error (see Section 3.2).

*Example 2:* A [7,4] Hamming Code ($n$ = 7 and $k$ =4)

$n$ = 7 *and* $k$ = 4 imply $2^4$ = 16 codewords, each 7-bit long [15]:

$C$ = {0000000, 0001011, 0010101, 0011110, 0100111, 0101100, 0110010, 0111001, 1000110, 1001101, 1010011, 1011000, 1100001, 1101010, 1110100, 1111111}.                         (3.2.2)

The message bits are the first 4 bits in each codeword, while the check bits are the remaining 7 - 4 = 3 bits, implying a code rate $R$ = 4/7 = 0.57.

The minimum distance $d$ (from Eq. 2) = 3. $u$, the maximum number of errors this code can detect, is equal to $d$ -1 = 2. Suppose in transmission, the codeword 0001011 changes to 1001011, i.e., the first bit flips. The error in the received word 1001011 is detectable because it is not in $C$ (see Eq. (3.2.2)). Similarly, if the codeword 0001011 changes to 1001010 during transmission, i.e., the first and the seventh bits flip (two errors), the receiver (Bob) can conclude that the received word 1001010 has errors, because it does not belong to $C$ in Eq. (3.2.2). Suppose now that, in addition to the first and the seventh bit-flips, there is a bit-flip in the second position, i.e., the received word is 1101010. This is a valid codeword in $C$, so Bob, the receiver, does not detect any error in the received word, even though 3 errors occurred. One can verify (if one so desires) that for any single bit-flip or any pair of bit flips in any codeword in $C$ (Eq. (3.2.2)), the received word is not in $C$. This is not true when three bit-flips (3 errors) occur (for example, the codeword 1110100 $\in C$ in Eq. (3.2.2) changes to 1111111 upon bit-flips in the $4^{th}$, $6^{th}$, and $7^{th}$ positions, but this word received by Bob is a valid codeword belonging to *C*; so the three-bit flip error is not detectable.

From property e) above, the number of errors *t* this code can correct is equal to 1. How the error is located and corrected will be described in Section 3.2.5.

### 3.2.2. Generating a Linear Code

To a given *[n,k]* linear code corresponds a *k* x *n* generating matrix *G*. In this section, we show how these codes are generated from the generator matrix *G*.

Let *G* be a *k x n* matrix. Let its *k* rows (each of *n* elements), denoted $\vec{x}_i$, *i = 1,2,...,k* be linearly independent, i.e., $\sum_{i=1}^{k} \gamma_i \vec{x}_i = \vec{0}$ only when $\gamma_i = 0$ for all *i =1,2,...k*. Then

$$C = RS(G) = \{\alpha_1 \vec{x}_1 + \alpha_2 \vec{x}_2 + \cdots + \alpha_k \vec{x}_k \,|\, \alpha_i \in \{0,1\}\} \subseteq F_2^n \qquad (3.2.3)$$

is the set of codewords generated by the *G* matrix; *RS(G)* denotes the row space of the matrix *G*, which is of rank *k*. Because $\alpha_i$ *(i = 1, 2,..., k)* can take on either value, 0 or 1, there are a total of $2^k$ linear combinations. Since $\vec{x}_i's$ are linearly independent, each different linear combination gives rise to a unique codeword. Thus, the matrix *G* generates a set of $2^k$ (unique) codewords, one of which is always $\vec{0}$ (corresponding to $\alpha_i = 0 \,\forall\, i$). Note that each of the *k* row vectors $\vec{x}_i$ itself constitutes a codeword; for example, the codeword $\vec{x}_j$ $(0 < j \le k)$ is a codeword generated from Eq. (3.2.3) by assigning $\alpha_i = 0 \,\forall\, i$, except *i = j*, for which $\alpha_j = 1$. The remaining codewords, $2^k - k$ -1, are generated when two of the $\alpha_i's$ in Eq. (3.2.3) are non-zero (leading to ${}^k\mathbb{C}_2$ possibilities or codewords), three of the $\alpha_i's$ in Eq. (3.2.3) are non-zero (leading to ${}^k\mathbb{C}_3$ possibilities or codewords), and so on, consistent with ${}^k\mathbb{C}_0$ *(=1)* + ${}^k\mathbb{C}_1$ (=k) + ${}^k\mathbb{C}_2$ + ${}^k\mathbb{C}_3$ +......+ ${}^k\mathbb{C}_k = 2^k$, where ${}^k\mathbb{C}_m$ denotes the number of combinations of *m* objects selected from *k* different objects and equals *k!/(m!(k-m)!)*.

Basically, a random assignment of 0 or 1 to each $\alpha_i$ *(i = 1,2,...,k)* in Eq. (3.2.3) constitutes a random *k*-bit string, which generates a codeword belonging to *C*. Calling this *k*-bit random string a "message" vector $\vec{m}$, the corresponding generated codeword $\vec{c}$ can then be written as

$$\vec{c} = \vec{m}\, G. \qquad (3.2.4)$$

In other words, the matrix *G* generates the codeword $\vec{c}$, which is an <u>*n*-bit encoding</u> of the *k*-bit "message" vector $\vec{m}$.

*Important note:* Because of the linear independence of the row vectors of the *G* matrix, different, equally valid forms of *G* are possible by manipulating its rows, e.g., interchanging rows, adding one or two row vectors to a third row vector, and so on. For example, from Eq. (3.2.3), it is evident that *C* remains unaltered upon interchange of $\vec{x}_1$ and $\vec{x}_2$; similarly, $\vec{x}_2$ replaced with $\vec{x}_1 + \vec{x}_2$ leaves *C* unaltered, noting that $\vec{x}_1 + \vec{x}_1 = \vec{0}$; in general, $\vec{x} + \vec{x} = \vec{0} \,\forall\, \vec{x} \in F_2^n$.

<u>*Example 1*</u>: For the binary parity check code of length *n* = 4 and *k* = 3 [15],

$$G = \begin{bmatrix} 1 & 0 & 0 & 1 \\ 0 & 1 & 0 & 1 \\ 0 & 0 & 1 & 1 \end{bmatrix} \qquad \text{(3 x 4 matrix)} \qquad (3.2.5)$$

Eq. (3.2.3) yields

$$C = \{\alpha_1(1001) + \alpha_2(0101) + \alpha_3(0011)| \ \alpha_i = 0 \text{ or } 1\} \subseteq F_2^4 \qquad (3.2.6)$$

$$= \{0000, 1001, 0101, 0011, 1100, 1010, 0110, 1111\},$$

which is the same as Eq. (3.2.1). One can verify that the three row vectors, $\vec{x}_1 = 1001$, $\vec{x}_2 = 0101$, and $\vec{x}_3 = 0011$ of the $G$ matrix above are linearly independent. Each one of these vectors is a codeword itself. The various linear combinations of these three vectors, $\vec{x}_1 + \vec{x}_2$, $\vec{x}_2 + \vec{x}_3$, $\vec{x}_3 + \vec{x}_1$, $\vec{x}_1 + \vec{x}_2 + \vec{x}_3$ yield the remaining 4 codewords, which are in addition to the codeword $\vec{0}$ ($\alpha_1 = \alpha_2 = \alpha_3 = 0$).

Suppose $\vec{m} = 011$ (one of the 8 three-bit vectors, representing the "message" bits). Then $\vec{c} = \vec{m} \, G$ gives $\vec{c} = 0110$, the encoding of $\vec{m} = 011$ by the generator $G$ (Eq. (3.2.5)) of the binary parity check code $C$.

*Example 2:* For the *[7,4]* Hamming Code (*n* = 7 and *k* =4) [15],

$$G = \begin{bmatrix} 1 & 0 & 0 & 0 & 1 & 1 & 0 \\ 0 & 1 & 0 & 0 & 1 & 1 & 1 \\ 0 & 0 & 1 & 0 & 1 & 0 & 1 \\ 0 & 0 & 0 & 1 & 0 & 1 & 1 \end{bmatrix} \qquad (3.2.7)$$

Using Eq. (3.2.3),

$$C = \{\alpha_1(1000110) + \alpha_2(0100111) + \alpha_3(0010101) + \alpha_4(0001011) \mid \alpha_i = 0 \text{ or } 1\} \subseteq F_2^7 , \qquad (3.2.8)$$

which gives the codespace $C$ in Eq. (3.2.2).

*Further Remarks*: Note from Eq. (3.2.3) one can generate different *[n,k]* linear codes $C$ by using different sets of $k$ linearly independent row vectors $\vec{x}_i$ that comprise the $k$ x $n$ generator matrix $G$; for example, if the $\vec{x}_i$'s in Eq. (3.2.6) are chosen to be $\vec{x}_1 = 1000$, $\vec{x}_2 = 0100$, $\vec{x}_3 = 0001$, then the generated linear code is {0000, 1000, 0100, 0001, 1100, 0101, 1001, 1101}. Clearly, the minimum weight $d = 1 \Rightarrow u = d$ -1 = 0 and $t$ = 0. This *[4,3]* linear code has a different value of $d$, and therefore different error detection and correction properties. Because linear codes characterized by the same pair of parameters, $n$ and $k$ can have different values of $d$, an *[n,k]* linear code is sometimes also written as an *[n,k,d]* linear code. Singleton [17] in 1964 provided a bound, which places an upper limit on the value of $d$ for any *[n,k]* linear code:

If $C$ is an *[n, k]* linear code, then

$$d \leq n - k + 1 \qquad \text{(Singleton Bound)} \qquad (3.2.9)$$

Basically, no matter how hard one tries, one cannot construct an *[n, k]* linear code with minimum distance $d$ greater than $n - k + 1$. This is important, since the number of errors $t$ an *[n, k]* linear code can correct gets bounded from Eq. (3.2.9) through the relationship: $t = \lfloor (d-1)/2 \rfloor$. Therefore, $t \leq \lfloor (n-k+1-1)/2 \rfloor = \lfloor (n-k)/2 \rfloor$.

Note that both of the above example codes, the binary parity check code of length $n$ =4, $k$ =3 for which $d$ = 2 and the *[7, 4]* Hamming code for which $d$ = 3 obey the above bound. A class of linear codes called the Generalized Reed-Solomon codes [15] (not discussed here) satisfy the equality, i.e., $d = n - k + 1$, and therefore provide for larger error correction relatively.

### 3.2.3  Dual Code

Error correction (described in greater detail in Section 3.2.5) for a given code *C* is accomplished by the application of a matrix, called the check matrix. This matrix is the generator of a linear code, which is the dual of the given code *C*. In what follows, we define the dual code and point out some of its properties useful in understanding the error-correction process.

The dual of a given code *C*, denoted $C^\perp$, is given by

$$C^\perp = \{ \vec{x} \in F_2^n \,|\, \vec{x}.\vec{c} = 0 \text{ for every } \vec{c} \in C \} \subseteq F_2^n \qquad (3.2.10)$$

*Some Properties*

1.  All elements of $C^\perp$ are orthogonal to all the elements of *C*, i.e., if $\vec{x} \in C^\perp$, then $\vec{x}G^T = \vec{0}$, where $G^T$ is the transpose of the generator matrix *G* of the code *C*.
2.  If *C* is an *[n, k]* linear code, then $C^\perp$ is an *[n, n-k]* linear code (dim $C^\perp$ = dim of the nullspace of *G* = *n-k*, since *G* has rank *k* [15]); note that $(C^\perp)^\perp$ = *C*.
3.  The generator of the dual code $C^\perp$ is an *(n-k) x n* matrix, denoted by *H*, and is called the check matrix.
4.  Property 1 then implies $HG^T$ =0 (the row vectors of *H* (which are elements of $C^\perp$) are orthogonal to row vectors of *G* (which are elements of *C*)).

*Example 1:* For the binary parity check code (Eqs. (1) and (3.2.6)), the check matrix *H* is a 1 x 4 matrix (Property 3 above), and from the relationship $HG^T$ =0 and Eq. (3.2.5), one easily obtains

$$H = [1111], \qquad (3.2.11)$$

which yields

$$C^\perp = RS(H) = \{\alpha_1 (1111) \,|\, \alpha_1 \in \{0,1\}\} = \{0000, 1111\} \subseteq F_2^4. \qquad (3.2.12)$$

Note that, in this example of the binary parity check code, $C^\perp \subset C$, i.e., *C* is <u>weakly self-dual</u> (a property we will encounter in the construction of the CSS codes (Section 3.3)). Even though $C^\perp$ is contained within *C*, any element of $C^\perp$ is still orthogonal to any element of *C*; e.g., the dot product 1111. 1010 ($\in C$) = 1 + 0 + 1 + 0 = 0.

*Example 2:* For the *[7,4]* Hamming Code, a check matrix *H* [15] is

$$H = \begin{bmatrix} 1 & 1 & 1 & 0 & 1 & 0 & 0 \\ 1 & 1 & 0 & 1 & 0 & 1 & 0 \\ 0 & 1 & 1 & 1 & 0 & 0 & 1 \end{bmatrix} \qquad (3.2.13)$$

Note that $HG^T = 0$ yields many equivalent solutions for $H$ related to each other by the manipulation of rows (adding one row to the other, etc) as discussed in Section 3.2.2. Eq. (3.2.13) is one of these solutions.

$C^{\perp} = RS(H) = \{\alpha_1\,(1110100) + \alpha_2\,(1101010) + \alpha_3\,(0111001) \ \ | \ \alpha_1 \in \{0,1\}\} \subseteq F_2^7$

$\qquad = \{0000000, 1110100, 1101010, 0111001, 0011110, 1010011, 1001101, 0100111\} \qquad (3.2.14)$

The reader can verify that the dot product of any codeword in $C^{\perp}$ (Eq. 3.2.14) with a codeword in $C$ (Eq. (3.2.2)) is equal to zero. One also observes that all the codewords in $C^{\perp}$ have even parity. Interestingly, they are also the elements of the $C$ code, i.e., $C^{\perp} \subset C$, i.e., the *[7,4]* Hamming code is also weakly self-dual.

### 3.2.4 Syndrome

In the error-correction process, it is the syndrome of the received word that helps ascertain the location of the errors that can be corrected. It is calculated using the check matrix described above[9]. Below we define the syndrome calculation for classical linear codes.

Let $C$ be an *[n, k]* linear code and $H$ the corresponding check matrix. Let $\vec{x} \in F_2^n$. Then

$$\vec{s} = H\vec{x}^T \in F_2^{n-k} \qquad (3.2.15)$$

is called the syndrome of $\vec{x}$; it is an element in $F_2^{n-k}$ and is represented by an *(n-k)* x 1 <u>column</u> vector. The syndrome $\vec{s} = H\vec{x}^T = 0$ if and only if $\vec{x} \in C$ [15]. If $\vec{x} \notin C$, $\vec{s} = H\vec{x}^T \neq 0$. Including the result $\vec{s} = 0$, there are $2^{n-k}$ possibilities for $\vec{s}$.

*Example:* For the binary parity check code (*n*=4, *k*=3), there are $2^{4-3}$ (= 2) possibilities only, column vectors [0] and [1]. Let $\vec{x} = 0110 \in C$. Then syndrome of $\vec{x}$ is $\vec{s} = H\vec{x}^T =$ [1111] [0110]$^T$ = [0], which is expected. But if we were to choose $\vec{x}$ such that $\vec{x} \notin C$, e.g., $\vec{x} = 0111$, one obtains $\vec{s} = [1]$, which is a non-zero result.

### 3.2.5 Detecting and Correcting Errors

As mentioned in Section 3.2.1, an *[n, k]* linear code $C$ <u>detects</u> $d-1$ errors and <u>corrects</u> $t = \lfloor (d\text{-}1)/2 \rfloor$ errors, where $d$ is the minimum weight of the code. Below we describe how syndrome determination (or <u>syndrome decoding</u>) by Bob for a received word helps detect and correct errors.

---

[9] In the quantum CSS codes (Section 3.3), the syndrome measurement is analogously defined.

When Bob receives a word $\vec{r}$ (a corrupted codeword), he can detect and correct errors (when possible) through a syndrome measurement on the received codeword. The syndrome of the received word $\vec{r}$ (see Section 3.2.4) is given by

$$\vec{s} = H\vec{r}^T \in F_2^{n-k} \tag{3.2.16}$$

The received word $\vec{r}$ may be written as

$$\vec{r} = \vec{c} + \vec{e}, \tag{3.2.17}$$

where $\vec{c} \in C$ and $\vec{e}$ (a possible error) $\in F_2^n$. We assume that the Hamming weight of $\vec{e}$, which is the number of errors (bit-flips), is small (less than or equal to $t$). Then

$$\vec{s} = H\vec{r}^T = H\vec{c}^T + H\vec{e}^T = H\vec{e}^T, \tag{3.2.18}$$

since $H\vec{c}^T = \vec{0}$. There are a total of $2^{n-k}$ possible syndromes, each corresponding to a different $\vec{r}$, or equivalently a different error vector $\vec{e}$, since the syndrome of $\vec{r}$ is the same as the syndrome of $\vec{e}$ (see Eq. (3.2.18)). $\vec{s} = \vec{0}$ implies $\vec{e} = \vec{0}$, i.e., no error has occurred. If $\vec{s} \neq \vec{0}$, an error has occurred, and $\vec{s}$ corresponds to one of the $2^{n-k} - 1$ possible error vectors, $\vec{e}$.

How does Bob correct the errors? Bob prepares a table of the syndrome/error relationship. Thus, when Bob receives $\vec{r}$ he calculates the syndrome $\vec{s}$ to which he matches the error $\vec{e}$ by looking up the syndrome/error table. He then computes $\vec{c} = \vec{r} - \vec{e}$ to obtain the corrected codeword.

*Example 1*: Binary Parity Check Code with length $n$ = 4 and $k$ =3.

For this code, $d = 2$, which implies that # of errors this code can detect is equal to 1 and the # of errors $t$ it can correct is equal to 0. $n = 4$ and $k = 3$ further imply $2^{n-k} = 2$, i.e., there are only two possibilities for $\vec{s}$. These are 1 x 1 column vectors, [0] and [1], with the former indicating no error has occurred and the latter indicating an error has occurred. If an error occurs, it cannot be corrected because the assumption that the weight of the error vector $\vec{e}$ is less than or equal to the number of errors $t$ the code can correct is violated.

Let $\vec{c}$ = 0110 (the transmitted codeword). Let $\vec{e}$ = 0001 be the error vector, which corresponds to the 4[th] bit flipping. Then $\vec{r}$ = 0111 (the received word).

$\vec{s} = H\vec{r}^T$ = [1111][0111]$^T$ = [1], i.e., Bob detects an error has occurred. However, he cannot correct it to $\vec{c}$ = 0110 with any certainty as he finds that the received word 0111 can correspond to any of the four possible original codewords: 1111, 0110 (the actual transmitted codeword), 0011, and 0101, assuming a single error has occurred.

*Example 2*: The binary *[7,4]* Hamming Code ($n$=7, k =4)

This code has $d = 3$ (see Eq. (3.2.2)), implying it can <u>detect 2 errors</u> and <u>correct 1 error</u>, i.e., $t = 1$. The generator matrix $G$ for this code is given by Eq. (3.2.7) and the check matrix $H$ is given by Eq. (3.2.13).

There are $2^{7-4}$ = 8 syndromes possible, unlike the binary parity check code of length $n$ = 4, where there were only 2 possible values, $\vec{s}$= [0] (indicating no error) and $\vec{s}$ =1, indicating an error has occurred (with no knowledge of its location within the codeword). In this example of the *[7, 4]* Hamming code, while $\vec{s}$ = $\vec{0}$ corresponds to no error, the remaining 7 syndrome vectors correspond to the 7 different single-bit errors that can be corrected. These 7 errors correspond to a bit 1 being in the 7 different positions possible (with the remaining 6 positions filled by 0's).

In order to correct a received codeword, Bob precalculates a table of syndromes and the corresponding error vectors $\vec{e}$:

| Syndrome | Error $\vec{e}$ |
|---|---|
| $\begin{bmatrix} 0 \\ 0 \\ 0 \end{bmatrix}$ | 0000000 (no error) |
| $\begin{bmatrix} 1 \\ 1 \\ 0 \end{bmatrix}$ | 1000000 |
| $\begin{bmatrix} 1 \\ 1 \\ 1 \end{bmatrix}$ | 0100000 |
| $\begin{bmatrix} 1 \\ 0 \\ 1 \end{bmatrix}$ | 0010000 |
| $\begin{bmatrix} 0 \\ 1 \\ 1 \end{bmatrix}$ | 0001000 |
| $\begin{bmatrix} 1 \\ 0 \\ 0 \end{bmatrix}$ | 0000100 |
| $\begin{bmatrix} 0 \\ 1 \\ 0 \end{bmatrix}$ | 0000010 |
| $\begin{bmatrix} 0 \\ 0 \\ 1 \end{bmatrix}$ | 0000001 |

Detection and correction when a single error has occurred

Suppose the transmitted codeword $\vec{c}$ = 0011110 (one of the 16 codewords $\in C$) and the received word $\vec{r}$ = 1011110 (the error is in the first bit which has flipped to 1). To detect and correct errors, Bob calculates $\vec{s} = H\vec{r}^T$; the result is $\begin{bmatrix} 1 \\ 1 \\ 0 \end{bmatrix}$. Bob finds from the above table the corresponding error to be $\vec{e}$ = 1000000 (its weight $\leq t$ = 1). He corrects the received word $\vec{r}$ to $\vec{r} - \vec{e}$ = 1011110 − 1000000 = 0011110, which was the transmitted codeword.

What happens if Bob's received codeword contains two errors?

Let $\vec{e}$ = 1000001, which is a case of two errors due to the first and the last bit flipping. Then, the received word is $\vec{r} = \vec{c} + \vec{e}$ = 0011110 + 1000001 = 1011111. Bob subjects his received word to a syndrome calculation: $\vec{s} = H\vec{r}^T$; the result is $\begin{bmatrix} 1 \\ 1 \\ 1 \end{bmatrix}$. Using the table above, he finds $\vec{e}$ = 0100000, which corresponds to a single-error ($2^{nd}$ bit flip) and corrects $\vec{r}$ to $\vec{r} - \vec{e}$ = 1011111 − 0100000 = 1111111, which is different from the transmitted codeword: 0011110, i.e., Bob corrects the received word to the wrong codeword without knowing it. Even though he correctly concludes that the received word is not correct, he is unable to correct it, simply because the # of errors in the received word exceeded $t$, the maximum number of errors the code can correct ($t$ = 1 here). In other words, the error correction fails when the number of errors exceeds $t$.

What happens if Bob's received codeword contains more than two errors?

Suppose the transmitted codeword is $\vec{c}$ = 0011110 and $\vec{e}$ = 1100001 (three errors). The received word $\vec{r} = \vec{c} + \vec{e}$ = 0011110 + 1000001 = 1111111, which is a valid codeword in $C$. $\vec{s} = H\vec{r}^T$ = 0, and Bob incorrectly concludes that no error has occurred. Thus, error detection fails, as expected because the number of errors exceeds $u$, the maximum number of errors that can be detected (for all codewords).

Bob's error correction algorithm may be summarized as follows:

1) Bob receives a word $\vec{r}$.
2) He calculates the syndrome $\vec{s} = H\vec{r}^T$ of $\vec{r}$.
3) If $\vec{s} = \vec{0}$, no error has occurred. If $\vec{s} \neq \vec{0}$, he finds the error vector $\vec{e}$ directly from the table of syndromes/error vectors. Since $\vec{r} = \vec{c} + \vec{e}$, he computes $\vec{c} = \vec{r} - \vec{e}$ to obtain the original (transmitted) codeword.

(If the # of errors exceeds $t$, the error correction algorithm fails; if the number of errors exceeds $u$, error detection fails).

## 3.2.6 Cosets

The concept of *coset* is important as its structure and properties are used in the design of the quantum CSS codes.

*Definition*: Let $\vec{x} \in F_2^n$ and $C$ be a linear code. We define the coset $\vec{x} + C$ to be

$$\vec{x} + C = \{\vec{x} + \vec{c} \,|\, \vec{c} \in C\} \tag{3.2.19}$$

*Example1:* For the binary parity check code of length $n = 4$, let $\vec{x} = 0111$. Then $\vec{x} + C = \{0111, 0100, 0010, 0001, 1110, 1101, 1011, 1000\}$. We now state an important theorem [15]:

<u>*Theorem 1*</u> : Let $\vec{x}_1, \vec{x}_2 \in F_2^n$. If $\vec{x}_2 \in \vec{x}_1 + C$, where $C$ is an *[n, k]* linear code, then

$$\vec{x}_2 + C = \vec{x}_1 + C \tag{3.2.20}$$

*Proof*: $\vec{x}_2 \in \vec{x}_1 + C$ implies $\vec{x}_2 = \vec{x}_1 + \vec{c}$, where $\vec{c} \in C$. Then $\vec{x}_2 + C = \vec{x}_1 + \vec{c} + C = \vec{x}_1 + C$, since $\vec{c} + C = C$ *(End of Proof)*.

*Theorem 1* has an important consequence: for each element $\vec{y} \in \vec{x}_1 + C$, the coset $\vec{y} + C$ is identical. Now each element $\vec{y} \in \vec{x}_1 + C$ is different. This leads to the conclusion that a set of $2^k$ different elements belonging to $F_2^n$ yield the same coset.

<u>*Theorem 2:*</u> Let $\vec{x}_1, \vec{x}_2 \in F_2^n$. If $\vec{x}_2 \notin \vec{x}_1 + C$, where $C$ is an *[n, k]* linear code, then

$$\vec{x}_2 + C \cap \vec{x}_1 + C = \emptyset. \tag{3.2.21}$$

*Proof*: $\vec{x}_2 \notin \vec{x}_1 + C$ implies $\vec{x}_2 \neq \vec{x}_1 + \vec{c} \,\forall\, \vec{c} \in C$. This further implies $\vec{x}_1 \neq \vec{x}_2 + \vec{c} \,\forall\, \vec{c} \in C$, i.e., $\vec{x}_1 \notin \vec{x}_2 + C$. Invoking Eq. (3.2.20), it follows $\forall\, \vec{y} \in \vec{x}_1 + C$, $\vec{y} \notin \vec{x}_2 + C$, i.e., $\vec{x}_2 + C \cap \vec{x}_1 + C = \emptyset$, if $\vec{x}_2 \notin \vec{x}_1 + C$ *(End of Proof)*.

*Theorem 2* has the consequence that all the $2^k$ elements of $\vec{x}_2 + C$ are different from the $2^k$ elements of $\vec{x}_1 + C$, i.e., the coset $\vec{x}_2 + C_2$ is completely disjoint from the coset $\vec{x}_1 + C$. Thus, the number of unique cosets of $C$ formed from $\vec{y}$ belonging to $F_2^n$ is equal to $2^n/2^k = 2^{n-k}$; another way of stating this result is that the

Number of cosets of $C$ in $F_2^n$ (also denoted $F_2^n/C$ ) = $2^n/2^k = 2^{n-k}$ (3.2.22)

<u>*Example 2*</u>: Let $C$ be the binary parity check code of length *n* = 4 and *k* =3; see Eqs. (3.2.1 and 6). From Eq. (3.2.22), the number of cosets we expect here is equal to 2. There are $2^4$ = 16 elements in $F_2^4$. Of these, 8 elements have even parity and comprise the $C$ code given in Eq. (3.2.1). Thus, the coset of C in these 8 even parity elements is the $C$ code itself. The other 8 elements of $F_2^4$ are the odd parity elements: 0001, 0010, 0100, 1000, 0111, 1011, 1101, 1110. Consider the coset 0111 + $C$ as in *Example 1* above:

0111 + *C* = {0111, 0100, 0010, 0001, 1110, 1101, 1011, 1000} (3.2.23)

This coset contains all the other 8 elements of $F_2^4$, and is the second (unique) coset we expected. The reader can verify the results, Eqs. (3.2.20) and (3.2.21), using Eqs. (3.2.1) and (3.2.23).

Now consider

$$\{0\} \subset C_2 \subset C_1 \subset F_2^n, \qquad (3.2.24)$$

where $C_1$ is an *[n, k₁]* linear code and $C_2$, a subspace of $C_1$, is an *[n, k₂]* linear code; clearly, $k_2 < k_1$. Replacing $F_2^n$ with $C_1$ and $C$ with $C_2$ in *Theorems* 1 and 2, we restate the theorems as

*Theorem 3*: Let $\vec{x}_1, \vec{x}_2 \in C_1$, an *[n, k₁]* linear code.

i)   If $\vec{x}_2 \in \vec{x}_1 + C_2$, where $C_2$ is an *[n, k₂]* linear code, then

$$\vec{x}_2 + C_2 = \vec{x}_1 + C_2 \qquad (3.2.25)$$

ii)  If $\vec{x}_2 \notin \vec{x}_1 + C_2$, where $C_2$ is an *[n, k₂]* linear code, then

$$\vec{x}_2 + C_2 \cap \vec{x}_1 + C_2 = \emptyset. \qquad (3.2.26)$$

Similar proofs apply. Also, Eq. (3.2.22) changes to

Number of cosets of $C_2$ in $C_1$ (also denoted $C_1 / C_2$) =   $2^{k_1 - k_2}$ \qquad (3.2.27)

Each coset is disjoint from the other in accordance with Eq. (3.2.26); they have no overlap.

*Example 3:* Let $\{0\} \subset C_2 \subset C_1 \subset F_2^4$, where

$C_1$ = {0000, 0011, 0101, 0110, 1001, 1010, 1100, 1111},

which is the binary parity check code of length *n* =4, and

$C_2 = C_1^{\perp}$ = {0000, 1111} $\subset C_1$.

$k_1$ = 3 and $k_2$ = 1, so $k_1 - k_2$ = 2. From Eq. (3.2.27), the number of cosets of $C_2$ in $C_1$ = $2^{3-1}$ = 4. These are indicated below:

$$\vec{v} = 0000 \text{ (or 1111)} \rightarrow \vec{v} + C_2 = \{0000, 1111\} \qquad (3.2.28a)$$

$$\vec{v} = 0011 \text{ (or 1100)} \rightarrow \vec{v} + C_2 = \{0011, 1100\} \qquad (3.2.28b)$$

$$\vec{v} = 0101 \text{ (or 1010)} \rightarrow \vec{v} + C_2 = \{0101, 1010\} \qquad (3.2.28c)$$

$$\vec{v} = 0110 \text{ (or 1001)} \rightarrow \vec{v} + C_2 = \{0110, 1001\} \qquad (3.2.28d)$$

The left-hand-side is the element *v* in $C_1$ which is added to $C_2$ to obtain the coset shown on the right-hand-side. The element in parenthesis is another element that yields the same coset in accordance with Eq. (3.2.25) (according to Eq. (3.2.25), there are $|C_2|$ (=2) different values of $\vec{v}$ that yield the same coset $\vec{v} + C_2$). Note that each of the 4 cosets above obeys Eq. (3.2.26) as well. Basically, the 8 elements of $C_1$ have been divided into 4 cosets, *with no element in any coset appearing in any other coset*. This is an important result, which is used in the construction of the quantum CSS codes to be discussed in Section 3.3; as we will see, it is this property of cosets, captured in Eq. (3.2.26), that will lead to the orthogonality of the constructed quantum codewords, which form the basis states of a Hilbert subspace, called the quantum codespace.

### 3.2.7 Some Other Useful Identities

We give below a couple of identities that will be useful in proving the error-correcting properties of the CSS codes in Section 3.3:

$$\sum_{\vec{v} \in C}(-1)^{\vec{v} \cdot \vec{u}} = |C| \qquad \text{if } \vec{u} \in C^{\perp} \tag{3.2.29a}$$

$$\sum_{\vec{v} \in C}(-1)^{\vec{v} \cdot \vec{u}} = 0 \qquad \text{if } \vec{u} \notin C^{\perp} \tag{3.2.29b}$$

where $C$ is a linear code and $C^{\perp}$ is its dual; $|C|$ (cardinality of $C$) denotes the number of elements (codewords) in code $C$; $|C| = 2^k$, where $k$ is the dimensionality of code $C$.

*Proof*:

The first identity, Eq. (3.2.29a), follows immediately from the fact that $\vec{v} \in C$ is orthogonal to $\vec{u} \in C^{\perp}$ (see Section 3.2.3). The second (nontrivial) identity is proved in the following way:

Let $C$ be an *[n,k]* linear code. Then, $C^{\perp}$ is an *[n, n-k]* linear code. We can express $\vec{v} \in C$ as $\vec{v} = \vec{\alpha}\, G$ (see Eq. (3.2.4)), where $G$ is the *k x n* generating matrix and $\vec{\alpha}$ is a *1 x k* row vector. Consequently, the left hand side of Eq. (3.2.29b) becomes

$$\sum_{\vec{\alpha} \in \{0,1\}^k}(-1)^{\vec{\alpha} G \cdot \vec{u}^T} = \sum_{\vec{\alpha} \in \{0,1\}^k}(-1)^{\vec{\alpha} \cdot G \vec{u}^T} \tag{3.2.30}$$

Now invoke a well-known general identity [20]:

$$\sum_{\vec{v} \in \{0,1\}^k}(-1)^{\vec{v} \cdot \vec{w}} = 0, \quad \vec{w} \neq \vec{0} \tag{3.2.31}$$

where $\vec{w}$ is also a string of length *k*. Comparing Eq. (3.2.30) with Eq. (3.2.31), we see that the right-hand-side of Eq. (3.2.30) is equal to zero, provided $G\vec{u}^T \neq 0$. But $G$ is the parity check matrix of $C^{\perp}$. So $G\vec{u}^T \neq 0$ implies that $\vec{u} \notin C^{\perp}$. Therefore, the right hand side of Eq. (3.2.30) is zero, provided $\vec{u} \notin C^{\perp}$ *(End of Proof).*

For completeness, we may add that if $\vec{w} = \vec{0}$, left-hand-side in Eq. (3.2.31) becomes $2^k$. Combining this result with Eq. (3.2.31),

$$\sum_{v \in \{0,1\}^k}(-1)^{v \cdot w} = 2^k\, \delta_w, \tag{3.2.32}$$

where $\delta_w$ is a Kronecker delta function (equal to 1 when *w* = 0, and equal to zero when *w* $\neq$ 0).

## 3.3 Quantum Error Correcting Codes

In QKD, the qubits traverse the quantum channel and therefore suffer errors due to their interaction with the environment, e.g. the fiber medium of transmission. The most general state of a qubit can be written as $|s> = a|0> + b|1>$, where the coefficients $a$ and $b$ are complex and the states $|0>$ and $|1>$ are the basis states of a *2*-dimensional complex Hilbert space (see Appendix A). A qubit's interaction with the environment (assuming it is not destroyed) is a unitary interaction described by a *2 x 2* unitary matrix, which in effect performs a "rotation" of the state vector $|s>$ in the two-dimensional Hilbert space. Because of the lack of the precise nature of the interaction and its duration, there are innumerable possibilities for the unitary matrix, and thus the final qubit state. In short, the interaction can generate an error of an arbitrary nature, whose effect on the qubit is described by a unitary matrix, the form of which we do not know.

Nevertheless, it is well known [15] that the most general 2 x 2 unitary matrix[10] can be written as a linear combination of the identity matrix *I* and the three Pauli matrices, $\sigma_x$, $\sigma_y$ and $\sigma_z$ (see Appendix B) which (along with the identity matrix) are unitary in nature:

$$U = tI + u\sigma_x + v\sigma_y + w\sigma_z , \qquad\qquad (3.3.1)$$

where $t$, $u$, $v$, and $w$ are complex numbers. The $x$ Pauli spin matrix, $\sigma_x$, causes bit flip, i.e., under the $\sigma_x$ transformation, the state $|s>$ changes to $a|1> + b|0>$; the $z$ Pauli matrix, $\sigma_z$, causes phase flip, i.e., under the $\sigma_z$ transformation, the basis state $|0>$ remains unchanged, but $|1>$ changes to $- |1>$; equivalently, the state $|s>$ changes to $a|0> - b|1>$, the transformation $i\,\sigma_y = \sigma_x\sigma_z$ (phase-flip followed by bit-flip) changes the state $|s>$ to $a|1> - b|0>$. Because Pauli matrices anticommute, $- i\,\sigma_y = \sigma_z\sigma_x$, which represents a bit-flip followed by phase-flip and produces the same final state, except for an overall negative sign, i.e., $-a|1> + b|0>$. Because of its most general form, Eq. (3.3.1) is representative of a continuum of possible errors on a given qubit.

After the qubit has undergone interaction described by Eq. (3.3.1), we want to identify and correct any error in a way that preserves the original state $|s>$, i.e., the original (unknown) coefficients $a$ and $b$ attached to the basis states $|0>$ and $|1>$ are preserved (clearly, a quantum measurement of $U|s>$ in the ($|0>$, $|1>$) basis, or for that matter any other basis, will destroy the original state). The preservation of the state $|s>$ in the quantum error correction process is achieved through judicious encoding, syndrome measurement (followed by error correction) and decoding.

The syndrome measurement, which preserves the unknown coefficients, $a$ and $b$, is sometimes called an *incomplete* measurement. A multi-qubit measurement is made which does not disturb the encoded quantum state, yet provides information on which physical qubits were affected, and what types of errors occurred. The errors detected are discrete, despite a continuum of possible errors implied by Eq. (3.3.1). The syndrome measurement causes the most general unitary operation (Eq. 3.3.1) to "collapse" into one of the four actions indicated by the operators *I* (no error), $\sigma_x$ (bit-flip), $\sigma_z$ (phase-flip), and $\sigma_x \sigma_z$ (phase flip followed by bit-flip), i.e, the syndrome measurement forces each qubit of the encoded state to yield (or "decide" for itself) a result of no change, bit-flip, phase-flip, or both phase-flip and bit-flip (in either order), and nothing else. This is in sharp contrast to classical coding, wherein the only error encountered is a bit-flip.

---

[10] In fact, any arbitrary 2 x 2 complex matrix can be written as a linear combination of such matrices.

Another difference with classical coding is that, whereas a classical bit can be duplicated, a qubit in a general state such as |s> cannot be duplicated (No-Cloning Theorem [10]). That is, an encoding scheme based on a repetition code such as |s> ⊗ |s> ⊗ |s> is forbidden. Rather, one utilizes elementary quantum gates, which represent unitary transformations, to perform encoding. In Section 3.3.1, we describe quantum encoding and decoding, followed by single error correction in Section 3.3.2; in particular, we illustrate syndrome measurement as an incomplete (or projective) measurement that leaves the structure of the quantum state unaffected during the quantum error correction process, and demonstrate the discretization of an arbitrary error upon syndrome measurement; in Section 3.3.3, we describe the CSS codes used in correcting more than a single error.

Henceforth, for convenience, we will also denote the *x*- and *z*- Pauli spin matrices by *X* and *Z*, respectively, which means, e.g., a phase-flip followed by bit-flip, would be represented by the product *XZ,* and so on.

### 3.3.1   Quantum Encoding and Decoding

Consider the general state of a single qubit:

$$|s> = a|0> + b|1>, \tag{3.3.2}$$

where the (complex) coefficients *a* and *b* are unknown, except for the fact that $|a|^2 + |b|^2 = 1$, i.e., the state |s> is normalized.  We create a three-qubit state by appending two qubits in the standard basis state |0>, i.e., create a state |s> ⊗ |0> ⊗ |0>. Writing it as |s00> for convenience,

$$|s00> = a|000> + b|100>. \tag{3.3.3}$$

Consider now the application of a CNOT gate, described in Appendix C.  Specifically, consider $CNOT_{12}$, where the indices indicate which two qubits it acts on; here the first qubit is the control bit and the second qubit is the target bit, so

$$CNOT_{12}|s00> = CNOT\ (a|000> + b|100>) = a|000> + b|110> \tag{3.3.4}$$

Similarly, if we apply $CNOT_{23}$ to the above state (Eq. (3.3.4)), we obtain

$$|s_C> = CNOT_{23}CNOT_{12}|s00> = a|000> + b|111>; \tag{3.3.5}$$

in CNOT$_{23}$, the second qubit is the control bit and the third qubit is the target bit; the subscript C in the state $|s_C>$ denotes the encoded state. In other words, by appending two qubits in the state $|0>$ and applying in succession the unitary transformations, CNOT$_{12}$ and CNOT$_{23}$, we have encoded a single qubit in the state $|s>$ into a triplet represented by $|s_C>$. Because coefficients $a$ and $b$ are arbitrary, we can interpret the above encoding (Eq. 3.3.5) to mean:

$$|0> \rightarrow |0_C> = |000> \qquad\qquad (3.3.6a)$$

$$|1> \rightarrow |1_C> = |111> \qquad\qquad (3.3.6b)$$

In general, applications of quantum gates can be used to create multi-qubit encoded states [10].

Decoding is achieved by performing the inverse of the encoding operations. Noting that the inverse of CNOT is CNOT, application of CNOT$_{23}$ followed by CNOT$_{12}$, to $|s_C>$ in Eq. (3.3.5) yields $a|000> + b|100>$, which can now be written as the product: $(a|0> + b|1>) \otimes |0> \otimes |0>$; the quantum state of the first qubit is thus $|s> = a|0> + b|1>$, the original state we started with.

In what follows, we illustrate single error corrections, using the above encoded state comprising three qubits. This will provide insight into the basics of the quantum error correction process.

### 3.3.2 Single Error Correction

Suppose Alice transmits the above encoded state on a noisy quantum channel. As discussed earlier, the noise can affect a qubit's state in a number of ways, but a measurement always forces the qubit to be found in a state obtained from the application of the identity operator (no error) or one of the three Pauli operators (occurrence of an error). Let us suppose that the noise affects only a single qubit of the (multi-qubit) encoded state. We will consider the three-qubit encoded state as described in Eq. (3.3.5).

### 3.3.2.1 An *X*-correcting (or bit-flip) code

Let us further suppose for now that $U = X$ (or $\sigma_x$), i.e., the error that occurs is a bit-flip error; a qubit in state $|0>$ changes to the state $|1>$, and vice versa. Since it is not known *a priori* which qubit gets flipped during transmission, there are three possibilities for the state corresponding to the received qubits:

$$|s'_C> = a|100> + b|011> \qquad\qquad (3.3.7a)$$

$$|s'_C> = a|010> + b|101> \qquad\qquad \grave{} \qquad (3.3.7b)$$

$$|s'_C> = a|001> + b|110> \qquad\qquad (3.3.7c)$$

*Error Detection (or Syndrome Measurement)*

We now perform a measurement which tells what error, if any occurred, on the quantum state. The result (or outcome) of the measurement is called the *syndrome*. For the above bit-flip case, there are four measurement outcomes, corresponding to the four projection operators[11]:

$$P_0 = |000> <000| + |111> <111| \quad \text{(no error)} \tag{3.3.8a}$$

$$P_1 = |100> <001| + |011> <011| \quad \text{(bit flip on 1st qubit)} \tag{3.3.8b}$$

$$P_2 = |010> <010| + |101> <101| \quad \text{(bit flip on 2nd qubit)} \tag{3.3.8c}$$

$$P_3 = |001> <001| + |110> <110| \quad \text{(bit flip on 3rd qubit)} \tag{3.3.8d}$$

If the first qubit flipped, then using $|s'_c>$ given in Eq. (3.3.7a), we find that $<s'_c|P_1|s'_c> = 1$ and $<s'_c|P_i|s'_c> = 0$ for $i = 0, 2, 3$, i.e., so the outcome of the measurement is certainly a bit flip on the first qubit. Also, the final state after measurement, given by $P_1|s'_c>/<s'_c|P_1|s'_c>$, remains unaltered, i.e., it is still $|s'_c>$ of Eq. (3.3.7a). Here the state vector $|s'_c>$ of the three qubits belongs to a $2^3$ (=8) - dimensional Hilbert space (Appendix A), but also lies (at the same time) in the 2-dimensional subspace spanned by vectors, $|100>$ and $|011>$. Similarly, if the second qubit had flipped, instead of the first one, then the state vector, (Eq, (3.3.7b)), would lie in the 2-dimensional space spanned by vectors, $|010>$ and $|101>$, with $<s'_c|P_2|s'_c> = 1$, and so on. There are four 2-dimensional spaces that the state vector $|s'_c>$ can lie in; one of these 2-dimensional subspaces (spanned by vectors $|000>$ and $|111>$) corresponds to the no error case; all these four subspaces are orthogonal to each other. The outcome of the measurement tells us with certainty i) whether an *X*-type error (bit-flip error) has occurred and if so ii), which qubit was affected.

One can assign numerical values (syndrome) to the outcomes of the measurement:  0 (no error), 1 (1st qubit flip), 2 (2nd qubit flip), 3 (3rd qubit flip). A syndrome measurement that actually yields these values can be effected through a quantum circuit made from CNOT gates; the syndrome value of 00 (=0), 01 (=1), 10 (=2), or 11 (=3) is read off from the two "extra" bits used in the measurement process [10, 19]. Thus, once the syndrome value is known, the affected is identified and the error correction subsequently performed (see below).

Note that the above (syndrome) measurement is an *incomplete* measurement on the system, as opposed to a complete measurement, which would involve knowing about the individual states of the three-qubit system.  The syndrome measurement clearly does not require us to do this, thus preserving the internal structure of the three-qubit encoded state, while providing information on the location of the error, to which the error correction is then applied.  As will be made even more evident, syndrome measurement is the projective effect of a quantum measurement, i.e., the measurement projects the state (with errors) onto a specific subspace of the Hilbert space, which identifies the location and type of the error.

---

[11] See Appendix D for some basics on projection operators.

*Error Correction*

The syndrome measurement indicates if an error has occurred or not. If it has, it tells us which qubit was affected, i.e., which qubit underwent a bit-flip. To correct the error, one simply applies $X$ (which is equal to $X^{-1}$) to the affected qubit to recover the original state.

## Another Example of Bit-flip Correction

Assume now that the unitary operation on a single qubit is of the following type:

$$U = cos\,\theta\,\mathbf{I} + i\,sin\,\theta\,X \qquad (3.3.9)$$

where $\theta$ is a real number. Suppose further that it is the third qubit of the encoded state $|s_C\rangle$ that $U$ acts on. The state $|s'_C\rangle$ one now obtains is given by

$$|s'_C\rangle = U_3|s_C\rangle = a\,(cos\,\theta\,|000\rangle + i\,sin\,\theta\,|001\rangle) + b(cos\,\theta\,|111\rangle + i\,sin\,\theta\,|110\rangle), \qquad (3.3.10)$$

where the subscript 3 indicates that $U$ acts on the third qubit. Now the probabilities of obtaining the outcomes corresponding to the measurements $P_0$, $P_1$, $P_2$, and $P_3$ are

$$p_0 = \langle s'_C|P_0|s'_C\rangle = cos^2\theta, \qquad (3.3.11a)$$

$$p_1 = \langle s'_C|P_1|s'_C\rangle = 0, \qquad (3.3.11b)$$

$$p_2 = \langle s'_C|P_2|s'_C\rangle = 0, \qquad (3.3.11c)$$

$$p_3 = \langle s'_C|P_3|s'_C\rangle = sin^2\theta. \qquad (3.3.11d)$$

Consequently, a measurement here does not give any outcome with 100% certainty as in the earlier example of bit-flip error. The outcome is either no error or bit flip in the third qubit, with respective probabilities, $cos^2\theta$ and $sin^2\theta$. If one gets the outcome $P_0$, the final state after measurement is $P_0|s'_C\rangle/\sqrt{(\langle P_0|s'_C|P_0\rangle)} = a|000\rangle + b|111\rangle$, which is the original state. As a result, we do not need any correction here (one way to look at it is that the projective effect of the syndrome measurement has *automatically* corrected any error the qubits may have been subjected to). But if we get the outcome $P_3$, the resulting state is $a|001\rangle + b|110\rangle$, and in this case, we know we need to correct the error by applying the $X$ operator to the third qubit. Again, in the syndrome measurement process, we do not need to know the coefficients $a$ and $b$ which are preserved in the measurement process; as a result, after error correction, the original quantum state is recovered.

Note an important difference between the two cases of unitary transformations considered above; when we assumed that the error was a bit-flip only (Eqs. (3.3.5) and (3.3.7a-c)), the state $|s'_c>$ was already in one of the four 2-dimensional subspaces picked out by the measurement (Eqs. (3.3.8a-d)); the measurement was used only to inform us about the error. But, when the error caused by the noise was of the *U* type in Eq. (3.3.9), the measurement caused the state vector $|s'_c>$ to be projected onto (or placed in) one of the two subspaces and it also told us which subspaces it had placed it in.

The case of *U* in Eq. (3.3.9) also illustrates how a problem of a continuum of errors (parametrized by $\theta$ here) is turned upon measurement into a simpler problem of correcting a discrete set of errors (no error or bit-flip error on the third qubit).

### 3.3.2.2  A Z correcting (or phase-flip) code

Suppose now that the state of a qubit subject to noise is transformed by the unitary operator, $U = Z$ (or $\sigma_Z$). Assume as before that the interaction changes the state of a single qubit. Then, the encoded state $|s_c> = a|000> + b|111>$ changes to $|s'_c> = a|000> - b|111>$, since $Z|0> = |0>$ and $Z|1> = -|1>$ (the result $|s'_c>$ is the same regardless of which qubit the Z operator acts upon). If we now perform a measurement with possible outcomes, $P_0$, $P_1$, $P_2$, $P_3$, as before, we obtain the outcome $P_0$ with probability $p_0 = <s'c|P_0|s'c> = 1$, i.e., the outcome is always a state with no error (the probabilities $p_1$, $p_2$, and $p_3$ are each zero). In other words, we are unable to detect a phase-flip error with the above measurement scheme. The measurement scheme (or the syndrome measurement) must be revised.

Consider the application of the Hadamard transform $H_d$ to each qubit of the encoded state, $|s_C>$:

$$|s^{(H)}_c> = H_d \otimes H_d \otimes H_d \, |s_c> = a|+++> + b|--->, \qquad (3.3.12)$$

since $H_d |0> = |+>$ and $H_d |1> = |->$ (see Appendix B). Also, $Z|+> = |->$ and $Z|-> = |+>$, i.e., *Z* acts on ($|+>$, $|->$) the same way that *X* acts on ($|0>$, $|1>$). Therefore, when the interaction described by the *Z* operator acts on a single qubit, say the second qubit in Eq. (3.3.12), then its label flips from + to − and vice versa, i.e., the encoded state $|s^{(H)}_c>$, after the application of the *Z* operator, becomes

$$|s'^{(H)}_c> = Z_2 \, | \, s^{(H)}_c > = a|+-+> + b|-+->, \qquad (3.3.13)$$

where the index 2 specifies that the *Z* operator acts on the second qubit. Because the *Z* operator in the ($|+>$, $|->$) basis acts like the *X* operator, the previous error detection and corrections scheme developed for the *X*-type of error applies to the error of the *Z*-type in the ($|+>$, $|->$) basis. Therefore, to detect and correct the *Z*-type error, we project the state $|s'^{(H)}_c>$ onto a different set of 2-dimensional subspaces, defined by the projection operators:

$$R_{0 =} \, |+++> <+++| + |--->  <---| \qquad (3.3.14a)$$

$$R_1 = |-++> <-++| + |+--> <+--|$$ (3.3.14b)

$$R_2 = |+-+> <+-+| + |-+-> <-+-|$$ (3.3.14c)

$$R_3 = |++-> <++-| + |-++> <-++|$$ (3.3.14d)

If the measurement provides an outcome $R_0$, no error has occurred. If the outcome is $R_1$, $R_2$, or $R_3$, then the first, second, or the third qubit, respectively, has undergone a phase-flip, and we apply the $Z$ operator to the appropriate qubit to correct the error, recalling that $Z|+> = |->$ and $Z|-> = |+>$; the corrected state is $|s^{(H)}_c>$ as in Eq. (3.3.12). We then apply the inverse of the Hadamard transform (which is the Hadamard transform itself) to each of the three qubits to obtain the corrected state in the (0, 1) basis, i.e., $|s_c> = a|000> + b|111>$.

### 3.3.2.3 The Shor code

When we combine the $X$-correcting code with the $Z$-correcting code, we obtain a scheme that protects against all types of errors, i.e., an arbitrary error, as described by Eq. (3.3.1).

We construct this code by first encoding a qubit using the phase flip code: $|0> \rightarrow |+++>$ and $|1> \rightarrow |--->$. Then we encode each of these qubits using the bit flip code: $|+> = 1\sqrt{2}(|0> + |1>) \rightarrow 1/\sqrt{2}(|000> + |111>)$; similarly, $|-> = 1\sqrt{2}(|0> - |1>) \rightarrow 1/\sqrt{2}(|000> - |111>)$. The result is a nine-bit code given by

$$|0_c> = 1/(2\sqrt{2})(|000> + |111> \otimes (|000> + |111>) \otimes (|000> + |111>),$$ (3.3.15a)

$$|1_c> = 1/(2\sqrt{2})(|000> - |111> \otimes (|000> - |111>) \otimes (|000> - |111>),$$

(3.3.15b)

i.e., the state $|s> = a|0> + b|1>$ is encoded to $|s_c> = a|0_c> + b|1_c>$ by the above scheme. This code was discovered by Peter Shor [18] and is called the Shor code.

One can start by considering the 27 different types of errors: $X_1$, $X_2$, $X_3$, $X_4$, $X_5$, $X_6$, $X_7$, $X_8$, $X_9$, $Z_1$, $Z_2$, $Z_3$, $Z_4$, $Z_5$, $Z_6$, $Z_7$, $Z_8$, $Z_9$, $X_1Z_1$, $X_2Z_2$, $X_3Z_3$, $X_4Z_4$, $X_5Z_5$, $X_6Z_6$, $X_7Z_7$, $X_8Z_8$, $X_9Z_9$, where the subscript indicates which qubit is affected; the product $XZ$ is simply the $Z$ operation followed by the $X$ operation. For example, if the error is $X_4$, then the fourth qubit is flipped, that is,

$$X_4|0_c> = 1/(2\sqrt{2})(|000> + |111> \otimes (|100> + |011>) \otimes (|000> + |111>),$$ (3.3.16a)

$$X_4|1_c> = 1/(2\sqrt{2})(|000> - |111> \otimes (|100> - |011>) \otimes (|000> - |111>),$$ (3.3.16b)

and so on. The resulting state is different for different qubit flips. But in the Z type of errors, the effects of $Z_1$, $Z_2$, and $Z_3$ are not distinguishable, as we saw in Section 3.3.2.2, and similarly, the effects of $Z_4$, $Z_5$, and $Z_6$ are not distinguishable, and the effects of $Z_7$, $Z_8$, and $Z_9$ are not distinguishable. For example,

$$Z_4|0_C> = 1/(2\sqrt{2})(|000> + |111> \otimes (|000> - |111>) \otimes (|000> + |111>) \tag{3.3.17}$$

where the effect of $Z_4$ introduces a negative sign between $|000>$ and $|111>$, the states corresponding to qubits, 4,5, and 6. But this negative sign could also have been introduced by the action of $Z_5$ (on the fifth qubit) or the action of $Z_6$ (on the sixth qubit). Thus, the effects of $Z_4$, $Z_5$, and $Z_6$ here are indistinguishable. To correct the above state (Eq. (3.3.17), it is sufficient to apply the Z operator to qubit 4, 5, or 6. Keeping this indistinguishibility in mind, the set comprising the error operators and the identity operator (no error) is:

$$X_1, X_2, X_3, X_4, X_5, X_6, X_7, X_8, X_9,$$

$$\{Z_1, Z_2, \text{ or } Z_3\}, \{Z_4, Z_5, \text{ or } Z_6\}, \{Z_7, Z_8, \text{ or } Z_9\},$$

$$X_1Z_1, X_2Z_2, X_3Z_3, X_4Z_4, X_5Z_5, X_6Z_6, X_7Z_7, X_8Z_8, X_9Z_9,$$

$$I \text{ (no error)} \tag{3.3.18}$$

Each operator corresponds to a 2-dimensional subspace of the $2^9$ - dimensional Hilbert space corresponding to the 9 qubits. All these subspaces are mutually orthogonal. Because the set of error operators includes the $I$ operator and the operators of the X, $Z$, and $XZ$ types, the Shor code can correct an arbitrary error described by Eq. (3.3.1), i.e.,

$$U_j |s_C> = t |s_C> + u X_j|s_C> + v Z_j |s_C> + w X_jZ_j |s_C>, \tag{3.3.19}$$

where $U_j$ denotes the operator $U$ of Eq. (3.3.1) acting on the jth qubit. Any measurement of the $U_j|s_C>$ state projects the state onto one of the two dimensional spaces corresponding to the 22 operators defined above.

In summary,

1) The action of environment on a qubit results in a continuum of possible errors as described by Eq. (3.3.1). A measurement, however, "collapses" any error into discrete types described by the actions of the Pauli spin operators.
2) The measurement is such that the contents of the original state are preserved, yet the error is detected; this is sometimes called an incomplete measurement.

In the next section, we describe a completely different type of a quantum code, called the Calderbank-Shor-Steane (CSS) code, which detects and corrects <u>more than one error.</u> This code is based on linear codes. The concept of syndrome measurement still applies, whereby the original state is preserved and arbitrary errors, "collapsable" into the discrete set of errors, are corrected. However, in order to understand how linear codes play a role, we first revisit the $X$ and $Z$ correcting codes, cast them in a different form and, subsequently, connect them to linear codes, which then provides a general mapping between linear codes and quantum codes. This mapping is subsequently employed in the development of the CSS codes.

### 3.3.3 Calderbank-Shor-Steane (CSS) Codes

Shor code described earlier can correct a single qubit error. But quantum computation can involve more than a single error. The inadequacy of the Shor code to correct more than one error was a matter of concern until Calderbank and Shor [6] and Steane [7] independently came up with a methodology to correct for more than a single error. They invoked the theory and mechanics of linear codes (Section 3.2) in a quantum setting to generate what are now termed as the Calderbank-Shor-Steane (CSS) quantum error-correcting codes.

#### 3.3.3.1 The *X*-correcting (or bit-flip) code revisited

Is the use of linear codes in CSS codes a radically different approach from the one used in the construction of the quantum codes in Section 3.3.2? Not quite so, if one revisits the *X*-correcting code and reexamines the (syndrome) measurement process based on the projection operators. The syndrome measurement therein corresponded to the outcomes, no bit-flip, first qubit flip, second qubit flip, and the third qubit flip, respectively. We now discuss a different way of making the syndrome measurement that yields the same four outcomes. Instead of using the projection operators $P_0$, $P_1$, $P_2$, and $P_3$ for measurement, we perform two measurements, first of the observable $Z_1Z_2$ ($\equiv Z \otimes Z \otimes I$) of the three qubit system, and the second of the observable $Z_2Z_3$ of the same three-qubit system. Each of these observables has an eigenvalue = $\pm$ 1, since the operator $Z$ has an eigenvalue +1 or -1 according to the qubit under consideration is in the state $|0>$ or $|1>$. As a result, each measurement provides 1 bit of information for a total of 2 bits of information, i.e., the four possible syndrome values, 00, 01, 10, 11, just as before, which correspond to the same four outcomes. Suppose, as an example, the measurement of observable $Z_1Z_2$ gives a value of +1; this then implies that neither of the first and the second qubits flipped (we are assuming only a single bit-flip here; otherwise, we would also have to consider the possibility of both the first and the second qubits undergoing bit-flips); similarly, if the measurement of observable $Z_2Z_3$ gives a value of -1, then either the second qubit flipped or the third qubit flipped. But from the first measurement, we know that the second qubit did not undergo bit-flip, so we conclude from the two measurements considered in conjunction that the third qubit flipped. Similarly, the other outcomes are ascertained from the other possible measurement values of $Z_1Z_2$ and $Z_2Z_3$ . Table 3.1 summarizes the results.

Table 3.1 Measured values of observables $Z_1Z_2$ and $Z_2Z_3$, which lead to different outcomes

| $Z_1Z_2$ | $Z_2Z_3$ | Outcome |
|---|---|---|
| +1 | +1 | No error |
| -1 | +1 | Bit-flip on $1^{st}$ qubit |
| -1 | -1 | Bit-flip on $2^{nd}$ qubit |
| +1 | -1 | Bit-flip on $3^{rd}$ qubit |

Basically, the measurement of $Z_1Z_2$ compares the bit values of the first two qubits, and measurement of $Z_2Z_3$ compares the bit values of the second and the third qubits[12]. This can be also seen from rewriting these observables (see Appendix C) as:

$$Z_1Z_2 = (|00\rangle\langle00| + |11\rangle\langle11|) \otimes I - (|01\rangle\langle01| + |10\rangle\langle01|) \otimes I$$

$$(3.3.20a)$$

$$Z_2Z_3 = I \otimes (|00\rangle\langle00| + |11\rangle\langle11|) - I \otimes (|01\rangle\langle01| + |10\rangle\langle01|) \qquad (3.3.20b)$$

In Eq. (3.3.20a), the identity operator acts on the third qubit, which is not measured; in Eq. (3.3.20b), the identity operator acts on the first qubit, which is not measured. $Z_1Z_2$ and $Z_2Z_3$ are projective measurements yielding

$$Z_1Z_2 \ |s'_c\rangle = g \ |s'_c\rangle, \qquad\qquad\qquad (3.3.21a)$$

where $g = \pm1$; $g = +1$ when $|s'c\rangle = a \ |000\rangle + b \ |111\rangle$ (no error, see Eq. (3.3.5)) or $|s'_c\rangle = a \ |001\rangle + b \ |110\rangle$ (bit-flip on third qubit; see Eq. (3.3.7c)); $g = -1$, when $|s'_c\rangle = a \ |100\rangle + b \ |011\rangle$ (bit-flip on first qubit; see Eq. (3.3.7a)) or $|s'_c\rangle = a \ |010\rangle + b \ |101\rangle$ (bit-flip on second qubit; see Eq. (3.3.7b)).

Similarly,

$$Z_2Z_3 \ |s'_c\rangle = g \ |s'_c\rangle, \qquad\qquad\qquad` \quad (3.3.21b)$$

where $g = \pm1$; $g = +1$ when $|s'c\rangle = a \ |000\rangle + b \ |111\rangle$ (no error, see Eq. (3.3.5)) or $|s'_c\rangle = a \ |100\rangle + b \ |011\rangle$ (bit-flip on first qubit; see Eq. (3.3.7a)); $g = -1$, when $|s'_c\rangle = a \ |010\rangle + b \ |101\rangle$ (bit-flip on second qubit; see Eq. (3.3.7b)) or $|s'_c\rangle = a \ |001\rangle + b \ |110\rangle$ (bit-flip on third qubit; see Eq. (3.3.7c)).

The results based on Eqs. (3.3.21a) and (3.3.21b) are consistent with the results in the table above.

### 3.3.3.2 Interpretation of the *X*-correcting code in terms of linear codes

We saw earlier that $|0\rangle$ was encoded into $|0_c\rangle = |000\rangle$ and $|1\rangle$ was encoded into $|1_c\rangle = |111\rangle$. These can, in fact, be regarded as two quantum codewords in a quantum error-correction scheme that allows for the detection and correction of a single bit-flip error.

Let us construct and examine a classical analog of this quantum error-correcting scheme. A classical analog is, clearly, a *binary repetition code of length 3*:

$$C = \{000, 111\}. \qquad\qquad\qquad (3.3.22)$$

This code satisfies the properties of a linear code (see Section 3.2.1) and is a *[3, 1]* linear code, with $k =1$ and $n = 3$, as it has two elements ($2^k = 2^1 = 2$), each of length $n =3$. It is, therefore, a subspace of dimensionality $k = 1$ in the binary vector space $F_2{}^n$ of dimensionality $n = 3$. $k = 1$ also indicates there are two "message" strings, which are 0 and 1, which get encoded into 000 and 111, respectively.

---

[12] Note that we could also have chosen $Z_1Z_3$ and $Z_2Z_3$ as the two observables; another equally valid choice here would also be $Z_1Z_2$ and $Z_1Z_3$; these two alternative choices lead to the same four possible outcomes listed in Table 3.1.

The generator matrix $G = [111]$ correctly generates the code $C = RS(G) = \alpha\ \vec{x}\ = \{000, 111\}$, as $\alpha \in (0, 1)$ and $\vec{x} = (111)$. $\vec{c} = \vec{m}\ G$, where $\vec{m}$ is the message vector, and $\vec{c}$ is the generated codeword; clearly, the message bits, 0 and 1, get encoded by matrix $G$ into the codewords 000 and 111, respectively.

The minimum weight $d$ of code $C$ above (binary repetition code of length 3) is 3, which means it can detect $d - 1 = 3 - 1 = 2$ errors. The number of errors it can correct is given by $t = \lfloor (d-1)/2 \rfloor = \lfloor (3-1)/2 \rfloor = 1$ (the reader can easily verify the above results by individually examining the codewords of $C$).

The parity check matrix $H$, satisfying $HG^T = 0$, is constructed to be

$$H = \begin{bmatrix} 1 & 1 & 0 \\ 0 & 1 & 1 \end{bmatrix} \qquad (3.3.23)$$

When Alice transmits a codeword $\vec{c}$, Bob receives the word $\vec{r} = \vec{c}\ + \vec{e}$, where $\vec{e}$ is a possible error. The syndrome $\vec{s}$ corresponding to $\vec{r}$ is given by $\vec{s} = H\ \vec{r}^T = H\ (\vec{c}^T + \vec{e}^T) = H\ \vec{e}^T$, since $H\ \vec{c}^T = 0$. Now there are $2^{n-k} = 2^{3-1} = 4$ such syndromes corresponding to the 4 possible errors, $\vec{e}$, as shown below:

$\qquad\qquad \underline{\vec{e}} \qquad\qquad\qquad\qquad\qquad\qquad\qquad \underline{\vec{s}}$

1)  000    (no error)  $\qquad\qquad\qquad\qquad \begin{bmatrix} 0 \\ 0 \end{bmatrix}$  (3.3.24a)

2)  100   (flip of the first bit)  $\qquad\qquad \begin{bmatrix} 1 \\ 0 \end{bmatrix}$  (3.3.24b)

3)  010   (flip of the second bit)  $\qquad\quad \begin{bmatrix} 1 \\ 1 \end{bmatrix}$  (3.3.24c)

4)  001   (flip of the third bit)  $\qquad\quad \begin{bmatrix} 0 \\ 1 \end{bmatrix}$  (3.3.24d)

Thus, a syndrome measurement $\vec{s}\ (= H\ \vec{r}^T)$ reveals the error, if any, in the received word. Consequently, the original codeword $\vec{c}$ is recovered by Bob through the correction: $\vec{c} = \vec{r} - \vec{e}$.

<u>How do all of the above results in the classical domain relate to the results obtained in the quantum error correction process</u>? If we recall, the quantum encoding is

$$|0> \rightarrow |0_C> = |000> \qquad (3.3.25a)$$

$$|1> \rightarrow |1_C> = |111>, \qquad (3.3.25b)$$

where $|0_C>$ and $|1_C>$ are two (quantum) codewords. Suppose Alice transmits three qubits in the state $|000>$ to Bob. When Bob receives these qubits (which may have an error), he can make a syndrome measurement (in the quantum domain), using the $H$ matrix of Eq. (3.3.23) modified for quantum measurement in accordance with the following rules:

1) Define operator $\sigma_z^{[r]}$ for the $r_{th}$ row of $H$ as

$$\sigma_z^{[r]} = \sigma_1 \otimes \sigma_2 \otimes \sigma_3 \,, \tag{3.3.26}$$

where $\sigma_i = Z$ (the z-Pauli spin matrix) or the *2 x 2* identity matrix *I* according as the $i_{th}$ bit in the row is 1 or 0 and redefine $H$ of Eq. (3.2.23) as    a column operator, $H_Q$  :

$$H \rightarrow H_Q = \begin{bmatrix} Z \otimes Z \otimes I \\ I \otimes Z \otimes Z \end{bmatrix} \equiv \begin{bmatrix} Z_1\,Z_2\,I_3 \\ I_1\,Z_2\,Z_3 \end{bmatrix} \tag{3.3.27}$$

2) i) Apply $H_Q$ to the received state, which yields an eigenvalue column vector, comprising +1 and -1 as the elements.

ii)  Apply a mapping of

$$1 \rightarrow 0, \text{-}1 \rightarrow 1, \tag{3.3.28}$$

where 0 and 1 are classical bits to obtain a syndrome corresponding to its classical analog.

Note that the structure of $H_Q$ , which follows from Eq. (3.2.26) (Rule 1), is consistent with the measurements $Z_1Z_2$ and $Z_2Z_3$ discussed earlier in Section 3.3.3.1. We now illustrate below the syndrome determination, using $H_Q$:

Suppose the received state is |001> (bit-flip on the third qubit). Then $H_Q$ |001> = $\begin{bmatrix} 1 \\ -1 \end{bmatrix}$ |001>, since $Z_1\,Z_2\,I_3$ |001> = + |001> and $I_1\,Z_2\,Z_3$ |001> = - |001>.  We now transform the eigenvalue column vector to its classical analog, using the mapping in Eq. (3.3.28):

$$\begin{bmatrix} 1 \\ -1 \end{bmatrix} \rightarrow \begin{bmatrix} 0 \\ 1 \end{bmatrix}, \tag{3.3.29}$$

This syndrome (see Eq. (3.3.24d )) is a third bit-flip for the classical code, consistent with the bit-flip on the third qubit in the quantum channel. Thus, after the above mapping, which yields a syndrome corresponding to the classical $H$ matrix upon which the quantum code is based, the bit-flip error in the quantum channel can be identified and thus corrected. Rules (1) and (2) above are a general result, which are also applied in the syndrome measurement, using CSS codes.

### Why Rules (1) and (2) are a general result?

The above rules and the mapping, Eq, (3.3.28), are understood if we contrast the arithmetic in the classical domain (which is modulo 2) with the arithmetic used in the quantum domain, to arrive at the syndromes. In the classical domain, the elements of the syndrome are obtained from multiplication of row elements of the parity check matrix $H$ with the corresponding elements of the column vector (the received word); in the quantum domain, we use products of Pauli operators and their eigenvalues, whose values (+1 and -1) depend upon the bit values of the qubits. The basic classical multiplication operations and their quantum analogs are given below:

| Classical Multiplication | Quantum Analog |
|---|---|
| 1 x 1 = 1 | $Z\|1> = (-1)\|1>$ |
| 1 x 0 = 0 | $Z\|0> = (+1)\|0>$ |
| 0 x 1 = 0 | $I\|1> = (+1)\|1>$ |
| 0 x 0 = 0 | $I\|0> = (+1)\|0>$ |

The first bit in the above classical operation case corresponds to the bit-value of an element of the $H$ operator, and thus defines, in accordance with Eq. (3.2.26), the quantum operator ($Z$ or $I$) to be used in the determination of the quantum syndrome, while the second bit defines the qubit state the operator acts upon. *One observes in the table above that whenever the result is 1 or 0 at the end of the (classical) multiplication process, the eigenvalue is (-1) or (+1), respectively, in the quantum domain. This correspondence, which we may call the sub-rule, is exactly the mapping, Eq. (3.3.28) discussed below.*

After the above multiplicative operations, the individual results are added up (modulo 2) in the classical domain to obtain the elements of the syndrome, while, in the quantum syndrome determination case, the arithmetic is a multiplication of eigenvalues (+1 or -1) as a result of the application of $H_Q$ to a quantum state; the basic results are summarized below:

| Classical Addition | Quantum Equivalent |
|---|---|
| 1 + 1 = 0 | (-1) x (-1) = +1 |
| 1 + 0 = 1 | (-1) x (+1) = -1 |
| 0 + 1 = 1 | (+1) x (-1) = -1 |
| 0 + 0 = 0 | (+1) x (+1) = +1, |

where to obtain the quantum equivalent, we have invoked the sub-rule, namely, that the classical bits 0 and 1 correspond to the eigenvalues +1 and -1, respectively (of course, the addition classically corresponds to multiplication in the quantum domain).

*If we now use the mapping of Eq. (3.3.28), we see that the above results in the quantum domain become exactly the same as the result in the classical domain, thus yielding a syndrome (a column vector) that is identified with its classical analog; thus the quantum error is identified.* Once it is identified, then it is corrected with the appropriate operator acting on the identified qubit.

Rules (1) and (2) above are of a general nature; Eq. (3.3.26) is generalized to an $n$-bit $H$ parity check matrix.  We will use these general results in the error correction process for the CSS codes (Section 3.3.3.3).

*Further Remarks*:

Note that the two quantum codewords $|0_C> = |000>$ and $|1_C> = |111>$ are orthogonal to each other in the $2^3 (=8)$-dimensional Hilbert space $\mathbb{C}^8$ (see Appendix A). They together comprise a $2^k$ ($k = 1$) dimensional subspace, as was also noted in Section 3.3.2.1.

### 3.3.3.3 Mapping rules for the *Z*-correcting (phase-flip) code

In the above section, we derived mapping rules relating the classical bit-error correcting code to the quantum bit-flip error correcting code. In the classical domain, we do not have the equivalent of the quantum phase-flip error correcting code; yet useful mapping rules can be obtained if one recognizes that the Hadamard transform application yields a structure similar to the bit-flip error correcting code. In the phase-flip code, Section 3.3.2.2, we noted that the pertinent two quantum codewords were obtained in the following way:

$$0 \rightarrow |0_C> = H_d^{\otimes 3} |000> = |+++> \tag{3.3.30a}$$

$$1 \rightarrow |1_C> = H_d^{\otimes 3} |111> = |--->, \tag{3.3.30b}$$

where the Hadamard transform, $H_d$, is applied to each of the three qubits. If we disregard the Hadamard transform momentarily, we then have a classical analog (code), which is exactly the same as the one for the quantum *X*-correcting code. The only difference with the *X*-correcting code case is that, in light of the fact that we are actually in the (+, -) basis due to the Hadamard transform, we replace the operator *Z* with operator *X* (see Appendix B), i.e. Rule 1 (Eq. 3.3.26) is redefined as

1') Define operator $\sigma_x^{[r]}$ for the $r_{th}$ row of *H* as

$$\sigma_x^{[r]} = \sigma_1 \otimes \sigma_2 \otimes \sigma_3, \tag{3.3.31}$$

where $\sigma_i = X$ (the *x*-Pauli spin matrix) or the *2 x 2* identity matrix *I* according as the $i_{th}$ bit in the row is 1 or 0

As a result of this new rule for the phase-flip code, $Z_1, Z_2,$ and $Z_3$ in Eq. (3.3.27) are replaced with $X_1, X_2,$ and $X_3$, respectively.

Again the results are general results and apply equally well to an *H* parity check matrix corresponding to *n*-bits.

### 3.3.3.3 Construction of CSS codes

A CSS quantum error-correcting code forms a subspace $\mathbb{C}^{2^k}$ of the Hilbert Space $\mathbb{C}^{2^n}$ (*k* less than *n*), where *n* is the number of qubits (see Appendix A); the subspace $\mathbb{C}^{2^k}$ corresponding to encoding of *k* qubits (into *n* qubits) is protected from errors in a small number $\delta$ ($\geq 1$) of these qubits. The errors can be bit-flip or phase-flip or arbitrary combination of these. The encoding is performed by appending to the *k* qubits, an additional set of *n-k* qubits, each in state |0>, and performing appropriate unitary transformations, similar to the transformations discussed in Section 3.3.1.

Just as the single-error correcting codes of previous section were shown to originate in linear codes, the CSS codes also find their origin in linear codes; in fact, construction of CSS codes follows the linear code theory very closely.  In particular, a CSS code *Q* on *n* qubits is derived from two error-correcting linear codes, an *[n, k₁]* linear code *C₁* and an *[n, k₂]* linear code *C₂*, with the latter contained in the former, i.e.,

$$\{0\} \subset C_2 \subset C_1 \subset F_2^n, \tag{3.3.32}$$

where $F_2^n$ is the binary vector space on *n* bits; $k_2$ and $k_1$ are the dimensionalities of the linear codes *C₁* and *C₂*, respectively (see Section 3.2); clearly, $k_2 < k_1$. It is assumed that both *C₁* and $C_2^\perp$ (the dual of *C₂*) can correct $\delta$ errors.

A set of basis states, called codewords, for the CSS code subspace $\mathbb{C}^{2^k}$ are then constructed from vectors $v \in C_1$ as follows[13]:

$$v \rightarrow |v + C_2> = \frac{1}{|C_2|^{\frac{1}{2}}} \sum_{w \in C_2} |v + w> \tag{3.3.33}$$

The vector $v + w$ in Eq. (3.3.33) is an element of the coset $v + C_2$ (see Section 3.2.6). If $v_1 - v_2 \in C_2$, then the coset $v_1 + C_2 =$ coset $v_2 + C_2$, i.e., the codewords corresponding to $v_1$ and $v_2$ are the same. If $v_1 - v_2 \notin C_2$, $\vec{v}_2 + C_2 \cap \vec{v}_1 + C_2 = \emptyset$, i.e., the codewords corresponding to $v_1$ and $v_2$ are distinct, with no elements in common. Invoking Eq. (A.6), these codewords are therefore orthogonal to each other.  From Eq. (3.2.27), we see that the number of such distinct codewords, which correspond to the (unique) cosets of $C_2$ in $C_1$, is $|C_1|/|C_2| = 2^{k_1}/2^{k_2} = 2^{k_1 - k_2}$. With the normalization constant, $|C_2|^{\frac{1}{2}}$, the codewords in Eq. (3.3.33) then form an orthonormal basis in a Hilbert space of dimension $2^k$, where $k = k_1 - k_2$.

How do the classical error-correcting properties of *C₁* and *C₂* enable us to make the corrections, bit flip and phase flip, respectively, in the quantum domain? Let us start with bit flip error detection and correction.

**Bit flip error Detection and Correction**

Suppose the bit-flip errors are described by an *n* bit vector $e_1$ (as in Section 3.2.5), with 1s where the bit flips occurred and zero everywhere else. If $|v + C_2>$ was the original state, then the corrupted state is

$$|v + C_2 + e_1> = \frac{1}{|C_2|^{1/2}} \sum_{w \in C_2} |v + w + e_1\rangle \tag{3.3.34}$$

---

[13] Henceforth, for convenience, we drop the arrow sign from a (binary) vector.

To detect where the bit flips occurred, one introduces an ancilla (separate measuring apparatus) which contains enough qubits to store the syndrome for the code $C_1$ (in the example of the single bit-flip error correcting code in Section 3.3.2.1, we needed two bits corresponding to a total of *4* syndrome values). These qubits are initially in the all zero state $|0>$. The $H_1$ parity check matrix is then applied, taking $|v + C_2 + e_1 > |0>$ to $|v + C_2 + e_1 > |H_1(v + C_2 + e_1)^T > = |v + C_2 + e_1 > |H_1 e_1^T >$, since $H_1 (v + C_2)^T = 0$. In other words, we obtain the state

$$|v + C_2 + e_1 > |H_1(v + C_2 + e_1)^T > = \frac{1}{|C_2|^{1/2}} \sum_{w \in C_2} |v + w + e_1\rangle |H_1 e_1^T > \qquad (3.3.35)$$

The ancilla is now measured, which yields the classical syndrome $H_1 e_1^T$. After this measurement, the ancilla is discarded, giving back the state, Eq. (3.3.34). But, from the knowledge of the syndrome $H_1 e_1^T$, we can now infer the error $e_1$, knowing $C_1$ can correct $\delta$ errors. Once the error $e_1$ is known, we can then perform the quantum error correction by applying the *X* operator to the appropriate qubits. The initial state, Eq. (3.3.33), is then obtained.

$|\underline{H_1 e_1}^T>$

Note that embedded within the ancilla measurement is a mapping from the quantum syndrome to the classical syndrome, which was discussed in Section 3.3.3.2 within the context of the quantum single error correction. This mapping in the context of the CSS codes involving *n* qubits is restated below.

Let $H_1$ denote the parity check matrix for code $C_1$. $H_1$ corrects bit-flips caused by the action of the *X* operators acting on different qubits. Following rule development for transitioning from the classical domain to the quantum domain (Section 3.3.3.2), we define operator $\sigma_z^{[r]}$ for the $r_{th}$ row of $H_1$ as follows:

$$\sigma_z^{[r]} = \sigma_1 \otimes \sigma_2 \otimes \sigma_3 \otimes \ldots \ldots \otimes \sigma_n , \qquad (3.3.36)$$

where *n* is the number of qubits on which it acts, and $\sigma_i = Z$ (the z-Pauli spin matrix) or 2 x 2 identity matrix *I* according as the $i_{th}$ bit in the row is 1 or 0.

To calculate the syndrome of the received *n*-qubit quantum codeword, we replace, as before, the rows of the $H_1$ matrix by the corresponding $\sigma_z^{[r]}$ operators, which act on the received *n*-qubit codeword to generate the syndrome *s*; this syndrome is an $n\text{-}k_1$ x *1* column vector; the value of the $r_{th}$ element of *s* is 0 or 1 according as the eigenvalue of $\sigma_z^{[r]}$ is 1 or -1. The syndrome permits identification of where the bit-flips occurred. These errors are subsequently corrected by applying the operator, *X* (the *x*-Pauli spin matrix).

**Phase flip error detection and correction**

Suppose the phase-flip errors are described by an *n* bit vector $e_2$ containing 1s where the phase-flips occurred and zero everywhere else. If $|v + C_2 >$ was the original state, then the corrupted state is

$$\frac{1}{|C_2|^{1/2}} \sum_{w \in C_2} (-1)^{(v+w).e_2} |v+w\rangle$$

Phase-flip refers to the error caused by the action of the $Z$ operator. Under this operation, $|0> \rightarrow |0>$ and $|1> \rightarrow -|1>$. Thus, wherever there is a 1 in the $n$ bit string $v+w$, a negative sign is picked up if the corresponding qubit undergoes a phase-flip, an action represented by a 1 in the error vector $e_2$. To detect and correct errors, we first apply Hadamard transform to each one of the $n$ qubits (Appendix C). This gives the state

$$\frac{1}{2^{n/2}|C_2|^{1/2}} \sum_{w \in C_2} \sum_{z} (-1)^{z.(v+w)} (-1)^{(v+w).e_2} |z\rangle, \tag{3.3.38}$$

where the sum is over all possible values of $n$-bit string $z$ (see Eq. F.2). Writing $z' = z + e_2$, the state is rewritten as

$$\frac{1}{2^{n/2}|C_2|^{1/2}} \sum_{z'} \sum_{w \in C_2} (-1)^{z'.(v+w)} \quad |z' + e_2\rangle \tag{3.3.39}$$

Now $z'$ either belongs to $C_2^{\perp}$ or not. If $z' \in C_2^{\perp}$, then $\sum_{w \in C_2} (-1)^{w.z'} = |C_2|$; if $z' \notin C_2^{\perp}$, then $\sum_{w \in C_2} (-1)^{w.z'} = 0$ (see Section 3.2.7). Using these results, the above state is written as

$$\frac{1}{2^{n/2}/|C_2|^{1/2}} \sum_{z' \in C_2^{\perp}} (-1)^{z'.v} |z' + e_2\rangle \tag{3.3.40a}$$

which just looks like a bit-flip error described by the error $e_2$. *Thus, we see that, for $\delta$ bit-flip errors to be corrected (in the Hadamard transformed state) or equivalently, $\delta$ phase-flip errors to be corrected, it is $C_2^{\perp}$ (and not $C_2$) that should be $\delta$-error correcting.*

For detection of the bit flip errors in Eq. (3.3.40a), we introduce, as before, an ancilla and apply the parity check matrix $H_2$ corresponding to $C_2^{\perp}$ to obtain $H_2 e_2$ (see Eq. (3.3.35)). We then correct the "bit flip error" $e_2$ above to obtain the state

$$\frac{1}{2^{n/2}/|C_2|^{1/2}} \sum_{z' \in C_2^{\perp}} (-1)^{z'.v} |z'\rangle \tag{3.3.40b}$$

To transform to the original state, we apply to each qubit the inverse of the Hadamard transform, which is the Hadamard transform itself (see Appendix B):

$$\tag{3.3.40c}$$

$$\frac{1}{2^{n/2}/|C_2|^{1/2}} \sum_{z' \in C_2^{\perp}} (-1)^{z'.v} \frac{1}{2^{n/2}} \sum_{z''} (-1)^{z'.z''} |z''\rangle$$

which can be rewritten as

$$\frac{1}{2^n/|C_2|^{1/2}} \sum_{z''} |z''\rangle \sum_{z' \in C_2^{\perp}} (-1)^{z'.(z''+v)}$$

Substituting $z = z'' + v$, the above state becomes

$$\frac{1}{2^n/|C_2|^{1/2}} \sum_{z} |v + z\rangle \sum_{z' \in C_2^{\perp}} (-1)^{z'.z}$$

Using Eqs. (3.2.29a-b), the right-hand sum becomes equal to $|C_2^{\perp}| = 2^n/|C_2|$ when $z \in C_2$ and is equal to zero when $z \notin C_2$. Consequently, the above expression simplifies to

$$\frac{1}{|C_2|^{1/2}} \sum_{z \in C_2} |v + z\rangle \qquad (3.3.40d)$$

which is the original, uncorrupted state; thus the phase-flip errors have been corrected.

<u>The Parity Check Matrix $H_2$ and the Syndrome</u>

From Eq. (3.3.40a), we see that it is the parity check matrix for code $C_2^{\perp}$ that needs to be employed to correct the "bit flip" errors in the Hadamard transformed basis (or phase-flip errors in the original basis). Let $H_2$ denote this parity check matrix. Then, the quantum equivalent, which is a $k_2$ x 1 column operator, is generated by the previous rules for transitioning from a classical domain to the quantum domain. In particular, Eq. (3.3.36) defines the $r_{th}$ element of this operator. The syndrome $s$ is then calculated by applying the $H_2$ operator to the given (Hadamard) transformed state (Eq. (3.3.40a).

Alternatively, in the original basis, obtained by another application of the Hadamard transform (Eq. (3.3.40c)), we see that the result $ZH_d|i\rangle$, where $|i\rangle = |0\rangle$ or $|1\rangle$, becomes $H_dZH_d|i\rangle = X|i\rangle$, where $X$ is the bit-flip operator (Eq. B.11a). The identity operator $I$ clearly remains unaltered ($H_d$ $I$ $H_d = I$). Thus, in the original basis, the $r_{th}$ element of $H_2$, which we denote by $\sigma_x^{[r]}$, is given by

$$\sigma_x^{[r]} = \sigma_1 \otimes \sigma_2 \otimes \sigma_3 \otimes \ldots \ldots \otimes \sigma_n , \qquad (3.3.41)$$

where $n$ is the number of qubits on which it acts, and $\sigma_i = X$ (the $x$-Pauli spin matrix) or 2 x 2 identity matrix $I$ according as the $i_{th}$ bit in the row is 1 or 0. $\sigma_x^{[r]}$ operators act on the received $n$-qubit codeword to generate the syndrome $s$, which is a $k_2$ x 1 column vector; the value of the $r_{th}$ element of $s$ is 0 or 1 according as the eigenvalue of $\sigma_x^{[r]}$ is 1 or -1. The syndrome permits identification of where the phase-flips occurred. These errors are subsequently corrected by applying the operator, $Z$ (the $z$-Pauli spin matrix).

**Example 1**:  We illustrate the construction of the above CSS quantum code, using simple linear codes such as the binary parity check code of length 4 (see Section 3.2). Although this code only detects an error (and does not correct it), the algebra involved is easier to deal with and facilitates understanding of the underlying concepts of the CSS code.

1) Let $C_1$ be the binary parity check code of length 4. This code is a *[4, 3]* linear code generated by

$$G_1 = \begin{bmatrix} 1 & 0 & 0 & 1 \\ 0 & 1 & 0 & 1 \\ 0 & 0 & 1 & 1 \end{bmatrix},$$ (3.3.42)

which is a *3 x 4* matrix ($k_1$ = 3, $n$ = 4). This generator matrix then yields

$C_1$ = {0000, 0011, 0101, 0110, 1001, 1010, 1100, 1111}. (3.3.43)

The corresponding check matrix $H_1$ (a $n$- $k_1$ $x$ $n$ matrix) satisfies $H_1 G_1^T = 0$, and is given by

$H_1 = [1111].$ (3.3.44)

$(U)H_1$ is also the generator of the dual code: $C_1^{\perp}$ = {0000, 1111}, which is a *[4, 1]* linear code. To generate a CSS code, we also need $C_2 \subset C_1$. In this example, we choose

$C_2 = C_1^{\perp}$ = {0000, 1111}, (3.3.45)

since $C_1^{\perp} \subset C_1$ (weakly self-dual); $k_2$ = 1. Having chosen $C_1$ and $C_2$, we now construct the quantum codewords $Q_i$ by constructing cosets in accordance with Eq. (3.3.33):

$$v = 0000 \text{ (or 1111)} \rightarrow Q_1 = \frac{1}{\sqrt{2}} \left[ |0000> + |1111> \right]$$ (3.3.46a)

$$v = 0011 \text{ (or 1100)} \rightarrow Q_2 = \frac{1}{\sqrt{2}} \left[ |0011> + |1100> \right]$$ (3.3.46b)

$$v = 0101 \text{ (or 1010)} \rightarrow Q_3 = \frac{1}{\sqrt{2}} \left[ |0101> + |1010> \right]$$ (3.3.46c)

$$v = 0110 \text{ (or 1001)} \rightarrow Q_4 = \frac{1}{\sqrt{2}} \left[ |0110> + |1001> \right]$$ (3.3.46d)

Note that there are four different (and thus orthogonal) quantum codewords ($|C_1|/|C_2|$ = 4); also there are $|C_2|$ (= 2) different values of $v \in C_1$, which yield the same coset $v + C_2$ (see Section 3.2.6) and thus the same quantum codeword.  These four codewords may be the encodings of the following sets of qubits:

$$|00> \rightarrow \; Q_1 = \frac{1}{\sqrt{2}} \left[ |0000> + |1111> \right] \tag{3.3.47a}$$

$$|01> \rightarrow \; Q_2 = \frac{1}{\sqrt{2}} \left[ |0011> + |1100> \right] \tag{3.3.47b}$$

$$|10> \rightarrow \; Q_3 = \frac{1}{\sqrt{2}} \left[ |0101> + |1010> \right] \tag{3.3.47c}$$

$$|11> \rightarrow \; Q_4 = \frac{1}{\sqrt{2}} \left[ |0110> + |1001> \right] \tag{3.3.47d}$$

(here $k = k_1 - k_2 = 3 - 1 = 2$, and $n = 4$).

<u>Bit Flip Error Detection and Correction</u>

Suppose qubits encoded to $Q_1$ are sent by Alice to Bob, and a single bit-flip on its way occurs due to the noise in the quantum channel; bit-flip is represented by the $X$ (the *x*-Pauli spin operator). Suppose the qubit in the first position is affected. Then, the received $Q_1$ codeword, containing the bit-flip error in the first position, can be written as:

$$Q_1^{(E)} = \frac{1}{\sqrt{2}} \left[ |1000> + |0111> \right] \tag{3.3.48}$$

*Syndrome Extraction*:  Now Eq. (3.3.36) applied to $H_1$ (which has a single row, see Eq. (3.3.44)) yields

$$\sigma_z^{[1]} = \; Z \otimes Z \otimes Z \otimes Z \tag{3.3.49}$$

Upon application to $Q_1^{(E)}$, one obtains

$$\sigma_z^{[1]} Q_1^{(E)} = (-1) \, Q_1^{(E)}, \tag{3.3.50}$$

because $\sigma_z$ acting on |0> or |1> gives +1 or -1, respectively.  The eigenvalue of $\sigma_z^{[1]}$ = -1 then translates to a syndrome *s* given by

$$s = [1], \qquad \text{(a 1 x 1 column vector)} \tag{3.3.51}$$

indicating an error has occurred. Note that the number of syndromes is $2^{n-k} = 2^{4-3} = 2$; thus $s = [0]$ or [1], where the former indicates no error. The binary code $C_1$ has a minimum distance (or minimum weight) $d$ = 2, which means the number of errors it can detect is $d$ -1 = 1, which is consistent with the result above. Likewise, the number of errors the code $C_1$ can correct is given by $\lfloor (d-1)/2 \rfloor$ = 0 Section 3.2.5).  As one can see by examination of the above expression for $Q_1^{(E)}$, there is no way one can correct it with any confidence to $Q_1$, since we do not know which bit flipped; as a result, all $Q_i's$ are equally possible upon a single qubit correction.

<u>Phase-flip Error Detection and Correction</u>

Suppose now the $Q_1$ codeword sent by Alice suffers a single phase-flip on its way to Bob; this happens through the action of the $z$-Pauli spin operator $Z$. We assume here that the phase of the first qubit of the encoded qubits has flipped. The received $Q_1$ codeword is then given by

$$Q_1^{(E')} = \frac{1}{\sqrt{2}} \left[ |0000> - |1111> \right] \tag{3.3.52}$$

*Syndrome Extraction*: One needs to use the parity check matrix $H_2$ for the code $C_2^\perp$. Now $C_2^\perp = (C_1^\perp)^\perp = C_1$. So,

$$H_2 = H_1 = [1111], \tag{3.3.53}$$

If we now apply Eq. (3.3.41) to Eq. (3.3.53), we obtain

$$\sigma_x^{[1]} = X \otimes X \otimes X \otimes X, \tag{3.3.54}$$

which, when applied to $Q_1^{(E')}$, yields

$$\sigma_x^{[1]} Q_1^{(E')} = \frac{1}{\sqrt{2}} \left[ |1111> - |0000> \right] = (-1) \, Q_1^{(E')}. \tag{3.3.55}$$

The eigenvalue of -1 then implies that the syndrome $s$ is [1] (a *1 x 1* column vector), implying a phase-flip error has occurred. Again, we cannot determine which qubit's phase actually underwent phase-flip (we can verify this by examining Eq. (3.3.52)).

<u>Alternatively</u>, one can first apply to $Q_1^{(E')}$ the Hadamard transform $H_d$:

$$H_d = \frac{1}{\sqrt{2}} \begin{bmatrix} 1 & 1 \\ 1 & -1 \end{bmatrix} \tag{3.3.56}$$

$(U)H_d$ transforms $|0>$ and $|1>$ to $|+>$ and $|->$, respectively, where $|+> = (1/\sqrt{2})[|0> + |1>]$ and $|-> = (1/\sqrt{2})[|0> - |1>]$. Denoting the Hadamard transformed received codeword by $Q_1^{(E')}$,

$$(Q_1^{(E')})' = H_d^{\otimes 4} \, Q_1^{(E')} = \frac{1}{\sqrt{2}} \left[ |++++> - |---->\right], \tag{3.3.57}$$

which becomes, upon substituting for $|+>$ and $|->$,

$$\left(Q_1^{(E')}\right)' = (1/(2\sqrt{2}))[|0001> + |0010> + |0100> + |1000> + |1110> + |1011> + |1101>] \tag{3.3.58}$$

(the bit strings here all have odd parity). Considering the error as a 'bit-flip' error (see Eq. (3.3.40a)), we can apply Eq. 3.3.54 with $X$ operator replaced with $Z$ operator. One still gets (-1) as the eigenvalue, as before. The 'bit-flip' error again is not correctable. In the event, the error was correctable, then after correction, one would again perform the Hadamard transform on each of the qubits to return to the original basis, in which the phase-flip error would be corrected, giving us back the original state $Q_1$. The effect of $\sigma_x^{[1]}$ on a single qubit belonging to the codeword can be summarized as follows: starting with $X$

|i>, where |i> is a single qubit state, we can rewrite it as $H_d H_d X H_d H_d$ |i> = $H_d Z H_d$ |i>, where the last $H_d$ is applied after the error correction (not possible here) to obtain the original state.

Note that the fact that $\left(Q_1^{(E')}\right)'$ has an eigenvalue of (-1) can also be seen from Eq. (3.3.57), because $Z$ acting on |+> transforms it to |-> and vice versa.

**Example 2   CSS codes based on the *[7,4]* Hamming code**

A *[7, 4]* Hamming code has $n$ = 7, $k$ = 4, and $d$ (minimum distance) = 3. As a result, it can detect 2 errors (= $d$ - 1 = 2), and correct 1 error ($\lfloor (d-1)/2 \rfloor$ = 1).  Thus a quantum code based on this code should be able to correct one error. Let $C_1$ be the *[7, 4]* Hamming code. The generator matrix $G_1$ for this code is given by (Eq. 3.2.7)

$$G_1 = \begin{bmatrix} 1 & 0 & 0 & 0 & 1 & 1 & 0 \\ 0 & 1 & 0 & 0 & 1 & 1 & 1 \\ 0 & 0 & 1 & 0 & 1 & 0 & 1 \\ 0 & 0 & 0 & 1 & 0 & 1 & 1 \end{bmatrix} \qquad \text{(a 4 x 7 matrix)} \tag{3.3.59}$$

$C_1$ = {0000000, 0001011, 0010101, 0011110, 0100111, 0101100, 0110010, 0111001, 1000110, 1001101, 1010011, 1011000, 1100001, 1101010, 1110100, 1111111}, (3.3.60)

which is a set of 16 elements. The corresponding check matrix $H_1$, satisfying $H_1 G_1^T = 0$, is given by

$$H_1 = \begin{bmatrix} 1 & 1 & 1 & 0 & 1 & 0 & 0 \\ 1 & 1 & 0 & 1 & 0 & 1 & 0 \\ 0 & 1 & 1 & 1 & 0 & 0 & 1 \end{bmatrix} \qquad \text{(a 3 x 7 matrix)} \tag{3.3.61}$$

$H_1$ is the generator of  $C_1^\perp$; where

$$C_1^\perp = \{0000000, 1110100, 1101010, 0111001, 0011110, 1001101, 1010011, 01000111\} \tag{3.3.62}$$

As one can easily verify that  $C_1^\perp \subset C_1$ (weakly self-dual). We choose $C_2$ = $C_1^\perp$, which means  $C_2^\perp = C_1$. Therefore, $H_2$, the check matrix for  $C_2^\perp$, is equal to $H_1$ given above. There are  $|C_1|/|C_2|$ = $2^4/2^3$ = 2 cosets of $C_2$ in $C_1$, implying there are **2 quantum codewords possible from Eq. (3.3.33)**:

$$v_1 = 0000000 \rightarrow Q_1 = \frac{1}{\sqrt{8}} [|0000000> + |1110100> + |1101010> + |0111001>$$
$$+ |0011110> + |1001101> + |1010011> + |01000111>] \tag{3.3.63a}$$

$$v_2 = 0001011 \rightarrow Q_2 = \frac{1}{\sqrt{8}} [|0001011> + |1111111> + |1100001> + |0110010>$$
$$+ |0010101> + |1011000> + |1000110> + |0101100>] \tag{3.3.63b}$$

(Note that $v_1, v_2 \in C_1$, but $v_{1-} v_2 \notin C_2$, as required; we could choose a different pair of $v_1$ and $v_2$, but the end result will always be the same, i.e., the generation of the above two codewords).

A Hilbert space of dimensionality 2 is protected, i.e., a single qubit is encoded into 7 qubits as given above in two possible ways to yield two quantum codewords (orthogonal to each other); these two quantum codewords may be regarded as the encodings of single qubit states |0> and |1>.

Bit-Flip Error Detection and Correction

Using Eq. (3.3.36), we get for the 3 rows of $H_1$ in Eq. (3.3.61):

$$\sigma_z^{[1]} = Z \otimes Z \otimes Z \otimes I \otimes Z \otimes I \otimes I \qquad (3.3.64a)$$

$$\sigma_z^{[2]} = Z \otimes Z \otimes I \otimes Z \otimes I \otimes Z \otimes I \qquad (3.3.64b)$$

$$\sigma_z^{[3]} = I \otimes Z \otimes Z \otimes Z \otimes I \otimes I \otimes Z. \qquad (3.3.64c)$$

$(U)\sigma_z^{[i]}$ acting on the received 7-qubit quantum codeword yields the $i$th element of the syndrome $s$, which is an 3 x 1 column vector (here $n$ - $k_1$ = 7 − 4 = 3); the $i$th element of $s$ is 0 or 1 according as the eigenvalue of $\sigma_z^{[i]}$ is +1 or -1, respectively.

For illustration purposes, we assume the quantum codeword under consideration is one of $Q_1$ and $Q_2$ given in Eqs. (3.3.63a) and (3.3.63b), respectively. If there is no error in the received word, one can verify by applying Eqs. (3.3.64a-c) that

$s = \begin{bmatrix} 0 \\ 0 \\ 0 \end{bmatrix}$. If a (single) error has occurred, then $s$ will have non-zero elements:

$\begin{bmatrix} 1 \\ 1 \\ 0 \end{bmatrix}$ (first-qubit flip), $\begin{bmatrix} 1 \\ 1 \\ 1 \end{bmatrix}$ (second-qubit flip), $\begin{bmatrix} 1 \\ 0 \\ 1 \end{bmatrix}$ (third-qubit flip), $\begin{bmatrix} 0 \\ 1 \\ 1 \end{bmatrix}$ (fourth-qubit flip), $\begin{bmatrix} 1 \\ 0 \\ 0 \end{bmatrix}$ (fifth-qubit flip), $\begin{bmatrix} 0 \\ 1 \\ 0 \end{bmatrix}$ (sixth-qubit flip), $\begin{bmatrix} 0 \\ 0 \\ 1 \end{bmatrix}$ (seventh-qubit flip).

First-qubit bit flip refers to the qubit in the first position flipping, and so on. One sees that for a different position qubit, the syndrome $s$ is different. Thus, by matching the syndrome determined for the received word to the table above, one knows in which position the bit-flip occurred. Once the position of the bit-flip error is known, one applies the bit-flip operator, $X$ (NOT gate), to the qubit in that position to correct the error.

Phase-Flip Error Detection and Correction

Here we employ $H_2$, the parity check matrix of $C_2^{\perp}$ to detect errors. As discussed earlier, $H_2 = H_1$ (see Eq. (3.3.361). Using Eq. (3.3.41), we transform the three rows of $H_2$ to

$$\sigma_x^{[1]} = X \otimes X \otimes X \otimes I \otimes X \otimes I \otimes I \qquad (3.3.65a)$$

$$\sigma_x^{[2]} = X \otimes X \otimes I \otimes X \otimes I \otimes X \otimes I \qquad (3.3.65b)$$

$$\sigma_x^{[3]} = I \otimes X \otimes X \otimes X \otimes I \otimes I \otimes X. \tag{3.3.65c}$$

$(U)\sigma_x^{[i]}$ acting on the received *7*-qubit quantum word yields the *i*th element of the syndrome *s*, which is an *3 x 1* column vector (here $n - k_1 = 7 - 4 = 3$); the *i*th element of *s* is 0 or 1 according as the eigenvalue of $\sigma_x^{[i]}$ is +1 or -1, respectively.

Because $H_2 = H_1$, the results for the syndrome *s* obtained by the application of Eqs. (3.3.65a-c) to the received word corresponding to the codeword $Q_1$ or $Q_2$ are identical to those for the bit-flip error case.

That is,

$s = \begin{bmatrix} 0 \\ 0 \\ 0 \end{bmatrix}$ (no phase-flip error), $\begin{bmatrix} 1 \\ 1 \\ 0 \end{bmatrix}$ (first-qubit phase flip), $\begin{bmatrix} 1 \\ 1 \\ 1 \end{bmatrix}$ (second-qubit phase flip), $\begin{bmatrix} 1 \\ 0 \\ 1 \end{bmatrix}$ (third-qubit phase flip), $\begin{bmatrix} 0 \\ 1 \\ 1 \end{bmatrix}$ (fourth-qubit phase flip), $\begin{bmatrix} 1 \\ 0 \\ 0 \end{bmatrix}$ (fifth-qubit phase flip), $\begin{bmatrix} 0 \\ 1 \\ 0 \end{bmatrix}$ (sixth-qubit phase flip), $\begin{bmatrix} 0 \\ 0 \\ 1 \end{bmatrix}$ (seventh-qubit phase flip).

As before, for a different position qubit undergoing phaseflip, the syndrome *s* is different. Thus, by matching the syndrome determined for the received word to the table above, one knows which qubit underwent phase-flip. Once this is known, one applies the phase-flip operator, the *Z* gate, to that qubit to correct the error.

**Simultaneous occurrence of bit-flip and phase-flip errors**

When a given codeword is subjected to both bit-flip and phase-flip errors, the corrupted CSS state is written as[14]

$$\frac{1}{|C_2|^{1/2}} \sum_{w \in C_2} (-1)^{(v+w).e_2} |v + w + e_1\rangle \tag{3.3.66}$$

One normally performs bit error correction first, using the $H_1$ matrix, as described above to get rid of the bit-flip error $e_1$; subsequently one performs Hadamard transform, followed by bit error correction in the transformed space, using $H_2$ instead of $H_1$, and Hadamard transform to convert back to the original space[15], which now has codewords with both the bit error and phase-flip errors corrected. This also includes correction of a bit-error followed by a phase-flip error on the *same* qubit, leading to the conclusion that arbitrary errors can also be corrected.

---

[14]Here simultaneous action of the bit-flip and phase flip-operators on the same (single) qubit is represented by the *XZ* operator (phase-flip followed by bit-flip as in Eq. (3.3.19)); we could have chosen to consider *ZX* (bit-flip followed by a phase-flip) instead, but that would only lead to an overall extra phase factor $(-1)^{e_1.e_2}$ in Eq. (3.3.66), which can be ignored.

[15] One could also use the operators based on the *x*-Pauli spin operator in the original basis, as discussed before, but it is practically more convenient to use Hadamard transforms (gates) and perform measurements using the *z*-Pauli spin operators.

### 3.3.4 The Generalized CSS Codes

The CSS codes developed in the previous section are generalized further by the addition of two parameters, denoted $x$ and $z$, which yield a family of equivalent codes, i.e., the new parameterized codes have the same error-correcting properties as the original codes. As we will show below, these codes form an orthonormal basis in the $2^n$ – dimensional Hilbert Space, a result that will be used in the Shor-Preskill security proof of the BB84 protocol.

The generalized CSS codes, denoted $CSS_{x,z}$, are given by

$$ v \rightarrow |v + C_2> = \frac{1}{|C_2|^{\frac{1}{2}}} \sum_{w \in C_2} (-1)^{w.z} |v + w + x > $$

(3.3.67)

he parameters $x$ and $z$ are arbitrary $n$-bit strings. The contents (the bit-strings) of the above basis states on the right-hand-side differ from the contents of the basis states in Eq. (3.3.33) by the additional bit-string, $x \in F_2^n$. We first note that upon a transformation: $y \rightarrow y + x$, where $x$ is a constant, the difference between any two $n$-bit elements remains unchanged, i.e., $y_1 - y_2 \rightarrow (y_1 + x) - (y_2 + x) = y_1 - y_2$. Consequently, the previous results (Section 3.2.6) hold: 1) elements within a coset $v + x + C_2$ remain unique 2) there are $2^k$ unique cosets of $C_2$ in $C_1 + x$ (no element of any coset belonging to the set of $2^k$ cosets is in any other coset belonging to this set); $k = k_1 - k_2$. Therefore, the codeword states defined by Eq. (3.3.65) form an orthonormal basis of a $2^k$ – dimensional subspace of the $2^n$ –dimensional Hilbert space.

All the previous arguments for bit and phase-flip errors apply, and the error-correcting properties, summarized earlier in Eqs. (3.3.34-3.3.40), are similarly reproduced. For example, $H_1 e_1^T$ in Eq. (3.3.35) now gets replaced by $H_1(e_1 + x)^T = H_1 e_1^T + H_1 x^T$ ,which is a "shifted" syndrome, i.e., different from $H_1 e_1^T$ by a constant amount equal to $H_1 x^T$. These "shifted" syndromes, thus, retain their uniqueness and comprise the full set of the $2^{n-k}{}_1$ syndromes. In phase-flip correction, besides the constant vector $x$ added to $v + w$ in Eq. (3.3.337), we also have the phase term $(-1)^{w.z}$ multiplying $(-1)^{(v+w+x).e_2}$ due to phase-flip. If we now multiply the entire codeword (or quantum state) by a constant phase $(-1)^{(v+x).z}$ (which does not change the contents of the codeword), and carry out the Hadamard transform as before, we arrive at a result similar to Eq. (3.3.40a); see Appendix E for details.

<u>$CSS_{x,z}$ forming an orthonormal basis in the $2^n$ - dimensional Hilbert space</u>

We observed above that adding a constant $n$-bit parameter $x$ did not alter the uniqueness properties of the cosets $v + C_2$, where $v \in C_1$; there are $2^{k_1 - k_2}$ such cosets. The question arises: how many more unique cosets can be generated by varying the parameter $x \in F_2^n$? There are a total of $2^n$ unique elements in $F_2^n$. The number of unique elements in the $2^{k_1-k_2}$ cosets = $2^{k_1-k_2}$ times $2^{k_2}$ (the number of elements in each coset) = $2^{k_1}$ , the cardinality of code $C_1$ (as expected). Let us suppose that for some parameter $x$ added to the $2^{k_1-k_2}$ cosets of $C_2$ in $C_1$, we obtain a different set of $2^{k_1-k_2}$ cosets; this set of $2^{k_1-k_2}$ cosets will not only be unique with respect to each other, but also with respect to the previous set of $2^{k_1-k_2}$ cosets (corresponding to $x = 0$). The number of unique elements in this set of $2^{k_1-k_2}$ cosets is also $2^{k_1}$. One, therefore, sees that the total number of such non-overlapping sets (each corresponding to a different value of $x$) in the binary vector space $F_2^n$ is $2^n/2^{k_1} = 2^{n-k_1}$. Equivalently, for each codeword defined by $v$ in Eq. (3.3.67), we have $2^{n-k_1}$ codewords that are orthogonal to each other. The codewords (or the basis vectors) corresponding to the $2^{k_1-k_2}$ values of $v$ and $2^{n-k_1}$ distinct values of parameter $x$ (that generate a different coset) then form a subspace of dimension: $2^{k_1-k_2} \times 2^{n-k_1} = 2^{n-k_2}$.

Are there other codewords (derivatives of Eq. (3.3.67)) that are orthogonal to the subspace defined above? We rewrite Eq. (3.367) for convenience as

$$|\psi_{v,x,z}> = \frac{1}{|C_2|^{\frac{1}{2}}} \sum_{w \in C_2} (-1)^{w.z} |v + w + x>$$

(3.3.68)

We further note that the subspace of dimensionality $2^{n-k_2}$ defined above did not take into account the phase factor $(-1)^{w.z}$, and therefore, corresponded to Eq. (3.3.67) with $z = 0$. Denoting this $z = 0$ parameter state by $|\psi_{v,x}>$,

$$|\psi_{v,x}> = \frac{1}{|C_2|^{\frac{1}{2}}} \sum_{w' \in C_2} |v + w' + x>$$

(3.3.69)

Consider now the scalar product:

$$<\psi_{v,x,z}|\psi_{v,x}> = \frac{1}{|C_2|} \sum_{w \in C_2} \sum_{w' \in C_2} (-1)^{z.w} <v + w + x|v + w' + x>$$

$$= \frac{1}{|C_2|} \sum_{w \in C_2} \sum_{w' \in C_2} (-1)^{z.w} \delta_{ww'}$$

$$= \frac{1}{|C_2|} \sum_{w \in C_2} (-1)^{z.w}$$

$$= \begin{cases} 1 & if\ z \in C_2^{\perp} \\ 0 & if\ z \notin C_2^{\perp} \end{cases}$$

(3.3.70)

where we have used the results, Eq. (A.6) and Eqs. (3.2.29a-b). The first result in Eq. (3.3.70) means that there are $|C_2^{\perp}| = 2^{n-k_2}$ distinct values of $z \in F_2^n$ for which $|\psi_{v,x,z}> = |\psi_{v,x}>$. The second (orthogonality) result, in conjunction with the first result, implies that there are $2^n/2^{n-k_2} = 2^{k_2}$ distinct states orthogonal to $|\psi_{v,x}>$, and therefore to the subspace of dimensionality $2^{n-k_2}$ described above. Consequently, the states $|\psi_{v,x,z}>$, described by the expression, Eq. (3.3.67), form an orthonormal

basis in the complex Hilbert space of dimensionality $2^{n-k_2}$ x $2^{k_2}$ = $2^n$. *In summary, the $2^{k_1-k_2}$ distinct values of $v_k$, the $2^{n-k_1}$ distinct values of $x$, and the $2^{k_2}$ distinct values of $z$ lead to $2^n$ (= $2^{k_1-k_2}$ x $2^{n-k_1}$ x $2^{k_2}$) different orthogonal states $|\psi_{v,x,z}>$ (Eq. (3.3.67)), which form an alternative orthonormal basis in the $2^n$ – dimensional Hilbert space.*

In Appendix A, the $2^n$ – dimensional Hilbert space was defined with respect to the basis vectors, $|j_1 j_2 \ldots j_{n-1} j_n>$, where the set of $j_1 j_2 \ldots j_{n-1} j_n$ strings ($j_i \in (0,1)$, $i = 1,2,.., n$) are the $2^n$ unique $n$-bit strings in the vector space $F_2^n$. Below we show the equivalence of the two bases via a result that will be used in the security proof of the BB84 protocol.

<u>*Equivalance of the $|\psi_{v,x,z}>$ basis and the $|j_1 j_2 \ldots j_{n-1} j_n>$ basis reexpressed*</u>

Consider

$$\sum_{v,x,z} |\psi_{v,x,z}> \ |\psi_{v,x,z}>$$

$$= \sum_{v,x,z} \frac{1}{|C_2|} \sum_{w \in C_2} \sum_{w' \in C_2} (-1)^{z.(w+w')} |v+w'+x> |v+w+x>$$

$$= \sum_{v,x} \frac{1}{|C_2|} \sum_{w \in C_2} \sum_{w' \in C_2} \sum_{z} (-1)^{z.(w+w')} |v+w'+x> |v+w+x>$$

$$= \sum_{v,x} \sum_{w \in C_2} |v+w+x> |v+w+x> \qquad (3.3.71)$$

where we have used the result[16], $\sum_z (-1)^{z.(w+w')} = 2^{k_2} \delta_{ww'}$. Now we show that

$$\sum_{v,x} \sum_{w \in C_2} |v+w+x> |v+w+x> = \sum_j |j> |j> \qquad (3.3.72)$$

where $j$ is a notation for the $n$-bit string $j_1 j_2 \ldots j_{n-1} j_n$ string, and the sum is over such possible strings, which are $2^n$ in number. As discussed above, $2^{k_1-k_2}$ distinct values of $v_k$ give rise to $2^{k_1-k_2}$ distinct cosets. Furthermore, the $2^{n-k_1}$ distinct values of $x$ generate distinct cosets, giving a total of $2^{k_1-k_2}$ x $2^{n-k_1}$= $2^{n-k_2}$ distinct cosets. Since each coset contains a distinct set of $2^{k_2}$ elements (see Section 3.2.6), there are a total of $2^{n-k_2}$ x $2^{k_2}$ = $2^n$ distinct elements (or distinct $n$-bit strings), i.e., the left-hand-side in the above equation is equal to its right-hand-side. It then follows

---

[16] There are $2^{k_2}$ distinct values of z. These correspond to binary strings belonging to $\{0,1\}^{k_2}$. Similarly, w, w' $\in C_2$ may be regarded as corresponding to $k_2$ bit-strings belonging to $\{0,1\}^{k_2}$. Applying Eqs. (31) and (32), the result follows.

$$\sum_{v,x,z} |\psi_{v,x,z}> \; |\psi_{v,x,z}> \; = \; \sum_{j} |j> |j>, \tag{3.3.73}$$

a result we will be employing in the BB84 proof. The reader may verify, as illustration, that all of the above properties of the generalized CSS codes are satisfied for the binary parity check code of length $n = 4$. For example, along with the 4 distinct values of $v$ (0011, 0101, 0110, 1001), one may choose 0000 and 0001 as the $2^{n-k_1} = 2^{4-3} = 2$ distinct values of $x$ and also 0000 and 0001 as the $2^{k_2} = 2^1 = 2$ distinct values of $z$.

## 3.4   The Shor-Preskill Proof of the BB84 Protocol

The Shor − Preskill proof [4] starts with consideration of the proof of Lo-Chau Protocol [5], which is an entanglement-based protocol, modifies it by replacing entanglement distillation using quantum computers with entanglement distillation using the CSS quantum error-correcting codes (Section 3.3). Subsequently, they reduce the modified Lo-Chau protocol to a protocol where entanglement is eliminated. Calling this protocol the CSS protocol, they provide further arguments to reduce it to a "Prepare and Measure" protocol, which they term the BB84 protocol. This reduction is described in the subsections below.

As a first step, however, it is important to understand the concept of generation of a key within the framework of an entanglement-based protocol, which is the starting point in the proof given by Shor and Preskill. To illustrate the concept, we initially assume the absence of any interference from Eve as well as any noise in the quantum channel. Starting with $n$ EPR pairs of qubits, where each EPR pair is an entangled pair of qubits (see Appendix G), Alice and Bob then obtain a random and private key that is $n$-bits long in the following way:

   a) Alice creates $n$ EPR pairs, $|\beta_{00}>^{\otimes n}$; $|\beta_{00}> = 1/\sqrt{2}(|00> + |11>)$
   b) She sends the second entangled qubit of each pair to Bob.
   c) She makes a measurement of her $n$ halves in the (0, 1) basis (also called the $Z$ basis).
   d) She obtains a random key of length $n$.
   e) Bob makes the measurement on his $n$ qubits and obtains the same result due to entanglement.
   f) Alice and Bob share a random, private key of length $n$.

When Eve and/or noise are present, the $n$ EPR pairs get corrupted. These EPR pairs are then subjected to a purification process [8] called entanglement distillation in which $k$ ($< n$) nearly perfect EPR pairs are obtained. Lo and Chau [5] were the first to give the unconditional security proof of the BB84 protocol. Unfortunately, the entanglement distillation they invoke in their proof involves the use of quantum computers. We summarize below the main ideas of their protocol:

### 3.4.1 Lo-Chau Protocol − The Main Concepts

   a) Alice creates $n$ EPR pairs, $|\beta_{00}>^{\otimes n}$; $|\beta_{00}> = 1/\sqrt{2}(|00> + |11>)$.
   b) She sends the second qubit of each of these pairs to Bob.

c) Before any measurement, Alice and Bob share imperfect EPR pairs because, for each pair, both the qubit Alice keeps and the qubit she sends to Bob are subject to errors (due to the channel noise and Eve's interference) and decoherence arising from interactions with the environment during storage.

d) Alice and Bob perform some local operations (unitary transformations) and classical communication (LOCC) on their halves of the imperfect pairs to distill a smaller number $k$ of perfect EPR pairs, in a process called entanglement distillation [8].

e) Alice and Bob make measurements on their halves of the $k$ distilled EPR pairs to obtain a random, private key of length $k$ ($< n$).

Because the entanglement-based protocol of Lo-Chau requires quantum computers for entanglement distillation, it is not a useful protocol from a practical standpoint. Shor and Preskill, in their quest to provide an unconditional security proof of the BB84 protocol, <u>modify</u> the Lo-Chau protocol such that entanglement distillation is provided via quantum error correction through the use of the CSS quantum error correcting codes (Section 3.3). The details of this protocol are described in Section 3.4.2 below.

### 3.4.2 Modified Lo-Chau Protocol

The steps of this protocol, which replaces entanglement distillation using quantum computers with entanglement distillation using the CSS codes described in Section 3.3, are given below:

1) Alice creates $2n$ EPR pairs, each in state $|\beta_{00}> = 1/\sqrt{2}(|00> + |11>)$.

2) Alice randomly selects $n$ of the $2n$ EPR pairs to serve as check for Eve's interference.

3) Alice selects a random $2n$-bit string $b$ and performs a Hadamard transform on the second qubit of each pair $i$ for which $b_i = 1$.

4) She sends the second qubit for each pair to Bob (in Step 3, if $b_i = 1$ and the second qubit of the $i^{th}$ EPR pair is $|0>$ ($|1>$), then Alice sends $|+>$ ($|->$) to Bob).

5) Bob receives the qubits and publicly announces the fact.

6) Alice announces the random string $b$ and the set of $n$ EPR pairs that will serve as checks.

7) Bob performs Hadamard transform on the qubits, where $b_i = 1$ (if $b_i = 1$ and the qubit Bob receives is $|+>$ ($|->$), he generates $|0>$ ($|1>$)).

8) Alice and Bob measure their $n$ check qubits in the Z basis and publicly share their results; for each check qubit, they should both measure $|0>$ or they should both measure $|1>$; an error is said to occur if their results disagree. If more than $t$ errors occur, they abort the protocol.

9) Alice and Bob measure their remaining $n$ qubits according to the check matrix for a predetermined $[n, k]$ quantum code capable of correcting $t$ errors, i.e., measure $\sigma_z^{[r]}$ for each row $r \in H_1$ and $\sigma_x^{[r]}$ for each row $r \in H_2$; $H_1$ is the parity check code for code $C_1$ capable of correcting $t$ bit-flips, and $H_2$ is the parity check code for code $C_2$ capable of correcting $t$ phase-flips (see CSS codes in Section 3.3). They share the results, compute the syndromes for the errors, and then correct their state, obtaining $k$ nearly perfect EPR pairs.

10) Alice and Bob measure the $k$ EPR pairs in the $Z$ basis to obtain a shared secret key.

Instead of the *n* EPR pairs, Alice starts with *2n* EPR pairs, of which one-half are used as check bits; the protocol is aborted if more than *t* errors occur as it is assumed that all the errors are caused by Eve's interference (this assumption is justified because there is no way to distinguish between errors caused by the channel and the errors caused by Eve; therefore, to be on the safe side, one assumes that all errors are caused by Eve). To detect Eve's presence, it is critical to subject half of the qubits sent to Bob to a Hadamard transform, so half the qubits are prepared in the (+, -) basis instead of the (0, 1) basis (**Step 3**); Bob subsequently performs a Hadamard transform on the appropriate received qubits (**Step 7**), so that all his measurements are also in the (0,1) basis. **Step 9** corresponds to entanglement distillation using the CSS codes. **Step 10** then follows from **Step 9**.

Clearly, the crux of this protocol is **Step 9**, whose interpretation we give below:

We assume here that the *n* EPR pairs are perfect, i.e., initially they have no errors. We write $|\beta_{00}\rangle = 1/\sqrt{2}(|00\rangle + |11\rangle)$ as $|\beta_{00}\rangle = 1/\sqrt{2}(|0_A 0_B\rangle + |1_A 1_B\rangle)$, where we have explicitly labeled the qubits that Alice keeps by A and the qubits she sends to Bob by B. We immediately see that $|\beta_{00}\rangle^{\otimes n} = (1/\sqrt{2})^n \sum_{j\in\{0,1\}^n} |j\rangle_A |j\rangle_B$. Furthermore, invoking the result, Eq. (3.3.73), Alice's *n* EPR state can be re-expressed as

$$|\beta_{00}\rangle^{\otimes n} = (\frac{1}{\sqrt{2}})^n \sum_{j\in\{0,1\}^n} |j\rangle_A |j\rangle_B = (\frac{1}{\sqrt{2}})^n \sum_{v,x,z} |\psi_{v,x,z}\rangle_A |\psi_{v,x,z}\rangle_B$$

(3.4.1)

where, as was shown in Section 3.3.4, the states $|\psi_{v,x,z}\rangle$ form an orthonormal basis; there are $2^{k_1-k_2}$ distinct values of *v*, $2^{n-k_1}$ distinct values of *x*, and the $2^{k_2}$ distinct values of *z*, which lead to a total of $2^n$ (= $2^{k_1-k_2}$ x $2^{n-k_1}$ x $2^{k_2}$) different orthogonal states $|\psi_{v,x,z}\rangle$. Thus, Alice has one ket and Bob has the other ket, in a superposition with equal probabilities. Thus, when Alice makes a syndrome measurement on her *n* qubits, using the parity check matrices $H_1$ and $H_2$ in the quantum domain, the measurement projects (or "collapses") her state (in the $2^n$ dimensional Hilbert space) onto the code subspace, CSS$_{x,z}$ of dimensionality $2^{k_1-k_2}$, spanned by the $2^{k_1-k_2}$ codewords $|\psi_{v,x,z}\rangle$ corresponding to fixed values of *x* and *z*; the parameters *x* and *z* are determined from the (measured) syndromes:

$$s_x = H_1 x^T,$$

(3.4.2a)

$$s_z = H_2 z^T$$

(3.4.2b)

The elements of the measured $(n-k_1)$ x 1 column vector $s_x$ form a random binary vector in $\{0,1\}^{n-k_1}$. The *x* vector Alice derives from this measurement can be any of the $2^{k_1}$ different *x* vectors, which are solutions to Eq. (3.4.2a); Alice can choose any of these $2^{k_1}$ values for *x* (see Section 3.3.4). Similarly, Alice selects the value of *z* from her measurement in Eq. (3.4.2b). Alice then communicates the values of *x* and *z* to Bob over the classical channel. Bob also makes measurement of $H_1$ and $H_2$ on his *n*-qubit state, which may be corrupted due to channel noise and/or Eve's interference. Because Bob's qubits are entangled with Alice's, Bob's (error) syndrome values are shifted by a constant amount determined by the values of *x* and *z* Alice sends (Section 3.3.4), i.e.,

$$s_x{'} = H_1 \, (x + e_1)^T \quad = H_1 \, x \quad + H_1 \, e_1^{\,T} \tag{3.4.3a}$$

$$s_z{'} = H_2 \, (z + e_2)^T \quad = H_2 \, z + \; H_2 \, e_2^{\,T} \tag{3.4.3b}$$

From knowledge of $x$ and $z$ and his syndrome measurements, Bob is able to determine the errors, $e_1$ and $e_2$, and correct for them.

Because of the above measurement, which fixes the values of $x$ and $z$, the right-hand-side of Eq. (3.4.1) collapses to

$$\sum_{v,x,z} | \psi_{v,x,z} >_A \; | \psi_{v,x,z} >_B \; \rightarrow \; \sum_{v} | \psi_{v,x,z} >_A \; | \psi_{v,x,z} >_B \tag{3.4.4a}$$

Subsequently, Alice and Bob perform error correction as described above (see Eqs. (3.4.3a-b)) and decode the right-hand-side:

$$\sum_{v} | \psi_{v,x,z} >_A \; | \psi_{v,x,z} >_B \; \rightarrow \; \sum_{j \in \{0,1\}^k} |j>_A \, |j>_B, \tag{3.4.4b}$$

where the right-hand-side, $k$ perfect EPR pairs, is the result of decoding, i.e., $| \psi_{v,x,z} >$, a state of $n$ qubits, decodes to a $k$-qubit quantum state $|j>$, $j \in \{0,1\}^k$. Since the number of errors statistically can exceed the quantum error correction limit of $t$ errors, the $k$ EPR pairs obtained are "nearly" perfect.

In **Step 10**, Alice and Bob measure the $k$ EPR pairs in the (0,1) basis to obtain a shared random key of length $k$.

Summarizing, Shor and Preskill modify the Lo-Chau protocol by replacing entanglement distillation using quantum computers with entanglement distillation using error-correction by the quantum CSS codes; these quantum codes are the generalized versions, *[n, k]* CSS$_{x,z}$ codes, characterized by certain random parameters, $x$ and $z$, and where, as before, $k$ qubits are encoded into $n$ qubits. These codes, also written as $| \psi_{v,x,z} >$, form an orthonormal basis. That is, the $n$ EPR states of Alice and Bob can be written as a superposition of entangled $| \psi_{v,x,z} >$ states corresponding to different sets of parameters, $v$, $x$ and $z$. However, syndrome measurements in accordance with the parity check matrices $H_1$ and $H_2$ then collapse the entangled states of Alice and Bob into entangled $| \psi_{v,x,z} >$ states corresponding to specific parameters $x$ and $z$ determined by the syndrome measurements. Subsequent error correction and decoding then lead to $k$ perfect EPR pairs, which upon measurement in the (0,1) basis yield a random $k$-bit key shared by Alice and Bob.

*The key point to note here is that even though in this modified protocol the n qubits of Alice and Bob (halves of the n EPR pairs) do not start out as code bits, the (quantum) syndrome measurement causes them to act as if they were the n-qubit encoding of k (key) bits; this encoding is performed by the CSS$_{x,z}$ quantum code.*

### 3.4.3 The CSS Protocol

We now reduce the modified Lo-Chau protocol to the CSS protocol described below. We first show that Alice really does not have to transmit Bob's halves of the 2$n$ EPR pairs; the requirement for entanglement goes away, i.e., Alice no longer has to send qubits in entangled states:

i) In **Step 2** of the modified Lo-Chau protocol, Alice randomly selects $n$ of the 2$n$ EPR pairs to serve as check bits. Subsequently, in **Step 8**, she makes the measurement of her check qubits in the (0, 1) basis. Clearly, <u>Alice can make measurement on her qubits up front</u> (before or just after transmitting to Bob his share of the qubits), i.e., she does not have to wait for Bob to receive his own qubits before making her own measurements.

ii) Making measurement of her $n$ EPR pair halves acting as check bits in **Step 8** is also equivalent to her creating $n$ random check bits since, for each of her halves of the EPR pairs, she obtains upon measurement a bit value of either 0 or 1.

iii) Thus, in view of i) and ii), instead of transmitting Bob's halves of the $n$ EPR pairs serving as a check for Eve's interference, Alice encodes $n$ qubits as |0> or |1> in accordance with randomly created check bits, and sends them to Bob.

As a result of these observations, **Steps 1, 2, 4** and **8** of the modified Lo-Chau protocol are modified to:

**1'**. Alice creates $n$ random check bits, and $n$ EPR pairs, $|\beta_{00}>^{\otimes n}$. She also encodes $n$ qubits as |0> or |1> according to the check bits.

**2'**. Alice randomly chooses $n$ positions (out of 2$n$) and puts the check qubits in these positions; she puts half of each EPR pair in the remaining positions.

**4'.** She sends the $n$ check qubits and the second qubit of each of the remaining $n$ EPR pairs to Bob.

**8'**. Bob measures the $n$ check qubits in the (0,1) basis and publicly shares the results with Alice.

Basically, half of the 2$n$ EPR pairs in **Step 1** serving as check qubits (for both Alice and Bob) in **Step 2** have been replaced with Alice simply creating $n$ qubits in **Step 1'** and sending these to Bob as part of **Step 4'**. Because the $n$ EPR pairs in **Step 2** serving as check qubits were randomly selected, the $n$ qubits being sent to Bob by Alice are put in random positions (out of the 2$n$ positions), with the rest being filled by the halves of the remaining $n$ EPR pairs. Since it is Alice who creates the $n$ check qubits and sends them to Bob, it is only Bob who makes the measurement in **Step 8'**.

We now make the following additional observations:

iv) In **Step 9**, measurement of the syndrome by Alice on her $n$ halves of the remaining $n$ EPR pairs (<u>which can also be done up front</u>) is similarly equivalent to her creating and transmitting $k$ halves of EPR pairs encoded into the $n$-qubit CSS$_{x,z}$ code, where $x$ and $z$ are two random vectors selected by her; these random numbers belong to $F_2^n$, i.e., it is equivalent to Alice creating $|\psi_{v,x,z}>$ summed over $v$ (right-hand-side of Eq. (3.4.4a)), where $x$ and $z$ (belonging to $F_2{}^n$) are randomly selected by her. Alice can then announce $x$ and $z$ to Bob.

v) Alice's final measurement in **Step** 10 gives her a specific (random) value of $v$, which we label as $v_k$; this value $v_k$ corresponds to the random $k$-bit key string $K$. This is equivalent to Alice starting with a random $k$-bit string $K$, and then encoding the $k$-qubit state $|K>$ into the $n$ qubit state $|\psi_{v_k,x,z}>$ (see Section 3.3.3).

In view of observations, iv) and v), Alice, instead of creating and sending halves of the remaining $n$ EPR pairs (not used for checking Eve's interference), can equivalently select random values of the $n$-bit string $x$, $n$-bit string $z$, and the $k$-bit string $K \in \{0,1\}^k$, and encode $|K>$ into $|\psi_{v_k,x,z}>$, where $v_k$ corresponds to the string $K$, and send the $n$-qubit encoded state $|\psi_{v_k,x,z}>$ to Bob. To enable Bob to decode the received quantum code, she also then sends the values of $x$ and $z$ over the classical channel. This results in further modifications:

**1''**. Alice creates $n$ random check bits, a random $k$-bit key $K$, and two random $n$-bit strings $x$ and $z$. She encodes $|K>$ into the $n$-qubit quantum state $|\psi_{v_k,x,z}>$ in accordance with the CSS$_{x,z}$ code[17]. She also encodes $n$ qubits as $|0>$ or $|1>$ according to the check bits.

**2''**. Alice randomly chooses $n$ positions (out of $2n$) and puts the check qubits in these positions and the CSS$x,z$ encoded $n$ qubits (which are in the quantum state $|\psi_{v_k,x,z}>$) in the remaining positions.

**4''**. She sends the resulting $2n$ qubits to Bob.

**6'**. Alice announces $b$, $x$, $z$, and which $n$ qubits are to provide check bits.

**9'**. Bob decodes the remaining $n$ qubits from the CSS$x,z$ code, $|\psi_{v_k,x,z}>$.

**10'**. Bob measures his qubits to obtain the shared secret key $K$.

Note that **Steps 8', 9'**, and **10'** contain only Bob's measurements as Alice's measurements - which could be all made up front - are equivalent to the creation of $n$ code qubits in conformity with the CSS$_{x,z}$ codes with randomly chosen $x$, $z$, and $v_k$ (see **Steps 1''** and **2''**) and the creation of $n$ check qubits in accordance with a random $n$-bit string. **Step 6** is modified to **Step 6'** to reflect the fact that values of $x$ and $z$ are also communicated along with string $b$ and the set of $n$ check qubits.

Basically, the $4n$ qubits of the $2n$ entangled EPR pairs have been replaced by the $2n$ qubits Alice creates and sends to Bob (underline{entanglement is eliminated}!). **Step 3** changes to **Step 3'**:

---

[17] There is a 1:1 correspondence between a random string $K \in \{0,1\}^k$ and $|\psi_{v_k,x,z}>$; note however that there are $\frac{2^{k_1}}{2^k} = 2^{k_2}$ different $v_k's$ belonging to $C_l$ that give the same $|\psi_{v_k,x,z}>$, and thus correspond to the same key $K$.

**3'.** Alice selects a random 2$n$-bit string $b$, and performs a Hadamard transform on each qubit $i$ for which $b_i = 1$.

Combining the modified steps (indicated with primed and multi-primed numbers) with the unmodified steps, the modified Lo-Chau protocol of the previous section is reduced to what is termed as the CSS protocol:

<u>CSS Protocol</u>

**1''**. Alice creates $n$ random check bits, a random $k$-bit key $K$, and two random $n$-bit strings $x$ and $z$. She encodes $|K>$ into the $n$-qubit quantum state $|\psi_{v_k,x,z}>$ in accordance with the CSS$_{x,z}$ code. She also encodes $n$ qubits as $|0>$ or $|1>$ according to the check bits.

**2''**. Alice randomly chooses $n$ positions (out of 2$n$) and puts the check qubits in these positions and the CSS$x,z$ encoded $n$ qubits (which are in the quantum state $|\psi_{v_k,x,z}>$) in the remaining positions.

**3'.** Alice selects a random 2$n$-bit string $b$, and performs a Hadamard transform on each qubit $i$ for which $b_i = 1$.

**4''**. She sends the resulting 2$n$ qubits to Bob.

**5.** Bob receives the qubits and publicly announces the fact.

**6'.** Alice announces $b$, $x$, $z$, and which $n$ qubits are to provide check bits.

**7.** Bob performs Hadamard transform on the qubits for which $b_i = 1$.

**8'.** Bob measures the $n$ check qubits in the (0, 1) basis and publicly shares the results with Alice. If more than $t$ errors occur, they abort the protocol.

**9'.** Bob decodes the remaining $n$ qubits from the CSS$x,z$ code, $|\psi_{v_k,x,z}>$; decoding includes error-correction here.

**10'**. Bob measures his qubits to obtain the shared secret key $K$.

Summarizing, in reduction of the modified Lo-Chau protocol to the CSS protocol, <u>the requirement for the two qubits of an EPR pair to be entangled goes way</u>; here Alice simply selects a random $k$-bit key $K$, and accordingly sends to Bob $k$ qubits encoded into $n$ qubits as a CSS$_{x,z}$ code (Section 3.3.4). Bob then makes error-corrections to the received quantum codeword and decodes it to obtain $k$-qubits, which are subsequently measured in the (0,1) basis to yield the $k$-bit key $K$. <u>*An important observation to make here is that encoding and decoding (which includes error correction) is still done in the quantum domain, i.e., quantum gates corresponding to the various required unitary transformations are still being used. Bob also needs to store the transmitted 2n encoded qubits while waiting for communication from Alice.*</u>

In what follows, we now describe a reduction of the CSS protocol to a "Prepare and Measure" protocol, in which qubits are prepared, sent, and measured one at a time. This then leads to the elimination of quantum encoding, decoding, and storage requirements.

### 3.4.4 Reduction to the BB84 Protocol (Shor-Preskill Version)

Let us suppose, instead of decoding the received $n$ qubits in **Step 9'** (which requires unitary transformations and thus the use of quantum computers), Bob simply makes a measurement of the received $n$ qubits (in the $Z$ basis). To understand what this means, let us start with the encoding by Alice of the state $|K\rangle$ (corresponding to the randomly selected $k$-bit key) to the $n$-bit quantum codeword, $|\Psi_{v_k,x,z}\rangle$:

$$|K\rangle \rightarrow \frac{1}{|C_2|^{\frac{1}{2}}} \sum_{w \in C_2} (-1)^{w.z} |v_K + w + x\rangle = |\Psi_{v_K,x,z}\rangle.$$

(3.4.5)

Alice subsequently sends this quantum state to Bob, who receives it with bit –flip error $e_1$ and phase-flip error $e_2$ (see Eq. (E.2)) :

$$\frac{1}{|C_2|^{1/2}} \sum_{w \in C_2} (-1)^{(v_k+w+x).e_2} (-1)^{w.z} |v_k + w + x + e_1\rangle$$

(3.4.6)

When Bob <u>measures</u> the above $n$-qubit quantum state, the state collapses to one corresponding to a fixed $w \in C_2$; this measurement then yields a classical bit string: $v_k + w + x + e_1$, from which Bob finds he can obtain the key $K$ after decoding <u>classically.</u> First, Bob subtracts $x$, using Alice's announcement (**Step 6'** of CSS protocol) to obtain $v_k + w + e_1$. Note that $v_k + w \in C_1$ because $v_K, w \in C_1$ ($w \in C_2$ but $C_2 \subset C_1$). So Bob, using the parity check matrix $H_1$, classically corrects the obtained string $v_k + w + e_1$ to $v_k + w$, which is a codeword in $C_1$.

To obtain the key $K$, Bob now computes the coset $v_k + w + C_2$, which is equal to coset $v_k + C_2$, because $w + C_2 = C_2$; coset $v_k + C_2$ is then identified with the bit-string $K$. <u>It is important to note here that $v_k \in C_1/C_2$ , which denotes (and reemphasizes) the fact that $v_k$ belongs to the set of $2^{k_1-k_2}$ distinct $v_k$'s that produce the $2^{k_1-k_2}$ distinct cosets of $C_2$ in $C_1$ (there is a 1:1 correspondence between the random string $K$ and $v_k \in C_1/C_2$ that yields a distinct coset $v_k + C_2$</u>); see Section 3.3.3.3. We have gone from a definition of key $K$ related to $|\Psi_{v_k,x,z}\rangle$ to a definition of key $K$ related to the coset $v_k + C_2$, but the two are equivalent; this is because the elements $v_k + w(\in C_2)$ of the coset $v_k + C_2$ are the bit-strings of the states $|v_k + w\rangle$ that form $|\Psi_{v_k,0,0}\rangle$ (see Eq. (3.3.33))[18], and $|\Psi_{v_k,0,0}\rangle$ corresponds to a specific key $K$, as in Eq. (3.4.5).

<u>*Important Remarks:*</u> It is very important to note here that as a consequence of the above measurement Bob makes, the phase-factors in the received quantum state are completely ignored. That is,

---

[18]This is the original CSS code without the $x$ and $z$ parameters.

1) Phase-flip errors are completely ignored; <u>the linear code $C_2^\perp$ is not used to correct any phase-flip errors, i.e., $C_2^\perp$ no longer plays a role as a phase-flip error correcting linear code.</u>

2) Because there is no phase-flip correction, <u>Alice does not have to reveal the value of $z$ to Bob</u> who does not need it; earlier, he needed this value to account for the "shift" in the measured syndrome for phase-flip errors due to the presence of $z$ in the quantum code, $CSS_{x,z}$.

As a result of the above quantum measurement by Bob, resulting in classical decoding to obtain the key $K$, **Steps 6'**, **9'**, and **10'** of the CSS protocol change to:

**6''.** Alice announces $b$, $x$, and which $n$ qubits are to provide as check bits.

**9''.** Bob measures the remaining $n$ qubits to obtain $v_k + w + x + e_1$ and subtracts $x$ before correcting it to $v_k + w$.

**10''.** Bob computes the coset $v_k + w + C_2 = v_k + C_2$ to obtain the $k$-bit string $K$.

<u>In summary, Bob does not need to do quantum error correction; the use of quantum computers by Bob to decode the received quantum codeword is eliminated. What about Alice? Can we remove Alice's quantum encoding?</u>

Since the component states, $|v_k + w + x + e_1\rangle$, in Eq. (3.4.6) occur with equal probability, Bob's measurement results in a collapse to $|v_k + w + x + e_1\rangle$ with a random $w \in C_2$ (the accompanying phase factor is not important). This is equivalent to Alice simply choosing at random $w \in C_2$ and preparing the state $|v_k + w + x\rangle$, and sending it to Bob; the state then suffers a bit-flip error $e_1$ en route to Bob; $v_k \in C_1/C_2$.

Another way to understand Alice's prepared state of $|v_k + w + x\rangle$ is the following: because Alice does not have to reveal the value of $z$, which Bob does not use, Alice is effectively sending in a mixed state, which is a random mixture of pure states, each defined by a specific value of $z$. This mixed quantum state is obtained by averaging over random values of $z$. Using the density matrix approach [10], the mixed state is given by

$$\rho_{v_{k},x} = \sum_z p_z |\psi_{v_{k},x,z}\rangle\langle\psi_{v_{k},x,z}| \,, \tag{3.4.7}$$

where $p_z$ is the probability of occurrence of $z$. Taking $z$ to be an $n$-bit string chosen randomly from $\{0,1\}^n$, $p_z = 1/2^n$. Substituting the expression for $|\psi_{v_{k},x,z}\rangle$ (see Section 3.3.4), we obtain

$$\rho_{v_{k},x} = \frac{1}{2^n |C_2|} \sum_{z \in \{0,1\}^n} \sum_{w,w' \in C_2} (-1)^{z.(w+w')} |v_k + w + x\rangle\langle v_k + w' + x| \tag{3.4.8}$$

Invoking the general identity, $\sum_{v \in \{0,1\}^k} (-1)^{v.w} = 2^k \delta_w$ (Eq. 3.2.32), where $\delta_w$ is a Kronecker delta function (equal to 1 when $w = 0$, and equal to zero when $w \neq 0$), the above equation simplifies to

$$\tag{3.4.9}$$

$$\rho_{v_k,x} = \frac{1}{|C_2|} \sum_{w \in C_2} |v_k + w + x> < v_k + w + x|$$

Thus, the mixed state above (averaged over $z$) is nothing but a mixed state of pure states $|v_k + w + x>$ averaged over random values of $w \in C_2$. Thus, all Alice has to do is randomly select $w \in C_2$ to create a state, $|v_k + w + x>$, where $x$ is randomly determined from $\{0,1\}^n$, and $v_k$ ($\in C_1/C_2$) corresponds to the randomly selected $k$-bit string $K$, which is the random key she wants to share with Bob.

To obtain the random key $K$, as mentioned earlier, Bob simply measures the received $n$ qubits to obtain the bit string: $v_k + w + x + e_1$ and decodes classically. Note that here there is no collapse of a general state, since Alice now prepared the $n$-qubits in the state $|v_k + w + x>$ corresponding to a specific (but randomly chosen) $w \in C_2$.

In the light of the above deductions/conclusions, **Step 1"** of the CSS protocol changes to:

**1"'**. Alice creates $n$ random check bits, a random $n$ bit string $x$, a random $v_k \in C_1/C_2$ and a random $w \in C_2$. She encodes $n$ qubits in the state $|0>$ or $|1>$ according to the string: $v_k + w + x$, and similarly $n$ qubits according to the check bits.

In view of the fact that $w$ is randomly selected from $C_2$, **Steps 1"'** and **9"** can be simplified further as follows:

a) There are $2^{k_1-k_2}$ distinct $v_k$'s and $2^{k_2}$ distinct $w$'s. Thus, there are $2^{k_1-k_2}$ x $2^{k_2} = 2^{k_1}$ distinct values of $v_k + w$, which are nothing but the $2^{k_1}$ elements of $C_1$ ( this is a fundamental property of cosets, see Section 3.2.6). In other words, $v_k(\in C_1/C_2)$ + $w(\in C_2)$ is an element $u \in C_1$. Thus, in **Step 1"'**, $v_k + w + x$ can be replaced with $u + x$, where $u \in C_1$. Now $u + x$ is a completely random string simply by virtue of $x$ being completely random.

b) Further, instead of creating and sending to Bob a state in accordance with $u + x$, and subsequently communicating to him the value of $x$ over the classical channel, Alice can alternatively create a state $|x>$ in accordance with the random $n$-bit string $x$, send it to Bob, and subsequently send him the value of $x - u$ over the classical channel. Bob then receives the state $|x + e_1>$, measures it to obtain the bit-string: $x + e_1$, and subtracts $x - u$ from it to obtain $u + e_1$, which is corrected to a codeword in $C_1$ (which most likely is $u$).

*Note that because the code bits (state $|x>$) Alice sends are random, there is no difference between the code bits and the random check bits; these can be blended together*.

All of the above leads to further modifications:

**1""**. Alice chooses a random $u \in C_1$ and creates $2n$ qubits in the state $|0>$ or $|1>$ according to $2n$ random bits; the coset $u + C_2$ is one of the $2^{k_1-k_2}$ distinct cosets of $C_2$ in $C_1$, and each such coset corresponds to a specific string $K \in \{0, 1\}^k$. Note that $k = k_1 - k_2$.

**2'''.** Alice randomly chooses $n$ positions (out of $2n$) and designates these as the check qubits, and the remainder as $|x>$.

**6'''.** Alice announces $b$, $x - u$, and which $n$ qubits are to provide check bits.

**9'''.** Bob measures the remaining $n$ qubits to get $x + e_1$, and subtracts $x - u$ to obtain $u + e_1$, which he corrects using code $C_1$, to obtain $u$.

**10'''.** Alice and Bob compute the coset $u + C_2$ to obtain the key $K$.

Other Points to Note

- Alice need not perform Hadamard (quantum) operations as in **Step 3** of the CSS protocol; she can instead encode her qubits directly in the (0,1) basis or in the (+,-) basis, depending upon the value of $b_i$. *Encoding and decoding are now done completely classically (**Step 7** is also removed; see below)*

- To eliminate the problem of quantum storage arising due to Bob having to store the received $2n$ qubits before receiving information on the classical channel, Bob decides to measure immediately the qubits as they are received one at a time, in the $Z$ or $X$ basis selected randomly (the problem of **storing** the received qubits goes away right here! **Step 7** is eliminated as Bob does not perform Hadamard transform). When Alice subsequently announces $b$, they keep those bits for which Bob's bases matched those of Alice's. The expectation is that half of the bits would be discarded in this matching process. Since it is only a high probability that half the number of sent qubits would be discarded, Alice actually sends a little more (say $\delta n$) than twice the previous number of qubits, i.e., sends $(4 + \delta) n$ qubits, one at a time.

The above analysis leads to a "Prepare and Measure" protocol, which is the Shor-Preskill version of the BB84 protocol state below; the primes on the numbers are removed and some steps are broken down into more than one step for clarity:

Other Points to Note

BB84 Protocol (Shor and Preskill Version)

1. Alice generates a bit string $d$ of $(4+\delta)$ $n$ random bits.
2. Alice chooses another random-bit string b of length $(4+\delta)$ $n$. For each bit of the string $d$, she creates a qubit in the $Z$ basis or the $X$ basis according to the bit values of the random string, $b$.
3. Alice sends the resulting $(4+\delta)$ $n$ qubits to Bob (one at a time).
4. Alice chooses a random codeword $u \in C_1$.
5. Bob receives the $(4+\delta)$ $n$ qubits, publicly announces this fact, and measures each in the $Z$ or $X$ basis at random.
6. Alice announces the string $b$ that determined the basis she used to encode each qubit in.
7. Bob discards the measured qubit values obtained if he measured in a basis different from the one that Alice prepared in. He tells Alice which measurements (but not their results) he discarded. Alice then discards the same set of bits.
   - With high probability, there are at least $2n$ bits left; if not, abort the protocol.

8. Alice randomly selects $2n$ bits from the remaining ($\geq 2n$) bits and announces which $2n$ bits she selected (but not their values).

9. Alice randomly selects $n$ of the $2n$ bits to use as check bits, and announces this selection of the $n$ bits and their bit values.

10. Bob compares the bit values he measured for the $n$ check bits selected by Alice and announces the bits where they disagree. If more than $t$ of these check bit values disagree, they abort the protocol.

11. Alice now has an $n$ bit string $x$, and Bob has an $n$-bit string $x + e_1$, where $e_1$ is the error caused by Eve's interference and/or channel noise.

12. Alice announces the $n$-bit string $x - u$.

13. Bob subtracts $x - u$ from his result $x + e_1$ to obtain $u + e_1$.

14. Bob corrects $u + e_1$, using code $C_1$, to obtain $u$.

15. Alice and Bob compute the coset $u + C_2$ in $C_1$, and determine the $k$-bit string belonging to $\{0,1\}^k$ the coset corresponds to; this $k$-bit string is their key $K$ of length $k$; the key is random because the codeword $u$ was chosen randomly in **Step 4**.

<u>Examining the Roles of $C_1$ and $C_2$</u>

1) $C_1$ plays the role of information reconciliation, i.e., error correction. If we examine the protocol above, we start with $u \in C_1$ (on Alice's side) and then correct Bob's result, $u + e_1$ using $C_1$ classically[19].

2) $C_2$, as was pointed earlier, does not play the role of an error (phase-flip) correcting code any more. Rather, it is very evident from **Step 15**, where $C_2$ first appears, that its role (in conjunction with $C_1$) is simply to provide privacy amplification quantified by $k = k_1 - k_2$, which defines the length of the final key (or the extent of privacy amplification); $k < n$.

The amount of privacy amplification needed is dependent upon the estimate of Eve's information on the key bits from all sources. If Eve is estimated to have $r$ bits of information, then the length of the final key, as noted earlier in Section 2.3, should be at least $n$-$r$. To reduce Eve's information further, then Alice and Bob must decide on the number of *extra* key bits they want to sacrifice. According to a theorem by Bennet et al [21], if they can shed another $s$ bits from the key they desire, then Eve's information about the $k$-bit string ($k = n$-$r$-$s$) is bounded by $2^{-s}/\ln 2 = e^{-s \ln 2}/\ln 2$, i.e., Eve's information drops exponentially with increasing $s$; for every extra bit sacrificed, the drop is by a factor of 2.

To perform error-corrections and privacy amplification, Alice and Bob must decide on the selection of the codes $C_1$ and $C_2$; these codes must satisfy the following properties:

---

[19] Again, because of the statistical nature of the errors, the size of the error $e_1$ in **Step 11** can exceed $t$, the threshold set by Alice and Bob, in which case the error correction will fail. But the probability of this happening is exponentially small, when $n$ is large; see the Random Sampling test in Ref. [10]; Eve's key information, as a consequence, is exponentially small.

a)   The code $C_1$ is an $[n, k_1]$ linear code $C_1$ that can correct $t$ errors (see Section 3.2); the parameter $t$ is determined from prior trial runs between Alice and Bob.

b)   The code $C_2$ is an $[n, k_2]$ linear code such that $C_2 \subset C_1$, with the requirement that the dual $C_2^{\perp}$ of the subcode $C_2$ is also $t$-error correcting; this is a requirement of the properties of CSS codes (Section 3.3.3.3) used in the reduction of the Lo-Chau entanglement-based protocol to the BB84 protocol.

c)   The dimensionalities $k_1$ and $k_2$ of the codes $C_1$ and $C_2$, respectively, are such that $k_1 - k_2$ equals the length $k$ of the desired key $K$; this is ascertained in advance from estimate of Eve's information.

Once codes $C_1$ and $C_2$ are selected, Bob performs error-correction using code $C_1$. Subsequently, both Alice and Bob compute the coset $u + C_2$ in $C_1$, which is then identified with one of the $2^k$ distinct cosets of $C_2$ in $C_1$; as a result of this identification, the key $K$ of length $k$ is obtained since each coset above corresponds to a distinct $k$ bit string in $\{0,1\}^k$; this entire process of selecting $C_1$ and $C_2$, computing the coset, and identifying it with a $k$-bit string, yielding a random key $K$ of length $k$, constitutes privacy amplification; the secret key rate is $k/n$.

**Example**: Suppose in the above protocol $n = 7$ and $t = 1$, i.e., the length $n$ of the bit string Alice and Bob each have at the end of **Step 10** is 7 and the maximum number of tolerable errors $t$ is 1 (from prior measurement trials). Likewise, suppose the estimated Eve's information on the key bits is $r = 1$ bit, which means they should reduce the length of the string they have by 1 bit at least. Since Eve's information is only an estimate, to be on the safe side, they decide to shed another 5 bits, i.e., $k = n - r - s = 7 - 1 - 5 = 1$. They now feel comfortable because Eve's information is now bounded (according to the theorem by Bennett et al [21]) by $2^{-r}/\ln 2 = 2^{-5}/\ln 2$ bits = 0.05 bit.

In short, they have decided to go from an initial string of length $n = 7$ to a string of length $k = 1$, which is the length of the final key $K$. But how do they achieve this reduction (or privacy amplification)?

With parameters $n = 7$, $k = 1$, and $t = 1$ at hand, they now look within their library of codes for

i)    A $[7, k_1]$ linear code $C_1$ that can correct a single error (see Section 3.2).

ii)   A $[7, k_2]$ linear code $C_2 \subset C_1$, with the requirement that the dual $C_2^{\perp}$ of the subcode $C_2$ is also single-error correcting.

iii)  Suitable dimensionalities $k_1$ and $k_2$ such that the relation: $k = k_1 - k_2$ is satisfied with $k = 1$.

They find that a *[7,4]* Hamming code as $C_1$ would work here very well, since this code is single-error correcting (which satisfies requirement i)) and is known to be weakly self-dual (see Example 2 in Section 3.3.3.3), i.e., $C_1^{\perp} \subset C_1$ (the dual of $C_1$ is a subcode of $C_1$). $C_2$ could then be chosen to be $C_1^\perp$ , which means $C_2^\perp = (C_1^{\perp})^{\perp} = C_1$, which then satisfies requirement ii) above of $C_2^\perp$ being able to correct a single error ($t$ = 1). Note that the dimensionality $k_1$ of code $C_1$ is 4, while the dimensionality $k_2$ of code $C_2$ (= $C_1^\perp$) is $n - k_1$ = 7- 4 = 3; this satisfies requirement iii) above, namely, $k = k_1 - k_2$ = 4 - 3 = 1.

Armed now with $C_1$ and $C_2$, Bob first performs error-correction using the $C_1$ code (see **Step 14** above) and then privacy amplification (in **Step 15**) by computing the coset $u + C_2$ in $C_1$, and identifying it with the classical bit string of length $k$ it is associated with ($u \in C_1$); this identified bit string is the key $K$. Alice does the same in **Step 15**. In the above example, the number of cosets of $u + C_2$ in $C_1$ is $2^k$; each coset is identified with a unique $k$-bit string that forms the key $K$. In the above example, $k$ = 1, which implies that the number of cosets is equal to 2 (see Eqs. (3.3.63a-b), which shows the contents of these cosets), one coset is associated with the bit value 0 and the other with the bit value 1 (this is a 1-bit string key $K$). The bit value for the key $K$ that Alice and Bob then get depends upon the 7-bit codeword $u \in C_1$ selected by Alice in **Step 4**.

In short, in the above example, a 7-bit codeword $u \in C_1$ is error-corrected by Bob (**Step 14**) using code $C_1$; subsequently, privacy amplification reduces the 7-bit string to a single bit string via the mechanism of computing the coset $u + C_2$ in $C_1$. The $C_1$ code was the [7,4] Hamming code and $C_2$ = $C_1^\perp$, which is [7,3] linear code. If Alice and Bob had decided, instead, on a key of length $k$ =2, they would have had to find a different pair of codes $C_1$ and $C_2$, satisfying all the code requirements, a) through c).

### 3.4.5 Comparison of the Shor-Preskill BB84 version with the the Standard Version (Section 2.4)

Other than the fact the Shor-Preskill version employs linear codes for information reconciliation and privacy amplification, there is really no fundamental difference in the structures of the traditional BB84 protocol (Section 2.3) and the Shor-Preskill version given above; in the former, one uses a suitable universal hashing function [10,23] after entropy estimation (Eve's information) and in the latter, one employs an appropriate linear code $C_2 \subset C_1$, where $C_1$ is used for information reconciliation.

The only issue that might arise in the Shor-Preskill version is the existence of suitable large error-correcting linear codes for practical QKD systems, where the key size may be very large (hundreds of thousands of bits, for example). Satisfying all the constraints, a) through c) might not be easy. Furthermore, the code $C_1$ might not be efficiently decodeable, i.e., be able to efficiently correct errors.

Does the Shor-Preskill proof apply to the standard version? Because the structures are so similar, the answer is *yes*. Nevertheless, there are assumptions that need to be stated.

### 3.4.6 Assumptions/Shortcomings within the Shor-Preskill Proof

1) Since the security proof is derived from the entanglement-based protocol, wherein Alice and Bob share an entangled qubit pair, meaning each of them has a single qubit (although entangled), the Shor-Preskill proof applies strictly to a perfect single-qubit source; if the qubit source is light as in practically conceived QKD systems, the security proof then strictly applies to a perfect single-photon source, which unfortunately is not currently available commercially; in practice, the qubit source used is a weak coherent laser pulse, which can contain more than one photon[20].

2) Implicit within the proof are also assumptions such as

> i) The random number generators used generate perfect random numbers; the proof provides no guidance on how to assess the loss of information to Eve on account of imperfect random number generators.

> ii) The classical channel is perfectly authenticated, while in reality this may not be true.

3) The proof is based only on correcting for errors due to eavesdropping and/or quantum noise.

4) The proof does not provide any means to estimate/bound Eve's information; a proper estimate of Eve's information is required to determine the amount of privacy amplification needed before any reasonable size key can be generated by a practical QKD system and declared secure.

5) The proof assumes perfect qubit state preparation by Alice and perfect (received) qubit state measurement by Bob. Imperfections, which exist in reality, can cause the error-rate to exceed the tolerable error limit, thus making a practical QKD system difficult to use.

6) In a practical system based on the Shor-Preskill version of the BB84 protocol, a difficulty would be in finding large linear codes (say, with $n$, the size of the code, of the order of $10^6$) satisfying the required properties for the generation of a predetermined key of length $k$ ($< n$). To overcome this problem, one might envision a practical implementation in which the message to be encrypted is broken down into a number of small messages, thus requiring only small size keys for encryption (one-time pad encryption) and thus small size (manageable) linear codes ($n < 10^4$) that would be relatively easily constructed with the above requirements. But the use of smaller size codes would tend to invalidate the security proof, which is based on the use of large size codes; for example, the fidelity of the keys generated by Alice and Bob would not be guaranteed. Assessing the full impact of using small size codes translates essentially to the theoretical problem of finite-key size corrections, which is an area for further research [3].

---

[20]The possibility of multi-photons in a weak laser pulse (representing a qubit) leads to a photon-number splitting attack in which Eve learns about the key by keeping one of the photons in the multi-photon pulse, and forwarding the rest of the (unaffected) photons to Bob (she measures her stored photons after Bob announces his measurement bases). While this attack is detected with the use of decoy states, and the protocol terminated at once, the fact remains that Shor-Preskill proof is simply inapplicable to such a scenario involving multi-photons.

Furthermore, finding efficiently decodeable codes would pose another difficult problem for such QKD systems to be viable.

# 4 Summary and Conclusions

In this report, we have described the motivation for the study of Quantum-Key-Distribution (QKD), the basic steps in generating a private, random key within the framework of QKD, and the popular BB84 protocol for QKD. Furthermore, we have addressed the issue of key security provided by this protocol, invoking the Shor-Preskill proof, which is the most widely cited proof in the literature. Specifically, we have provided all the details of the proof. We started with the entanglement-based protocol of Lo and Chau for which the security proof exists, and then reduced it step-by-step to the "Prepare and Measure" protocol, arriving at a version of BB84 protocol, which is different from the standard version. We explicitly showed (and demonstrated via an example) that error-correction and privacy amplification, the two key steps in prrivate key generation are accomplished using linear codes, instead of standard error-correcting algorithms such as the Cascade error-correcting algorithm and 2-universal hashing functions. Structurally, the two versions, the Shor-Preskill version (from the reduction) and the standard version are the same. There is no fundamental difference in the two protocols. Therefore, the security proof provided by the Shor - Preskill reduction applies to the standard version as well. Nevertheless, there are shortcomings as well as assumptions made in the proof that should be taken into account any secure practical QKD design.

Since the reduction of the entanglement-based protocol to the BB84 protocol involves the use of the Calderbank-Shor-Steane (CSS) quantum error-correcting codes, and it is critical to understand them in the reduction to the Shor-Preskill BB84 version, we have provided an in-depth understanding of the mechanics and properties of these codes. Furthermore, because these codes are constructed from linear codes, we have also provided a detailed review of their pertinent properties, especially of *cosets*, as a prelude to CSS codes.

# 5 Acknowledgment

The author thanks Gerry Baumgartner for reading the manuscript and helpful comments.

# Appendix A   Multi-Qubit States and Hilbert Space

i) The general state $|s>$ of a qubit is written as $a|0>$ and $b|1>$, where the arbitrary coefficients $a$ and $b$ are complex. When the coefficients satisfy the equation: $|a|^2 + |b|^2 = 1$, the state $|s>$ is said to be *normalized*. Henceforth, unless otherwise stated, we will always assume a given quantum state to be normalized.

 The states $|0>$ and $|1>$ are linearly independent; in fact, they form an orthonormal basis of a *2-dimensional* complex Hilbert space $\mathbb{C}^2$, i.e., they satisfy the scalar product:

$$<i|j> = \delta_{ij}, \qquad\qquad (A.1)$$

where $i$ and $j$ are each 0 or 1, and $\delta$ is the Kronecker delta function. In matrix representation, these basis states are traditionally represented by *2 x 1* column vectors:

$$|0> = \begin{bmatrix} 1 \\ 0 \end{bmatrix}, \quad |1> = \begin{bmatrix} 0 \\ 1 \end{bmatrix}. \qquad\qquad (A.2)$$

The orthonormality is easily verified if one recalls that the *bra*, $<i|$, is a row vector, a transpose of the column vector corresponding to the *ket*, $|i>$ (see Eq. (A.2)), with its elements complex conjugated.

ii) Consider now a qubit in the state $|0>$. Append another qubit in the same state $|0>$ to obtain a two-qubit system represented as $|0> \otimes |0>$, a tensor product. For convenience, this product of two states is also written as $|00>$.  In general, taking two such individual qubits, each one of which can be in either $|0>$ or $|1>$, we obtain four different possible 2-qubit states:  $|00>$, $|01>$, $|10>$, and $|11>$. These states form an orthonormal basis of a 4-dimensional complex Hilbert space $\mathbb{C}^4$, i.e.,

$$<i_1 i_2|j_1 j_2> = <i_1|\,j_1> <i_2|\,j_2> = \delta_{i_1 j_1}\, \delta_{i_2 j_2} \qquad\qquad (A.3)$$

where $i_1, i_2, j_1, j_2$ can each take on the values, 0 and 1, and the $\delta$'s are the Kronecker delta-functions, and where we have invoked the basic result, Eq. (A.1). Any time two quantum states differ in the bit-string content (00, 10, 01, or 11), they are orthogonal to each other. The most general two-qubit state is a linear combination of these four-basis states, with arbitrary complex coefficients.

Employing the general tensor product rule in the matrix notation,

$$\begin{bmatrix} a \\ b \end{bmatrix} \otimes \begin{bmatrix} c \\ d \end{bmatrix} = \begin{bmatrix} ac \\ ad \\ bc \\ bd \end{bmatrix}, \qquad\qquad (A.4)$$

where $\begin{bmatrix} a \\ b \end{bmatrix}$ and $\begin{bmatrix} c \\ d \end{bmatrix}$ are arbitrary *2 x 1* column vectors, and each element of the first column vector multiplies the entire second vector,

$$|00> = \begin{bmatrix} 1 \\ 0 \end{bmatrix} \otimes \begin{bmatrix} 1 \\ 0 \end{bmatrix} = \begin{bmatrix} 1 \\ 0 \\ 0 \\ 0 \end{bmatrix}. \tag{A.5a}$$

$$|01> = \begin{bmatrix} 1 \\ 0 \end{bmatrix} \otimes \begin{bmatrix} 0 \\ 1 \end{bmatrix} = \begin{bmatrix} 0 \\ 1 \\ 0 \\ 0 \end{bmatrix}, \tag{A.5b}$$

$$|10> = \begin{bmatrix} 0 \\ 1 \end{bmatrix} \otimes \begin{bmatrix} 1 \\ 0 \end{bmatrix} = \begin{bmatrix} 0 \\ 0 \\ 1 \\ 0 \end{bmatrix}, \tag{A.5c}$$

$$|11> = \begin{bmatrix} 0 \\ 1 \end{bmatrix} \otimes \begin{bmatrix} 0 \\ 1 \end{bmatrix} = \begin{bmatrix} 0 \\ 0 \\ 0 \\ 1 \end{bmatrix} \tag{A.5d}$$

The orthonormality, Eq. (A.3), is easily verified.

<u>$n$-qubit system</u>

The above concept is generalized to an $n$-qubit system, whose most general state is defined in a $2^n$ – dimensional complex Hilbert space $\mathbb{C}^{2^n}$ spanned by $2^n$ basis states, $|j_1j_2....j_{n-1} j_n>$, where each $j_i$, $i=1,..,n$ takes on the value 0 or 1. These basis states form an orthornormal set as each such state corresponds to a different $n$-bit string:

$$<i_1i_2....i_n|j_1j_2...j_n> = <i_1|j_1> <i_2|j_2> ....... <i_n|j_n> = \delta_{i_1 j_1} \delta_{i_2 j_2} .........\delta_{i_n j_n} \tag{A.6}$$

# Appendix B  Single Qubit Operators

<u>Pauli Spin Matrices</u>

The three Pauli spin operators, $\sigma_x$, $\sigma_y$, and $\sigma_z$, which commonly occur in quantum mechanics, are stated below in their matrix representation:

$$\sigma_x = \begin{bmatrix} 0 & 1 \\ 1 & 0 \end{bmatrix}, \sigma_y = \begin{bmatrix} 0 & -i \\ i & 0 \end{bmatrix}, \sigma_z = \begin{bmatrix} 1 & 0 \\ 0 & -1 \end{bmatrix} \tag{B.1}$$

($\sigma_x$ and $\sigma_z$ are also sometimes denoted by $X$ and $Z$, respectively). These are unitary matrices, i.e.,

$$\sigma_x{}^+ \sigma_x = \sigma_y{}^+ \sigma_y = \sigma_Z{}^+ \sigma_Z = I. \tag{B.2}$$

Also, the reader can verify that these matrices anticommute, i.e.,

$$\sigma_x \sigma_y + \sigma_y \sigma_x = \sigma_y \sigma_z + \sigma_z \sigma_y = \sigma_z \sigma_x + \sigma_x \sigma_z = 0. \tag{B.3}$$

<u>Bit-flip Operator</u>

The $x$-Pauli spin operator acting on a qubit causes a bit-flip:

$$\begin{bmatrix} 0 & 1 \\ 1 & 0 \end{bmatrix} \begin{bmatrix} 1 \\ 0 \end{bmatrix} = \begin{bmatrix} 0 \\ 1 \end{bmatrix}, \begin{bmatrix} 0 & 1 \\ 1 & 0 \end{bmatrix} \begin{bmatrix} 0 \\ 1 \end{bmatrix} = \begin{bmatrix} 1 \\ 0 \end{bmatrix}, \tag{B. 4a}$$

i.e.,

$$X|0> = |1> , X|1> = |0>, \tag{B.4b}$$

where

$$|0> = \begin{bmatrix} 1 \\ 0 \end{bmatrix}, \ |1> = \begin{bmatrix} 0 \\ 1 \end{bmatrix} \tag{B.4c}$$

are the two basis vectors  for a qubit state (see Eq. (A.2); they constitute the ($|0>$, $|1>$) basis or, what is termed in the shorter notation,  the (0,1) basis. Another basis called the (+, -) basis will be defined below as a consequence of the Hadamard transform.

<u>Phase-flip Operator</u>

The $z$-Pauli spin matrix acting on a qubit causes a phase-flip in the following way:

$$\begin{bmatrix} 1 & 0 \\ 0 & -1 \end{bmatrix} \begin{bmatrix} 1 \\ 0 \end{bmatrix} = \begin{bmatrix} 1 \\ 0 \end{bmatrix}, \begin{bmatrix} 1 & 0 \\ 0 & -1 \end{bmatrix} \begin{bmatrix} 0 \\ 1 \end{bmatrix} = - \begin{bmatrix} 0 \\ 1 \end{bmatrix}, \tag{B. 5}$$

i.e.,

$$Z|0> = (+1) \; |0> \; , \; Z|1> = (-1) \; |1>.$$

(B.6)

<u>Hadamard Transform</u>

1) This is a transform that transforms the (0, 1) basis into the (+, -) basis and vice versa. Its matrix representation is

$$H_d = \frac{1}{\sqrt{2}} \begin{bmatrix} 1 & 1 \\ 1 & -1 \end{bmatrix}$$

(B.7)

It is unitary ($H_d^{-1} = H_d^{+}$) and its inverse is equal to itself, i.e., $H_d^{-1} = H_d$. One can verify that

$$H_d \; |0> = \; |+>,$$

(B.8a)

$$H_d \; |1> = \; |->,$$

(B.8b)

$$H_d \; |+> = \; |0>,$$

(B.8c)

$$H_d \; |-> = \; |1>,$$

(B.8d)

where

$$|+> = \frac{1}{\sqrt{2}} \begin{bmatrix} 1 \\ 1 \end{bmatrix}$$

(B. 9a)

and

$$|-> = \frac{1}{\sqrt{2}} \begin{bmatrix} 1 \\ -1 \end{bmatrix}.$$

(B. 9b)

Using Eq. (B.1) and Eqs. (B.9a-b), one easily verifies that

$$X \; |+> = \; |+>,$$

(B.10a)

$$X|-> = - \; |->,$$

(B.10b)

where |+> and |-> are |+> and |-> are now the eigenstates of the $x$-Pauli spin operator, with eigenvalues +1 and -1, respectively. Furthermore,

$$Z|+> = \; |->,$$

(B.10c)

$$Z|-> = \; |+>,$$

(B.10d)

i.e., the Z operator acts as the "bit-flip" operator under the Hadamard transform.

Another way of stating the above results is that under the Hadamard transform, the operator $Z$ changes to operator $X$ and vice versa. Consider, for example, Eq. (B.8a): $H_d \; |0> = \; |+>$. Then, Z $H_d$ |0> = Z |+> = |->, using Eq. (B.10c). Applying $H_d$ on both sides, we get $H_d \; Z \; H_d$ |0> = $H_d$ |-> = |1>, using

Eq. (B.8d). The right-hand-side is a bit-flipped state. Similarly, one can show that $H_d Z H_d |1> = |0>$. This leads to the conclusion that the operator

$$H_d Z H_d = X, \tag{B.11a}$$

the bit-flip operator, i.e., the $Z$ operator has changed to the $X$ operator under the Hadamard transform. Similarly, starting with Eq. (B.8a) and using Eqs. (B.10a-b), one can show that

$$H_d X H_d = Z. \tag{B.11b}$$

The results, Eqs. (B.11a-b), are explicitly verified below using the matrix notation:

$$H_d Z H_d = \frac{1}{\sqrt{2}}\begin{bmatrix} 1 & 1 \\ 1 & -1 \end{bmatrix}\begin{bmatrix} 1 & 0 \\ 0 & -1 \end{bmatrix}\frac{1}{\sqrt{2}}\begin{bmatrix} 1 & 1 \\ 1 & -1 \end{bmatrix} = \begin{bmatrix} 0 & 1 \\ 1 & 0 \end{bmatrix} = X \tag{B.12a}$$

$$H_d X H_d^{-1} = \frac{1}{\sqrt{2}}\begin{bmatrix} 1 & 1 \\ 1 & -1 \end{bmatrix}\begin{bmatrix} 0 & 1 \\ 1 & 0 \end{bmatrix}\frac{1}{\sqrt{2}}\begin{bmatrix} 1 & 1 \\ 1 & -1 \end{bmatrix} = \begin{bmatrix} 1 & 0 \\ 0 & -1 \end{bmatrix} = Z \tag{B.12b}$$

# Appendix C   Matrix Representation of Operators for Multi-Qubit Systems

As an example, consider the operator $Z \otimes Z$ acting on a two-qubit system described by the column vectors: Eqs. (A5a-d). Its matrix representation in the *4*-dimensional space is a *4 x 4* matrix given by

$$Z \otimes Z = \begin{bmatrix} 1 & 0 \\ 0 & -1 \end{bmatrix} \otimes \begin{bmatrix} 1 & 0 \\ 0 & -1 \end{bmatrix} = \begin{bmatrix} 1 & 0 & 0 & 0 \\ 0 & -1 & 0 & 0 \\ 0 & 0 & -1 & 0 \\ 0 & 0 & 0 & 1 \end{bmatrix}, \tag{C.1}$$

where we have used the general tensor product rule: given two operators *T* and *V* given by

$$T = \begin{bmatrix} a & b \\ c & d \end{bmatrix} \text{ and } V = \begin{bmatrix} e & f \\ g & h \end{bmatrix}, \tag{C.2}$$

then

$$T \otimes V = \begin{bmatrix} ae & af & be & bf \\ ag & ah & bg & bh \\ ce & cf & de & df \\ cg & ch & dg & dh \end{bmatrix}, \tag{C.3}$$

which is obtained by multiplying each element of *T* with the entire second matrix *V*.

### $Z_1Z_2$

For convenience, $Z \otimes Z$ may also be written as $Z_1Z_2$, where the indices indicate the qubits the $Z$ operator acts upon; here it is the first and the second qubit in a two-qubit system. The matrix representation of $Z_1Z_2$ is then given in Eq. (C.1). We now relate $Z_1Z_2$ to the operators |00> <00|, |01> <01|, |10> <10>, and |11> <11|, which are projection operators in a space defined for a 2-qubit system (see Appendix D also). Let us consider their matrix representations. Using the rule above, then

$$|00> <00| = \begin{bmatrix} 1 \\ 0 \\ 0 \\ 0 \end{bmatrix} [1000] = \begin{bmatrix} 1 & 0 & 0 & 0 \\ 0 & 0 & 0 & 0 \\ 0 & 0 & 0 & 0 \\ 0 & 0 & 0 & 0 \end{bmatrix}, \tag{C.4a}$$

$$|01> <01| = \begin{bmatrix} 0 \\ 1 \\ 0 \\ 0 \end{bmatrix} [0100] = \begin{bmatrix} 0 & 0 & 0 & 0 \\ 0 & 1 & 0 & 0 \\ 0 & 0 & 0 & 0 \\ 0 & 0 & 0 & 0 \end{bmatrix}, \tag{C.4b}$$

$$|10\rangle\langle10| = \begin{bmatrix} 0 \\ 0 \\ 1 \\ 0 \end{bmatrix} [0010] = \begin{bmatrix} 0 & 0 & 0 & 0 \\ 0 & 0 & 0 & 0 \\ 0 & 0 & 1 & 0 \\ 0 & 0 & 0 & 0 \end{bmatrix}, \qquad\qquad (C.4c)$$

$$|11\rangle\langle11| = \begin{bmatrix} 0 \\ 0 \\ 0 \\ 1 \end{bmatrix} [0001] = \begin{bmatrix} 0 & 0 & 0 & 0 \\ 0 & 0 & 0 & 0 \\ 0 & 0 & 0 & 0 \\ 0 & 0 & 0 & 1 \end{bmatrix}. \qquad\qquad (C.4d)$$

Upon examining the matrix structure of the above projection operators (Eqs. (C.4a-d)) and the $Z_1Z_2$ operator in Eq. (C.1), one sees that

$$Z_1Z_2 = |00\rangle\langle00| + |11\rangle\langle11| - (|01\rangle\langle01| + |10\rangle\langle10|) \qquad\qquad (C.5)$$

<u>The Controlled-Not Gate: CNOT</u>

This is one of the most basic quantum gates used in quantum computation. It acts on two qubits, and its matrix representation is

$$CNOT = \begin{bmatrix} 1 & 0 & 0 & 0 \\ 0 & 1 & 0 & 0 \\ 0 & 0 & 0 & 1 \\ 0 & 0 & 1 & 0 \end{bmatrix}. \qquad\qquad (C.6)$$

When applied to a pair of qubits (Eqs. (A.5a-d)), it has the following effect:

$$CNOT\ |00\rangle = |00\rangle, \qquad\qquad (C.7a)$$

$$CNOT\ |01\rangle = |01\rangle, \qquad\qquad (C.7b)$$

$$CNOT\ |10\rangle = |11\rangle, \qquad\qquad (C.7c)$$

$$CNOT\ |11\rangle = |10\rangle. \qquad\qquad (C.7d)$$

That is, when the first qubit is 0, the two-qubit state is unchanged, but, if the first qubit is 1, the bit value of the second qubit in the two-qubit state is flipped. In other words, the second qubit is controlled by the first qubit. The first qubit is called the "control qubit" and the second qubit is called the "target qubit".

One can also verify that the CNOT operation is unitary ($(CNOT)^{-1} = CNOT^+$); note also that $(CNOT)^{-1} = CNOT$ because $CNOT^+ = CNOT$.

# Appendix D  Projection Operators

1) <u>One-dimensional subspace</u>

Let $|r>$ be a state vector in an $n$-dimensional Hilbert space. Then the operator $P = |r> <r|$ is a projection operator that projects an arbitrary state $|\psi>$ defined in the $n$-dimensional Hilbert space onto the one-dimensional space spanned by $|r>$. Not that in the matrix notation, $<r|$ is a 1 x $n$ row vector and $P$, therefore, is an $n$ x $n$ matrix.

Let states $|c_i>$, $i$ = 1, 2,…, $n$ comprise an orthonornmal basis that includes the state $|r>$. $|c_1> = |r>$. Let $|\psi> = a_1|c_1> + a_2|c_2> +…….+ a_n|c_n>$, where, say, $|c_1> = |r>$ and all $c_i$'s are complex in general. Application of $P$ on the state $|\psi>$ gives

$$P|\psi> = |r> <r| \psi> = a_1 <r|c_1> + a_2<r|c_2> +…….+ a_n<r|c_n> = a_1|c_1> \qquad \text{(D.1)}$$

due to the orthogonality of the basis states $|c_i>$. The result $a_1|c_1>$ is called the projection of the state $|\psi>$ (due to the action of $P$) onto the one-dimensional space spanned by $|r>$; $a_1$ is the component of $|\psi>$ along the basis state $|c_1>$ $(=|r>)$.

*Example 1*

Let $|s> = a|0> +b|1> = \begin{bmatrix} a \\ b \end{bmatrix}$ in the 2-dimensional Hilbert space spanned by $|0>$ and $|1>$; $|0>$ and $|1>$ form an orthonormal basis (see Eqs. (A.1) and (A.2)). Consider another orthonormal basis formed by the states $|+> = \frac{1}{\sqrt{2}} \begin{bmatrix} 1 \\ 1 \end{bmatrix}$ and $|-> = \frac{1}{\sqrt{2}} \begin{bmatrix} 1 \\ -1 \end{bmatrix}$ (see Eqs. (B.9a) and (B.9b)). The reader can verify that $<+|+> = <-|-> = 1$, while $<+|-> = <-|+> = 0$. Let us consider the projection of $|s>$ onto the one-dimensional space spanned by $|+>$, i.e., let $|r> = |+>$. Then $P_+ = |r> <r| = |+> <+| = ½ \begin{bmatrix} 1 \\ 1 \end{bmatrix}$ [1 1] $= ½ \begin{bmatrix} 1 & 1 \\ 1 & 1 \end{bmatrix}$. Then $P_+ |s> = ½ \begin{bmatrix} 1 & 1 \\ 1 & 1 \end{bmatrix} \begin{bmatrix} a \\ b \end{bmatrix} = ½ (a+b) \begin{bmatrix} 1 \\ 1 \end{bmatrix} = \frac{1}{\sqrt{2}} (a+b) \frac{1}{\sqrt{2}} \begin{bmatrix} 1 \\ 1 \end{bmatrix} = \frac{1}{\sqrt{2}} (a+b) |+>$, i.e., the projection operator $P_+$ projects $|s>$ onto $|+>$, as desired; its component along $|+> = \frac{1}{\sqrt{2}} (a+b)$. Similarly, defining $P_- = |-> < -|$, we find $P_- |s> = \frac{1}{\sqrt{2}} (a-b) |->$. These projection results also imply that one can rewrite $|s> = a|0> + b|1>$ as $|s> = \frac{1}{\sqrt{2}} (a+b) |+> + \frac{1}{\sqrt{2}} (a-b) |->$. Noting that $|a|^2 + |b|^2 = 1$ (the given state $|s>$ is normalized), one can verify that $|\frac{1}{\sqrt{2}} (a+b)|^2 + |\frac{1}{\sqrt{2}} (a-b)|^2 = 1$, as expected.

2) <u>Two-Dimensional Subspace</u>

Let $|r_1>$ and $|r_2>$ be two orthogonal states in an $n$-dimensional Hilbert Space. Together they span a 2-dimensional subspace. The operator that projects a given state onto this two-dimensional subspace is written as

$$P = |r_1> <r_1| + |r_2> <r_2|. \tag{D.2}$$

Consider, as before, an arbitrary state $|\psi> = a_1|c_1> + a_2|c_2> + \ldots\ldots + a_n|c_n>$, where, say, $|c_1> = |r_1>$ and $|c_2> = |r_2>$, all $a_i$'s are complex and the $|c_i>$'s form an orthonormal set. Now

$$P|\psi> = (|c_1> <c_1| + |c_2> <c_2|) \, (a_1|c_1> + a_2|c_2> + \ldots\ldots + a_n|c_n>) = a_1|c_1> + a_2|c_2> \tag{D.3}$$

The components along the basis states $|c_1>$ and $|c_2>$ are projected out by the $P$ operator, i.e., the right-hand-side is the component vector projected out in the two dimensional subspace spanned by the vectors $|c_1>$ and $|c_2>$ (this is analogous to the 3-dimensional Cartesian space where, for example, a given 3-dimensional vector may be projected onto a two-dimensional space spanned by Cartesian unit vectors, $\hat{x}$ and $\hat{y}$, or $\hat{y}$ and $\hat{z}$, or $\hat{z}$ and $\hat{x}$). Note that if we had chosen $|r_1>$ as $|c_1>$ and $|r_2>$ as $|c_3>$, then we would have obtained $a_1|c_1> + a_3|c_3>$, and so on.

Note also that if all the coefficients $a_i$ above were zero, except for $a_1$ and $a_2$, then the right-hand side in Eq. (D.3) is just the original state $|\psi>$. The state $|\psi>$, in this case, is actually already lying in the *2*-dimensional space spanned by basis states $|c_1>$ and $|c_2>$.

One can similarly define projection operators that project a given state onto a subspace that is of dimensionality greater than two (but equal to or less than $n$).

3) <u>Measurement and Outcome Probabilities</u>

a) A general measurement $M$ is represented by an ordered set of projection operators: $M = (P_1, P_2, \ldots, P_N)$, representing $N$ different outcomes, such that $\sum_i P_i = I$, the identity operator.

In Eq. (D.1), we can project an arbitrary state into $n$ different one-dimensional spaces, using a different basis state $|c_j>$ ($j = 1,2,..,n$) for $|r>$, so $N = n$ here. Note that the $\sum_i P_i = I$ is satisfied because $\sum_i P_i \, |\psi> = |\psi>$. This result also implies that the sum of the probabilities of the outcomes equals 1, as seen below.

b) If the initial state of the system is $|\psi>$ and the measurement $M$ is performed, the probability of the $i$th outcome is

$$p_i = <\psi|P_i|\psi> \tag{D.4}$$

$\sum_i p_i = \sum_i <\psi|Pi|\psi> = <\Psi|\sum_i Pi|\psi> = <\Psi|I|\Psi> = 1$, i.e., the probabilities $p_i$ add up to 1.

c) If the initial state of the system is $|\psi\rangle$ and the $i$th outcome occurs, the final state $|s_i\rangle$ of the system is given by

$$|s_i\rangle = \frac{P_i|\psi\rangle}{\langle\psi|P_i|\psi\rangle} \tag{D.5}$$

That is, the system stays in the state projected out after the measurement. The denominator normalizes the projected state.

### *Example 2*

Consider a 2-qubit system described by a $2^2$ (=4) - dimensional Hilbert space spanned by $|00\rangle$, $|10\rangle$, $|01\rangle$ and $|11\rangle$ (see Eqs. (A.5a-5d)).

Consider now the projection operators:

$$P_1 = |00\rangle\langle00| + |11\rangle\langle11| \tag{D.6a}$$

$$P_2 = |01\rangle\langle01| + |10\rangle\langle10| \tag{D.6b}$$

(one can express them in the matrix notation, using the results: Eqs. (C.4a-d), if one wishes, but we can do without it here). Operator $P_1$ has the effect of projecting any 2-qubit state onto a subspace spanned by the basis vectors $|00\rangle$ and $|11\rangle$. Similarly, operator $P_2$ has the effect of projecting any 2-qubit state onto a subspace spanned by basis vectors $|01\rangle\langle01|$ and $|10\rangle\langle10|$.

Suppose the 2-qubit system is described by the state:

$$|\psi\rangle = 1/\sqrt{2}(|00\rangle + |11\rangle) \tag{D.7}$$

If we now perform the measurement, whose outcomes are described by the above two projection operators (Eqs. (D.7a-b)), we obtain

$$p_1 = \langle\psi|P_1|\psi\rangle = \tfrac{1}{2}(\langle00| + \langle11|)(|00\rangle\langle00| + |11\rangle\langle11|(|00\rangle + |11\rangle)) = 1 \tag{D.8a}$$

$$p_2 = \langle\psi|P_2|\psi\rangle = \tfrac{1}{2}(\langle00| + \langle11|)(|01\rangle\langle01| + |10\rangle\langle10|(|00\rangle + |11\rangle)) = 0 \tag{D.8b}$$

using the orthonormality relation, (Eq. (A.3)). Clearly, the first outcome will definitely occur. The final state of this pair of qubits is

$$|s_1\rangle = \frac{P_1|\psi\rangle}{\langle\psi|P_1|\psi\rangle} = |\psi\rangle \tag{D.9}$$

i.e., the measurement has not disturbed the state $|\psi>$. It certainly would have disturbed the state if we had measured the individual states of the two qubits. These qubits might be two photons and the measurement[21] may be to determine whether the two photons have the same polarization (outcome $P_1$) or opposite polarizations (outcome $P_2$) in the horizontal-vertical basis, i.e., the (0,1) basis.

*Further Remarks:* The above measurement that left the 2-photon (qubit) state undisturbed is an example of an incomplete measurement, in contrast to a complete measurement which would reveal the individual states of the two photons (clearly, a measurement of the individual photon's polarization will result in the final state being in either the $|00>$ or the $|11>$ state).

Note that the incomplete measurement described above discriminates between the two subspaces specified by the above projection operators: $P_1$ and $P_2$. In the example considered, the state $|\psi>$ was already lying in the subspace spanned by $|00>$ and $|11>$.

---

[21] In practice, this joint measurement on the two photons (as compared to spin ½ particles) may be hard to perform.

# Appendix E  Phase Flip Error Correcting Properties of CSS$_{x,z}$

The CSS$_{x,z}$ code corresponding to $v \in C_1$ (the original state) is represented as

$$\frac{1}{|C_2|^{1/2}} \sum_{w \in C_2} (-1)^{w.z} |v + w + x\rangle \qquad \text{(E.1)}$$

A phase flip error represented by $e_2$ changes it to

$$\frac{1}{|C_2|^{1/2}} \sum_{w \in C_2} (-1)^{(v+w+x).e_2}(-1)^{w.z} |v + w + x\rangle \qquad \text{(E.2)}$$

as in Eq. (3.3.37). Multiplying with a constant phase $(-1)^{(v+x).z}$ throughout yields

$$\frac{1}{|C_2|^{1/2}} \sum_{w \in C_2} (-1)^{(v+w+x).e_2}(-1)^{(v+w+x).z} |v + w + x\rangle \qquad \text{(E.3)}$$

Applying Hadamard transform to each of the qubits in the above expression, we then obtain

$$\frac{1}{2^{n/2}|C_2|^{1/2}} \sum_{w \in C_2} \sum_{z'} (-1)^{(v+w+x).z'}(-1)^{(v+w+x).e_2}(-1)^{(v+w+x).z} |z'\rangle \qquad \text{(E.4)}$$

where $z' \in \{0,1\}^k$. This expression simplifies to

$$\frac{1}{2^{n/2}|C_2|^{1/2}} \sum_{w \in C_2} \sum_{z'} (-1)^{(v+w+x).(z' + z + e_2)} \; |z'\rangle \qquad \text{(E.5)}$$

Setting $z'' = z' + z + e_2$, we get

$$\frac{1}{2^{n/2}|C_2|^{1/2}} \sum_{w \in C_2} \sum_{z''} (-1)^{(v+w+x).z''} \; |z'' + z + e_2\rangle. \qquad \text{(E.6)}$$

where $z'' \in \{0,1\}^k$.

Again invoking the results, Eqs. (3.2.29a-b), Eq. (E.6)  becomes

$$\frac{1}{2^{n/2}/|C_2|^{1/2}} \sum_{w \in C_2^{\perp}} (-1)^{(v+x).z"} |z" + z + e_2\rangle \qquad \text{(E.7)}$$

Thus, as before, we correct the phase-flip errors as bit-flip errors in the Hadamard transformed space using $H_2$, which is the parity check matrix for $C_2^{\perp}$. Subsequently, we transform back to the original basis by applying the Hadamard transform once again to obtain an error-free original state.

# Appendix F  Hadamard transform applied to n qubits

In Appendix B, we described the Hadamard transform acting on a single qubit. Its effect on a single qubit expressed via Eqs. (B.8a-b) and Eqs. (B.9a-b) can be reexpressed as

$$H_d|i> = \frac{1}{\sqrt{2}} \sum_z (-1)^{i.z} |z> \qquad \text{(F.1)}$$

where $|i> = |0>$ or $|1>$ and $z \in \{0,1\}$, and $i.z$ is a bitwise inner product modulo 2. Now Hadamard transform applied to $n$-qubits can be written as

$$H_d|i_1> \otimes H_d|i_2> \otimes \ldots \otimes H_d|i_n>$$
$$= \frac{1}{\sqrt{2}} \sum_{z_1} (-1)^{i_1.z_1} |z_1> \frac{1}{\sqrt{2}} \sum_{z_2} (-1)^{i_2.z_2} |z_2> \cdots \frac{1}{\sqrt{2}} \sum_{z_n} (-1)^{i_n.z_n} |z_n>$$

where $z_i \in \{0,1\}$, $i = 1,2,\ldots,n$. This expression is rewritten as

$$H_d^{\otimes n}|i_1\, i_2 \ldots i_n> = \frac{1}{2^{n/2}} \sum_{z_1} \sum_{z_2} \ldots \sum_{z_n} (-1)^{i_1.z_1} (-1)^{i_2.z_2} \ldots (-1)^{i_n.z_n} |z_1 z_2 \ldots z_n>$$
$$= \frac{1}{2^{n/2}} \sum_z (-1)^{i.z} |z> \qquad \text{(F.2)}$$

where, in the above compact notation, $H_d^{\otimes n}$ (tensor product of $n$ $H_d$ operators) acts on the $n$ qubits,  $i$ is the given $n$-bit string on the left, $z \in \{0,1\}^n$, and $i.z$ is the bitwise inner product of $i$ and $z$, modulo 2.

# Appendix G  Einstein-Podolsky-Rosen (EPR) Pair

EPR pair of qubits is an entangled pair of qubits written as

$$|\beta_{00}> = 1/\sqrt{2}(|00> + |11>) \qquad\qquad (G.1)$$

The two qubits are either in state |00> or state |11> with equal probability. So if one measures the first qubit and finds that it is in state |0>, then the second qubit is also in state |0>, and vice versa. Similarly, if the first qubit is found to be in state |1>, then the second qubit is also in state |1>, and vice versa. Thus, each qubit is entangled with the other.

If Alice keeps the first qubit and sends the second qubit to Bob, which experiences bit-flip and phase-flip errors, the above entangled state changes to

$$|\beta_{01}> = 1/\sqrt{2}(|01> + |10>) \qquad \text{(bit-flip error)} \qquad (G.2)$$

$$|\beta_{10}> = 1/\sqrt{2}(|00> - |11>) \qquad \text{(phase-flip error)} \qquad (G.3)$$

$$|\beta_{11}> = 1/\sqrt{2}(|01> - |10>) \qquad \text{(bit-flip and phase-flip errors)} \qquad (G.4)$$

Note that the qubits in these states are also entangled. Together with the state $|\beta_{00}>$, they form what is called the Bell basis.